\documentclass[11pt,a4,twoside]{article}
\usepackage[dvips,a4paper,left=2.7cm,right=2.7cm,top=2.2cm,bottom=2.2cm,headsep=1em]{geometry}
\usepackage{fancyhdr}
\usepackage{titlesec}
\usepackage[latin1]{inputenc}
\usepackage{amsmath,amsfonts,mathrsfs,amsbsy}
\usepackage{rotating}
\usepackage[font={small}]{caption}
\usepackage{natbib}
\usepackage{lmodern,slantsc}
\usepackage{maplestd2e}

\DefineParaStyle{Maple Output} \DefineParaStyle{Maple Output}
\DefineCharStyle{2D Math} \DefineCharStyle{2D Output}
\usepackage[breaklinks,
colorlinks=true, linkcolor=blue, citecolor=black, urlcolor=blue,
pdfborder={0 0 0}]{hyperref}
\usepackage{tikz}

\def\vr{\mathbf r}

\def\v0{\boldsymbol{0}}

\newcommand*\mystrut[1]{\vrule width0pt height0pt depth#1\relax}
\newlength{\FigureHeight}
\newlength{\FigureHeightHalf}

\numberwithin{equation}{section}
\titleformat{\section}
{\large\bfseries}{\thetitle.}{0.5em}{}
\titleformat{\subsection}
{\normalfont\itshape\filcenter}{\normalfont\thetitle.}{0.5em}{}
\begin{document}
\setcounter{page}{1}

\title{\vspace{-3em}Is the log-law a first principle result from Lie-group\\ invariance
analysis?\\ {\Large\emph{A comment on the Article by Oberlack
(2001)}}}
\author{M. Frewer$\,^1$\thanks{Email address for correspondence:
frewer.science@gmail.com}$\:\,$, G. Khujadze$\,^2$ \& H.
Foysi$\,^2$\\[0.75em]
\small $^1$ Tr\"ubnerstr. 42, 69121 Heidelberg, Germany\\
\small $^2$ Chair of Fluid Mechanics, Universit\"at Siegen, 57068
Siegen, Germany}
\date{{\small\today}}
\clearpage \maketitle \thispagestyle{empty}

\vspace{-3em}\begin{abstract} \noindent The invariance method of
Lie-groups in the theory of turbulence carries the high
ex\-pectation of being a first principle method for generating
statistical scaling laws.~The purpose of this comment is to show
that this expectation has not been met so far. In particular for
wall-bounded turbulent flows, the prospects for success are not
promising in view of the facts we will present herein.

Although the invariance method of Lie-groups is able to generate
statistical scaling laws for wall-bounded turbulent flows, like
the log-law for example, these invariant results yet not only fail
to fulfil the basic requirements for a first principle result, but
also are strongly misleading. The reason is that not the
functional structure of the log-law itself is misleading, but that
its invariant Lie-group based derivation yielding this function is
what is misleading. By revisiting the study of
\href{http://journals.cambridge.org/action/displayAbstract?fromPage=online&aid=64901&fileId=S0022112000002408}
{Oberlack (2001)}~{\it [M. Oberlack, A~unified approach\linebreak
for symmetries in plane parallel turbulent shear flows. J. Fluid
Mech. 427, pp. 299-328]}\linebreak we will demonstrate that all
Lie-group generated scaling laws derived therein do not convince
as first principle solutions. Instead, a rigorous derivation
reveals complete arbitrariness rather than uniqueness in the
construction of invariant turbulent scaling laws. Important to
note here is that the key results obtained in
\href{http://journals.cambridge.org/action/displayAbstract?fromPage=online&aid=64901&fileId=S0022112000002408}
{Oberlack (2001)} are based on several technical errors, which all
will be revealed, discussed and corrected. The reason and
motivation why we put our focus solely on
\href{http://journals.cambridge.org/action/displayAbstract?fromPage=online&aid=64901&fileId=S0022112000002408}
{Oberlack (2001)} is that it still marks the core study and
central reference point when applying the method of Lie-groups to
turbulence theory. Hence it is necessary to shed the correct light
onto that~study.

Nevertheless, even if the method of Lie-groups in its full extent
is applied and interpreted correctly, strong natural limits of
this method within the theory of turbulence exist, which, as will
be finally discussed, constitute insurmountable obstacles in the
progress of achieving a significant breakthrough.

\vspace{0.5em}\noindent{\footnotesize{\bf Keywords:} {\it
Symmetries, Lie groups, Scaling laws, Turbulence, Law of the wall,
Log-law}}$\,$;\\ {\footnotesize{\bf PACS:} 47.10.-g, 47.27.-i,
05.20.-y, 02.20.Qs, 02.20.Tw}
\end{abstract}

\section{Introduction\label{S1}}

Predicting the mean temporal evolution and spatial structure of
turbulent fluid flow still remains the Holy Grail of systematic
turbulence research for more than a century now. Statistical
scaling laws play a key role in this context. To date, aside from
several inequalities \citep{Constantin94,Constantin99}, the only
exact statistical scaling equality that has been rigorously
derived from {\it first principles}, i.e. by only making sole use
of the Navier-Stokes equations without any major assumptions or
approximations, is, so far, Kolmogorov's 4/5-law for the
third-order longitudinal structure function of the velocity field
\citep{Kolmogorov41.3,Frisch95,Davidson04}. However its exact
validity is only restricted to the asymptotic regime of infinite
Reynolds number within the inertial range of an idealized
unbounded flow, that of statistically homogeneous isotropic flow.

For wall-bounded flows, which inherently give raise to
inhomogeneous turbulent boundary layers, the situation is even
more worse: All attempts to derive scaling laws from {\it
first\linebreak principles} were doomed to failure. Every
derivation done so far necessarily involves such\pagebreak[4]
strong assumptions and approximations that the initially attempted
linkage to the Navier-Stokes equations with its associated
geometrical boundary conditions is lost. However, when withdrawing
the claim of dealing with first principle derivations a vast
functional range of different scaling laws exist.

Unfortunately this lack of mathematical rigor split the turbulence
community into different schools each favoring their own beliefs
in never-ending debates on holding the correct scaling behavior
within fully developed turbulent boundary layers. In particular
when regarding the issue of the functional form for the mean
velocity distribution in the overlap region between the inner
(near-wall) and outer regions. Even in the asymptotic limit of
large Reynolds numbers, theoretical analysis does not predict a
unique statistical law as how the mean velocity profile should
scale within this thin layer from the wall. Two contrasting laws
coexist in the literature: One is the classical log-law where its
coefficients are believed to be universal, i.e. independent of the
Reynolds number and first deduced by \cite{Karman30}, while the
second one is a power-law, where, opposed to universality, its
coefficients are believed to depend weakly on the Reynolds number
\citep{Barenblatt93a,Barenblatt93b,Cipra96,George97,George07,Barenblatt14}.
Usually the log-law has been emphasized over the power-law,
sometimes to the exclusion of the latter
\citep{Zagarola97,Osterlund00,Lindgren04,Jimenez12,Jimenez13}.
However in recent years it shows that maybe also the opposite is a
reasonable possibility, but nevertheless, the question which
scaling law ultimately applies is definitely not resolved yet
\citep{Dallas09,Marusic10,Gad11}.

This situation provokes a different perspective in deriving and
interpreting statistical scaling laws for high-Reynolds-number
turbulent boundary layer flows. Due to an overall lack of
mathematical rigor the still-ongoing debate would greatly benefit
from a twofold change in attitude:

\begin{itemize}
\item[i)] From an attitude which would regard all scaling laws as
only being approximative models to the exact but unknown
Navier-Stokes scaling behavior. Going with the insight of George
Box that essentially ``all models are wrong, but some are useful"
\citep{Box87}, the adequate question to ask is: \emph{How wrong do
statistical scaling laws have to be to not be useful?}
\item[ii)] From an attitude which shares the position of Paul
Feyerabend in that ``all methodologies, even the most obvious
ones, have their limits" \citep{Feyerabend75}. Nevertheless, in
extending or in using new methodologies to construct a theoretical
framework for generating turbulent scaling laws ``anything goes"
\citep{Feyerabend75}, however only under the premiss, of course,
that mathematical consistency is always ensured.
\end{itemize}

Exactly in this context we want to place our comment regarding the
method of Lie-group invariance analysis when applied to the theory
of turbulence. Although this method carries the high expectation
of being a first principle method for generating turbulent scaling
laws, we will show that it faces the same analytical problems as
any other method when trying to generate such laws {\it a priori}.
In other words, within the theory of turbulent flows the invariant
method of Lie-groups needs to be reduced from its high expectation
level of being a first principle method down to a method, which
also, like any other analytical method in turbulence theory, is
only capable to generate turbulent scaling laws {\it a
posteriori}, i.e. on basis of physical experience by making use of
assumptions and approximations. In this sense the invariant method
of Lie-groups has the immanent property, too, to generate next to
useful also {\it non}-useful turbulent scaling laws, which, if no
physical or numerical experiment can be matched to them, need to
be discarded.

The aim of our comment is twofold. At first to draw attention to
the fact that the first-principle approach in \cite{Oberlack01},
namely to generate statistical scaling laws for {\it wall-bounded}
turbulent flows based on the ``maximum symmetry principle", is
heavily misleading.~Moreover, our analysis will show that the
results as they are derived in \cite{Oberlack01} cannot be
reproduced. In one of the two necessary steps to generate the Lie
symmetries in \cite{Oberlack01} we obtain a different intermediate
result, leading thus to a diverse and more general symmetry in
particular for the mean velocity field. Secondly, to clarify that
although the methodology of Lie-group symmetry analysis
\citep{Ovsiannikov82,Stephani89,Fushchich93,Olver93,Ibragimov94,Bluman96,Hydon00,Cantwell02}
is a powerful and indispensable tool to generate non-trivial
solutions for nonlinear differential equations, it nevertheless
brings along strong natural limits when applied to the theory of
turbulence in general. These limits need all to be recognized in
order to correctly assess the feasibility of this method in future
work.

This study is organized as follows: Section \ref{S2} revisits in
detail the results given in \cite{Oberlack01}. We will show that
all statistical scaling laws derived in \cite{Oberlack01} are
based on an incorrectly calculated invariance transformation,
which misleadingly lead, for example, to the log-law as a specific
result for the mean streamwise velocity profile. In contrast to
these published and as we believe incorrect results, our newly
recalculated and cross-validated invariance transformation does
not allow for such a specification. Rather, it results in complete
arbitrariness in the scaling behavior for the mean streamwise
velocity profile, thus showing that the method of Lie-groups is
unable to generate a functionally unique set of turbulent scaling
laws from first principles as it is misleadingly claimed in
\cite{Oberlack01}.

Section \ref{S3} then concludes by generally addressing the limits
of the Lie-group methodology in turb\-ulence research, in
particular when genera\-ting statistical multi-point scaling laws
beyond the one-point case. Next to the limits as statistical
unclosedness and the presence of boundary conditions, the limits
are further driven by the property of intermittency. The issue is
that all multi-point functions are highly sensitive to
intermittent effects, while a Lie-group based invariance analysis
performed upon them, which standardly (up to some exceptions) only
returns a finite dimensional Lie-group, is effectively unable to
account for these effects. Yet, it should be clear that we do not
criticize the method of Lie-groups itself, being a very useful
mathematical tool indeed, when only applied to the right problems.

\vspace{-0.28em}
\section{Phantom turbulent scaling laws as first principle solutions\label{S2}}

As a representative example for a wall-bounded turbulent scaling
law, we will for brevity herein only focus on the invariant
Lie-group based derivation of the log-law for the mean streamwise
velocity profile in the inertial region. In the conclusion of this
section, however, we will briefly compare this log-law derivation
also to its corresponding and equally valid power-law derivation.

At this point we already would like to emphasize that it is not
the log-law itself to be criticized here, which, as we know,
undoubtedly acts as a useful scaling law in the inertial region,
but only its particular derivation and interpretation originating
from the invariance method of Lie-groups. All the more so, as the
claim exists of dealing here with a derivation from first
principles \citep{Oberlack01,Oberlack10}.

Currently there are two different and independent Lie-group based
derivations existing in the literature to generate the log-law.
The first and older derivation is based on the deterministic
fluctuating Navier-Stokes equations \citep{Oberlack01}, while the
second and more recent derivation is based on the infinite
statistical hierarchy of the multi-point velocity correlation
functions \citep{Oberlack10}. Although both derivations claim to
generate the log-law a priori from first principles, i.e. {\it
directly} from the Navier-Stokes equations without any major
assumptions and approximations, it is only the former derivation
in \cite{Oberlack01} that possibly can claim this.

The reason is that the derivation in \cite{Oberlack10} is based on
a statistical description in which the system of equations is
obviously not closed$\,$\footnote[2]{Furthermore, in
\cite{Frewer14.1}, which is supported by the study
\cite{Frewer15.1}, it is shown that the log-law derivation in
\cite{Oberlack10} is essentially based on an unphysical
equivalence transformation.}, that is, the invariance ana\-ly\-sis
in \cite{Oberlack10} is based on a system with more unknown
functions than equations, while in \cite{Oberlack01} the
corresponding invariance analysis at {\it first sight} is
presumably based on an opposite system having more equations than
unknown functions, due to augmenting the unclosed fluctuating
equations by the so-called ``velocity product equations"
[Eq.$\,$(2.12), p.$\,$302].

At first sight it thus seems that \cite{Oberlack01} has a decisive
advantage over \cite{Oberlack10} when taking the {\it perspective
of an invariance analysis} into which the ``velocity product
equations" are incorporated. Because, on the level of the
fluctuating Navier-Stokes equations it seems that with this
approach a natural invariant closure constraint has been achieved,
since ``the purpose of (2.12) [the velocity product equations]
regarding the symmetry properties of plane shear flows is quite
different" [p.$\,$302] and ``that (2.12) is crucial to find
self-similar mean velocity profiles consistent with the second
moment and all higher-order correlation equations" [p.$\,$307],
where ``the major difference between the classical turbulence
modelling approach and the present procedure is the treatment of
equation (2.12)" [p.$\,$309].

However, these statements are misleading. As we will demonstrate,
this additional set of ``velocity product equations" are
completely redundant, not only from the perspective of the
fluctuating equations themselves, but also from the perspective of
any invariance analysis performed upon them. The immediate
consequence is that, instead of the log-law, a completely
indifferent result is obtained, being incapable to gain any
mathematical insight into the statistical solution manifold of the
underlying dynamical process. Since this is also the case for all
other velocity moments, it simply shows that for {\it
wall-bounded} turbulent flows a first principle result based on
the invariance method of Lie-groups is yet still missing. The
obvious reason is that the closure problem of turbulence cannot be
bypassed, also then not when utilizing the analytical invariance
method of Lie-groups.

\subsection{Revisiting the results derived in Oberlack (2001)\label{S2.1}}

If we restrict to non-rotating flows, the transport equations
which were analyzed for invariance in \cite{Oberlack01} are
[Eqs.$\,$(2.7b), (2.11) and (2.12), respectively]
\begin{align}
\mathscr{C} =&\;\frac{\partial u^\prime_k}{\partial x_k}=0,\label{131228:1918}\\
\mathscr{N}_i=&\;\frac{\partial u^\prime_i}{\partial
t}+\bar{u}_1\frac{\partial u^\prime_i}{\partial x_1}+\delta_{i1}
u_2^\prime\frac{d\bar{u}_1}{dx_2}-\delta_{i1}\left(K+\nu\frac{d^2\bar{u}_1}{d
x^2_2}\right)\nonumber\\
& +\;\delta_{i2}\frac{d\bar{p}^*}{d x_2}+\frac{\partial u^\prime_i
u^\prime_k}{\partial x_k}+\frac{\partial p^\prime}{\partial
x_i}-\nu\frac{\partial^2 u^\prime_i}{\partial x_k^2}=0,\label{131228:1921}\\
\mathscr{P}_{ij}=&\;\mathscr{N}_i u^\prime_j +\mathscr{N}_j
u^\prime_i=0,\label{131228:1928}
\end{align}
denoting the continuity $(\mathscr{C})$, the momentum
$(\mathscr{N}_i)$ and the above-discussed ``velocity product"
$(\mathscr{P}_{ij})$ equations  for the {\it fluctuating} fields
of the Reynolds decomposed Navier-Stokes equations. All overbared
quantities denote the corresponding averaged fields. Since the
analysis is restricted to stationary parallel shear flows, the
above set of fluctuating transport equations are augmented by the
defining restrictions for the mean fields
\begin{equation}
\frac{\partial \bar{u}_1}{\partial t}=\frac{\partial
\bar{u}_1}{\partial x_1}= \frac{\partial \bar{u}_1}{\partial x_3}=
\frac{\partial \bar{p}^*}{\partial t}=\frac{\partial
\bar{p}^*}{\partial x_1}=\frac{\partial \bar{p}^*}{\partial
x_3}=\bar{u}_2=\bar{u}_3=0. \label{131228:1919}
\end{equation}
Note that in \cite{Oberlack01} (from Eqs.$\,$(2.7a-b), p.$\,$301
onwards) not the original mean pressure field
$\bar{p}=\bar{p}(x_1,x_2)$ is used, but instead the effective
(lower dimensional) mean pressure field $\bar{p}^*=\bar{p}^*(x_2)$
is used by decomposing the original field as $\bar{p}=-K\cdot
x_1+\bar{p}^*$, where $K$ is the mean constant pressure gradient
to drive the flow in the streamwise $x_1$-direction.

In order to allow for a possible Reynolds number dependence in the
scaling law coefficients, the search in \cite{Oberlack01} to
determine all invariant point-transformations for system
(\ref{131228:1918})-(\ref{131228:1919}) was hence extended to the
class of equivalence transformations, in which the viscosity
$\nu\sim 1/Re$ is not treated as a parameter, but, next to the
space-time coordinates, as an own independent variable.

The corresponding infinitesimal invariance operator
\citep{Stephani89,Olver93,Ibragimov94,Bluman96,Hydon00,Cantwell02}
to generate these Lie-point equivalence transformations for the
complete system (\ref{131228:1918})-(\ref{131228:1919}) is thus
given by the following scalar structure
\begin{equation}
X=\xi_t\partial_t+\xi_{x_i}\partial_{x_i}+\xi_\nu\partial_\nu
+\eta_{u^\prime_i}\partial_{u^\prime_i}+\eta_{p^\prime}\partial_{p^\prime}
+\eta_{\bar{u}_1}\partial_{\bar{u}_1}+\eta_{\bar{p}^*}\partial_{\bar{p}^*}
+ Y, \label{141125:2224}
\end{equation}
where $Y$ is the prolongation in the infinitesimals for all
derivatives appearing in system
(\ref{131228:1918})-(\ref{131228:1928}) {\it including} the
constraints (\ref{131228:1919}):
\begin{align}
Y&=\,\eta_{u^\prime_{i,t}}\partial_{u^\prime_{i,t}}+
\eta_{u^\prime_{i,j}}\partial_{u^\prime_{i,j}}+\eta_{u^\prime_{i,jk}}
\partial_{u^\prime_{i,jk}}+\eta_{p^\prime_{i}}\partial_{p^\prime_{i}}\nonumber\\
&\hspace{2.55cm}+\,\eta_{\bar{u}_{1,t}}\partial_{\bar{u}_{1,t}}
+\eta_{\bar{u}_{1,i}}\partial_{\bar{u}_{1,i}}+
\eta_{\bar{u}_{1,jk}}\partial_{\bar{u}_{1,jk}}+
\eta_{\bar{p}^*_{t}}\partial_{\bar{p}^*_{t}}+
\eta_{\bar{p}^*_{i}}\partial_{\bar{p}^*_{i}}. \label{140618:0050}
\end{align}
The prolongation $Y$ appears due to the already predetermined
transformation rules for all relevant derivatives, as e.g. for
$u^\prime_{i,t}:=\partial_t u^\prime_i$ or
$u^\prime_{i,j}:=\partial_{x_j} u^\prime_i$, etc., given then by
the yet unknown and still to be determined transformation rules
for the corresponding dependent variables with respect to their
independent variables. In other words, all components of $Y$ in
(\ref{140618:0050}) can be explicitly expressed as functions of
both the unknown infinitesimals for the independent variables
$\xi_t$, $\xi_{x_i}$, $\xi_\nu$, and dependent variables
$\eta_{u^\prime_i}$, $\eta_{p^\prime}$, $\eta_{\bar{u}_1}$,
$\eta_{\bar{p}^*}$. The general and explicit formula for a
prolongation of any differential order in terms of its independent
and dependent infinitesimals can be found, for example, in the
above cited literature.

Now, according to \cite{Oberlack01}, the desired equivalence
transformations $X$ (\ref{141125:2224}) for the considered system
(\ref{131228:1918})-(\ref{131228:1919}) are determined by solving,
in respect to the constraints (\ref{131228:1919}), the following
corresponding invariance conditions [Eqs.$\,$(3.12a-c), p.$\,$305]
\begin{align}
X\mathscr{C}\mid_{\mathscr{C}=0} &=0,\label{140618:0005}\\
X\mathscr{N}_i\mid_{\mathscr{N}_i=0} &=0,\label{140618:0006}\\
X\mathscr{P}_{ij}\mid_{\mathscr{P}_{ij}=0} &=0.
\label{140618:0007}
\end{align}
In order to explicitly yield these invariant transformations, the
calculation in \cite{Oberlack01} was performed in two steps. In
the first step, by solving according to the flow assumptions
(\ref{131228:1919}) the invariance conditions (\ref{140618:0005})
and (\ref{140618:0006}), the resulting general equivalence
transformation of subsystem
(\ref{131228:1918})-(\ref{131228:1921}) augmented by
(\ref{131228:1919}) got calculated in infinitesimal form as
[Eq.$\,$(3.14)\footnote[2]{Note that Eq.$\,$(3.14) in
\cite{Oberlack01} is the general result for the {\it non}-rotating
case; the rotating case $\Omega\neq 0$ is only considered later in
Sec.$\,$3.4, p.$\,$313.}, p.$\,$307]:
\vspace{-0.25em}
\begin{equation}
\left. \begin{aligned} \xi_{x_1} =&\;
a_1(\nu)x_1+a_2(\nu)x_3+f_1(t,\nu),\\
\xi_{x_2} =&\; a_1(\nu)x_2+a_3(\nu),\\
\xi_{x_3} =&\;
a_1(\nu)x_3-a_2(\nu)x_1+f_2(t,\nu),\\
\xi_t =&\; a_4(\nu)t+a_5(\nu),\\
\xi_{\nu} =&\; \left[ 2a_1(\nu)-a_4(\nu)\right]\nu,\\[-0.4em]
\eta_{u^\prime_1} =&\; \left[
a_1(\nu)-a_4(\nu)\right]u^\prime_1+a_2(\nu)u^\prime_3+\frac{\partial
f_1}{\partial t}-g_1(x_2,\bar{u}_1,\bar{p}^*,\nu),\\
\eta_{u^\prime_2} =&\; \left[
a_1(\nu)-a_4(\nu)\right]u^\prime_2,\\[-0.2em]
\eta_{u^\prime_3} =&\; \left[
a_1(\nu)-a_4(\nu)\right]u^\prime_3-a_2(\nu)\left[
u^\prime_1+\bar{u}_1\right]+\frac{\partial f_2}{\partial t},\\
\eta_{p^\prime} =&\; 2\left[ a_1(\nu)-a_4(\nu)\right ]p^\prime
-x_1\left[\frac{\partial^2f_1}{\partial
t^2}+\left[a_1(\nu)-2a_4(\nu)\right] K\right]\\[-0.15em]
 & - x_3\left[\frac{\partial^2f_2}{\partial
t^2}-a_2(\nu)K\right]-g_2(x_2,\bar{u}_1,\bar{p}^*,\nu)+f_3(t,\nu),\\
\eta_{\bar{u}_1} =&\; \left[
a_1(\nu)-a_4(\nu)\right]\bar{u}_1+g_1(x_2,\bar{u}_1,\bar{p}^*,\nu),\\
\eta_{\bar{p}^*} =&\; 2\left[
a_1(\nu)-a_4(\nu)\right]\bar{p}^*+g_2(x_2,\bar{u}_1,\bar{p}^*,\nu).
\end{aligned}
~~~ \right \} \label{131228:1839}
\end{equation}

\noindent This result of this first step can be obtained and
confirmed e.g. by using the computer algebra system (CAS) based
symmetry-software-packages GeM \citep{Cheviakov07}, SADE
\citep{Filho11}, or the latest DESOLV-II package \citep{Vu12}.
However, it should be noted that in first instance the latter
DESOLV-II package of \cite{Vu12} is not giving the invariance
(\ref{131228:1839}) in its general form. Instead it restricts the
result (\ref{131228:1839}) to
\begin{equation}
a_2(\nu)=0,\;\;\; g_1(x_2,\bar{u}_1,\bar{p}^*,\nu)=F(x_2,\nu).
\label{131228:1955}
\end{equation}
In the following we will, for reasons of simplicity, only consider
this special restriction (\ref{131228:1955}), which, as a
consequence, will only simplify but not change the general
statement and message of this study. Hence, irrespective of this
choice, whether to restrict (\ref{131228:1839}) by
(\ref{131228:1955}) or not, the first step in \cite{Oberlack01}
provides a reproducible result.

In clear contrast now to the second step, where the remaining
invariance condition (\ref{140618:0007}) for the system of the
``velo\-city product equations" (\ref{131228:1928}) got {\it
incorrectly} recognized in \cite{Oberlack01} as a symmetry
breaking condition within the already analyzed subsystem
(\ref{140618:0005})-(\ref{140618:0006}). This lead (amongst
others) to the following wrong {\it key restriction}
[Eq.$\,$(3.18), p.$\,$308]:
\begin{equation}
g_1(x_2,\bar{u}_1,\bar{p}^*,\nu)=b_1(\nu). \label{131228:1959}
\end{equation}
However, our recalculation of this second step only shows complete
redundancy in the invariant condition for the ``velocity product
equations" (\ref{131228:1928}), rather than a symmetry-breaking
mechanism in the invariances admitted by
(\ref{131228:1918})-(\ref{131228:1921}). That is to say, reduction
(\ref{131228:1959}) is {\it not} supported by our recalculation
(see second part of Appendix \ref{SB}, and entire Appendix
\ref{SC}).

The essential error in \cite{Oberlack01}, not to recognize this
redundancy, already resides in the idea to consider the invariance
conditions of system (\ref{131228:1918})-(\ref{131228:1928}) in
the {\it uncoupled} infinitesimal form
(\ref{140618:0005})-(\ref{140618:0007}). This allowed to split the
invariance analysis into two separate and independent steps. As
was shown above, (\ref{140618:0005}) and (\ref{140618:0006}) were
solved in a first step, followed by solving (\ref{140618:0007}) in
a second step using the results then from the first step. This
reasoning, however, is incorrect in that it's not complete. Since
(\ref{131228:1918})-(\ref{131228:1928}) represents a strongly
coupled system, the corresponding invariance conditions may not be
decoupled as expressed in (\ref{140618:0005})-(\ref{140618:0007}),
but must rather be solved as a coupled, or more precisely, as a
combined functional vector, namely as
\begin{equation}
X\mathscr{T}_{n}\mid_{\mathscr{T}_{n}=0}=0,\;\; \text{with}\;\;
\mathscr{T}_{n}:=(\mathscr{C},\mathscr{N}_i,\mathscr{P}_{ij}),
\label{131228:1947}
\end{equation}
in order to not only guarantee for consistency, but to also allow
for the most general solution, i.e. to allow for the minimal
restricted invariant solution manifold in that system. As a
result, the coupled invariance condition (\ref{131228:1947})
immediately shows that all restrictions coming from the ``velocity
product equations" (\ref{140618:0007}) are essentially redundant
to (\ref{140618:0005}) and (\ref{140618:0006}), and thus not
symmetry-breaking as incorrectly claimed in \cite{Oberlack01}. In
other words, condition (\ref{140618:0007}) has no impact on
evaluating (\ref{140618:0005}) and (\ref{140618:0006}). Because,
splitting equation (\ref{131228:1947}) into its relevant
components \vspace{-1.2em}
\begin{align}
X\mathscr{C}\mid_{\mathscr{T}_n=0} &=0,\label{131228:1950}\\
X\mathscr{N}_i\mid_{\mathscr{T}_n=0} &=0,\label{131228:1951}\\
X\mathscr{P}_{ij}\mid_{\mathscr{T}_n=0} &=0\label{131228:1952},
\end{align}
the last condition (\ref{131228:1952}), which includes
(\ref{140618:0007}), is now satisfied identically when evaluating
its left-hand side
\vspace{-0.8em}
\begin{align}
X\mathscr{P}_{ij}\mid_{\mathscr{T}_n=0}=&\;\Big[ \mathscr{N}_i X
u^\prime_j+ \mathscr{N}_j X u^\prime_i+u^\prime_j
X\mathscr{N}_i+u^\prime_i
X\mathscr{N}_j\Big]\Big\vert_{\mathscr{T}_n=0}\nonumber\\
=&\;\mathscr{N}_i\mid_{\mathscr{T}_n=0}\cdot \Big[X
u_j^\prime\Big]\Big\vert_{\mathscr{T}_n=0}+
\mathscr{N}_j\mid_{\mathscr{T}_n=0}\cdot \Big[X u^\prime_i\Big]
\Big\vert_{\mathscr{T}_n=0}\nonumber\\
&\; + u^\prime_j\cdot
\big[X\mathscr{N}_i\big]\big\vert_{\mathscr{T}_n=0}+
u^\prime_i\cdot
\big[X\mathscr{N}_j\big]\big\vert_{\mathscr{T}_n=0}\nonumber\\
\equiv&\; 0,\label{160905:1938}
\end{align}
by using the condition (\ref{131228:1921}) for the first two
summands, since
$\mathscr{N}_i\mid_{\mathscr{T}_n=0}=\mathscr{N}_i\mid_{\mathscr{N}_i=0}\equiv
0$, and the condition (\ref{131228:1951}), which includes
(\ref{140618:0006}), for the last two summands.

Hence, the correct equivalence transformation for system
(\ref{131228:1918})-(\ref{131228:1919}) is thus either given by
the unrestricted result (\ref{131228:1839}), or by its restriction
(\ref{131228:1955}), {\it but not} if it's restricted by
(\ref{131228:1959}), which, according to \cite{Oberlack01}, would
also go along with two other wrongly enforced restrictions
$f_1(t,\nu)=b_1(\nu)\, t+b_2(\nu)$ and $f_2(t,\nu)=b_3(\nu)$
[Eq.$\,$(3.18), p.$\,$308].~To provide an overview, Table
\ref{tab1} lists the final resulting set of all infinitesimal
generators.~It {\it explicitly} shows and summarizes the
substantial difference between the final results given in
\cite{Oberlack01} and our on the restriction (\ref{131228:1955})
based re-evaluation.~In addition, Appendix \ref{SB} {\it
explicitly} demonstrates that the key invariant transformation for
the mean velocity field is definitely a to-the-restrictions
(\ref{131228:1919}) compatible symmetry for the equations
(\ref{131228:1918})-(\ref{131228:1921}), and, again, that all
remaining equations (\ref{131228:1928}) are redundant and thus
obviously not symmetry~breaking.

\begin{table}
\begin{center}
\def~{\hphantom{0}}
\begin{tabular}{p{6.5cm} c | c p{6.8cm}}\hline\\[0.0em]
$\xi_{x_1}=a_1(\nu)x_1+b_1(\nu)t+b_2(\nu),$ & & &
$\xi_{x_1}=a_1(\nu)x_1+f_1(t,\nu),$\\[0.5em]
$\xi_{x_2}=a_1(\nu)x_2+a_3(\nu),$ & & &
$\xi_{x_2}=a_1(\nu)x_2+a_3(\nu),$\\[0.5em]
$\xi_{x_3}=a_1(\nu)x_3+b_3(\nu),$ & & &
$\xi_{x_3}=a_1(\nu)x_3+f_2(t,\nu),$\\[0.5em]
$\xi_t=a_4(\nu)t+a_5(\nu),$ & & & $\xi_t=a_4(\nu)t+a_5(\nu),$
\\[0.5em]
$\xi_{\nu} =\left[ 2a_1(\nu)-a_4(\nu)\right]\nu,$ & & & $\xi_{\nu}
=\left[ 2a_1(\nu)-a_4(\nu)\right]\nu ,$\\[0.5em]
$\eta_{u^\prime_1}=\left[ a_1(\nu)-a_4(\nu)\right]u^\prime_1 ,$ &
& & $\eta_{u^\prime_1}=\left[
a_1(\nu)-a_4(\nu)\right]u^\prime_1+\partial_t
f_1-F(x_2,\nu),$\\[0.5em]
$\eta_{u^\prime_2}=\left[ a_1(\nu)-a_4(\nu)\right]u^\prime_2 ,$ &
& & $\eta_{u^\prime_2}=\left[
a_1(\nu)-a_4(\nu)\right]u^\prime_2 ,$\\[0.5em]
$\eta_{u^\prime_3}=\left[ a_1(\nu)-a_4(\nu)\right]u^\prime_3 ,$ &
& & $\eta_{u^\prime_3}=\left[ a_1(\nu)-a_4(\nu)\right]u^\prime_3
+\partial_t f_2 ,$\\[0.5em]
$\eta_{p^\prime}=2\left[ a_1(\nu)-a_4(\nu)\right ]p^\prime
\vspace{0.5em}\newline\phantom{xxxx}
+g_2(x_2,\bar{u}_1,\bar{p}^*,\nu)$
\vspace{0.5em}\newline\phantom{xxxx}
$-x_1\left[a_1(\nu)-2a_4(\nu)\right]K+f_3(t,\nu),$ & & &
$\eta_{p^\prime}=2\left[ a_1(\nu)-a_4(\nu)\right
]p^\prime-x_1\partial^2_t f_1 \vspace{0.5em}\newline\phantom{xxxx}
-g_2(x_2,\bar{u}_1,\bar{p}^*,\nu)-x_3\partial^2_t f_2$
\vspace{0.5em}\newline\phantom{xxxx}
$-x_1\left[a_1(\nu)-2a_4(\nu)\right]K+f_3(t,\nu),$\\[0.5em]
$\eta_{\bar{u}_1} =\left[ a_1(\nu)-a_4(\nu)\right]\bar{u}_1
+b_1(\nu),$ & & & $\eta_{\bar{u}_1} =\left[
a_1(\nu)-a_4(\nu)\right]\bar{u}_1 +F(x_2,\nu),$\\[0.5em]
$\eta_{\bar{p}^*} =2\left[ a_1(\nu)-a_4(\nu)\right]\bar{p}^*$
\vspace{0.5em}\newline\phantom{xxxx}
$-g_2(x_2,\bar{u}_1,\bar{p}^*,\nu),$ & & & $\eta_{\bar{p}^*}
=2\left[ a_1(\nu)-a_4(\nu)\right]\bar{p}^*$
\vspace{0.5em}\newline\phantom{xxxx}
$+g_2(x_2,\bar{u}_1,\bar{p}^*,\nu).$
\end{tabular}
\caption{Comparison of the final resulting set of infinitesimal
generators between \cite{Oberlack01} [Eq.$\,$(3.19), p.$\,$308]
(left-hand side) and our re-evaluation ((\ref{131228:1839})
restricted by (\ref{131228:1955})) (right-hand side).}
\label{tab1}
\end{center}
\vspace{-0.4em}\hrule
\end{table}

This change in the result is decisive as it now leads to a
completely different picture in \cite{Oberlack01} when generating
invariant scaling laws for the 1D mean streamwise velocity profile
$\bar{u}_1=\bar{u}_1(x_2)$ in the limit of infinite Reynolds
number ($\nu\rightarrow 0$). Considering the derivation of the
log-law, for which, according to \cite{Oberlack01}, the friction
velocity breaks the scaling symmetry of the mean velocity profile
as $a_1=a_4\neq 0$, the corresponding invariant surface condition
is then no longer given as
\begin{equation}
\frac{dx_2}{a_1x_2+a_3}=\frac{d\bar{u}_1}{b_1},
\end{equation}
which then would lead to the misleading result [Eq.$\,$(3.29),
p.$\,$312]
\begin{equation}
\bar{u}_1(x_2)=d_2\log(x_2+d_1)+C,\;\; \text{with}\;\;
d_1=a_3/a_1,\;\;d_2=b_1/a_1. \label{131228:2000}
\end{equation}
Instead, one obtains the more general and correct invariant
surface condition
\begin{equation}
\frac{dx_2}{a_1x_2+a_3}=\frac{d\bar{u}_1}{F(x_2)},
\label{141024:1838}
\end{equation}
which in contrast to (\ref{131228:2000}) leads to complete
arbitrariness in the result
\begin{equation}
\bar{u}_1(x_2)=\int\!\!\frac{F(x_2)}{a_1x_2+a_3}dx_2+C,
\label{131228:2001}
\end{equation}
as the function $F(x_2)$ is not closer specified, except for only
being at least once integrable. Hence the derivation of the
invariant log-law (\ref{131228:2000}) is completely misleading in
\cite{Oberlack01} as the correct equivalence transformation of the
system (\ref{131228:1918})-(\ref{131228:1919}) does not allow for
such a unique functional specification. It rather induces the
generation of any desirable function for the mean velocity profile
as given in (\ref{131228:2001}), making this result thus
essentially useless. The same holds true not only for the log-law,
but for {\it all} invariant scaling laws which were derived in
\cite{Oberlack01}, including, for example, the exponential scaling
law in the wake region of turbulent boundary layer flow which were
later extensively utilized in the papers by \cite{Khujadze04} and
\cite{Guenther05}.

Considering, for example, the following statements made in
\cite{Oberlack01} such as~that

\vspace{0.5em}\noindent ``In the case of the logarithmic law of
the wall, the scaling with the distance from the wall arises as a
result of the analysis and has {\it not} been assumed in the
derivation." [p.$\,$299],

\vspace{0.5em}\noindent or ``... important to note that group
theoretical arguments very much guide the finding ... where the
mean velocity profiles are applicable." [p.$\,$306],

\vspace{0.5em}\noindent or ``This is an assumption in the
classical derivation [von-Kármán-derivation] of the log law of the
wall {\it but} is a result of the present analysis
[Oberlack-(2001)-derivation]." [p.$\,$312],

\vspace{0.5em}\noindent or ``The theory is fully algorithmic and
{\it no} intuition is needed to find a self-similar mean velocity
profile." [p.$\,$321],

\vspace{0.5em}\noindent we have to conclude, in retrospect, that
all these statements are {\it not} correct and are even misleading
if the {\it correct} result (\ref{131228:2001}) is not being put
forward. That is, without considering the generalized result
(\ref{131228:2001}), the study in Oberlack (2001) basically gives
a wrong impression for using the method of Lie groups in
turbulence theory. Because, when performing the analysis of
Lie-groups for turbulent flows correctly, in the end one still has
to make strong assumptions and to use more than a strong intuition
in order to get from a general expression as (\ref{131228:2001})
the correct statistical scaling law in the way as it would be
proposed by a corresponding physical or numerical experiment.

For example, Oberlack (2001) claims that under certain conditions
which are assumed to be valid close to the wall, either the
algebraic-law ``(3.27)" [Eq.$\,$(3.27), p.$\,$312] or the
log-\linebreak law ``(3.29)" [Eq.$\,$(3.29), p.$\,$312] emerges as
a functionally unique result when employing the method of
Lie-groups, namely by adjusting only a finite set of group
parameters towards compatible values. However, our re-evaluated
and correct result for the algebraic-law conditions ($a_1\neq
a_4\neq 0$ and $F(x_2)\neq0$)
\begin{equation}
\left.
\begin{aligned}
&\hspace{2.25cm}\frac{dx_2}{a_1 x_2+a_3}=
\frac{d\bar{u}_1}{(a_1-a_4)\bar{u}_1+F(x_2)}\\[0.5em]
\Leftrightarrow\quad\; &
\bar{u}_1(x_2)=(a_1x_2+a_3)^{1-a_4/a_1}\cdot\left(\int
\frac{F(x_2)}{(a_1x_2+ a_3)^{2-a_4/a_1}} dx_2+C\right),
\end{aligned}
~~~ \right \} \label{141024:1826}
\end{equation}
and the correct result (\ref{141024:1838})-(\ref{131228:2001}) for
the log-law conditions ($a_1=a_4\neq 0$ and $F(x_2)\neq0$)

\begin{equation}
\frac{dx_2}{a_1x_2+a_3}=\frac{d\bar{u}_1}{F(x_2)}\quad\;
\Leftrightarrow\quad\;
\bar{u}_1(x_2)=\int\!\!\frac{F(x_2)}{a_1x_2+a_3}dx_2+C,
\label{141024:1842}
\end{equation}
both show that this is not the case. The reason is that
(\ref{141024:1826}) as well as (\ref{141024:1842}) each form a
completely indifferent result: Any arbitrary (integrable) function
$F(x_2)$ can be chosen to represent after its integration the law
of the wall.~That means, from the perspective~of Lie-group theory,
neither the algebraic-law nor the log-law is a special or
privileged scaling law; any desirable function can be derived from
this theory, e.g. choosing $F(x_2)=\text{const.}$ in
(\ref{141024:1826}) or (\ref{141024:1842}) gives of course the
algebraic-law ``(3.27)" or the log-law ``(3.29)" respectively. But
this choice for $F(x_2)$ is not privileged, one can also choose
e.g.~a complicated hypergeometric

\newpage\noindent function, then
leading to a law-of-the-wall which behaves not like a pure
power-law or a log-law, but as an integrated hypergeometric
function, thus ultimately having infinitely many different but
equally privileged laws of the wall.

Hence, neither the power-law ``(3.27)" nor the log-law ``(3.29)"
is a first-principle result from Lie-group theory. In Oberlack
(2001) both these laws were just derived under wrong conditions,
in that a mathematical correct theory has been unfortunately
misapplied. In this regard the reader should note that all these
problems initially already exist in the earlier publication
\cite{Oberlack99.1} which only then got generalized in
\cite{Oberlack01}.

In summary, the open problem is that turbulence theory should
predict a certain scaling function for the one-dimensional mean
velocity profile $\bar{u}_1(x_2)$ {\it a priori}, which in our
opinion, even for a particular flow regime, has not been achieved
yet. For example, choosing the sophisticated method of Lie-groups
to construct e.g. a scaling law in the vicinity of the wall, will
only give the useless result (\ref{141024:1826}) or
(\ref{141024:1842}), which in each case is just an alternative yet
more complicated representation for the unknown one-dimensional
function $\bar{u}_1(x_2)$. One thus gained nothing in using
Lie-groups, one just shifted the problem from one unknown function
$\bar{u}_1(x_2)$ to another unknown function $F(x_2)$.~However, in
\cite{Oberlack01} the result for the law-of-the-wall is that
$F(x_2)$ {\it must} be a constant function in the wall-normal
coordinate $x_2$, which definitely {\it is} an assumption, and
thus forms a contradiction to all relevant statements made
in~\cite{Oberlack01}, in particular to the one that ``the theory
is fully algorithmic and {\it no} intuition is needed to find a
self-similar mean velocity profile." [p.$\,$321].

In this sense Lie-group theory offers no answer, nor does it give
any prediction {\it a priori} in how turbulence should scale.~This
failure simply reflects the classical closure problem of
turbulence, which, also with Lie-group theory, cannot be
bypassed.~However, using this~method to nevertheless get such an
answer would be the same as guessing it, and if one knows what to
expect {\it a posteriori} then, of course, one can manually
arrange everything backwards, and pretend that theory is
predicting this result. But such an approach has nothing to do
with~science.

This concludes the section on deriving a specific turbulent
scaling law when using the invariance method of Lie-groups. It
showed that although this method carries the high expectation of
being a first principle method in turbulence theory, a convincing
result is yet still missing in the literature to date --- and
remains very questionable if such an analytical result from first
principles is achievable at all.

While the invariance analysis just presented is set in the broader
context of an analysis on Lie-groups in general, we would like to
close this investigation by pointing out all natural limits this
method faces when using it within the theory of turbulence.

\vspace{-0.25em}
\section{The limits of Lie-group methodology in the theory of
turbulence\label{S3}}

\vspace{-0.25em}\noindent Regarding the limits in using the
invariance theory of Lie-groups for differential equations in
turbulence research, the main fact should be faced that this
analytical approach experiences effectively the same
closure-problem issues as the corresponding statistical transport
equations for the correlation moments of the flow fields. Changing
even to a possible nonlocal integral and {\it formally} closed
framework, e.g.~to the higher statistical level of probability
densities in order to allow for a {\it formally} complete and
fully determined statistical description, as in the framework of
the infinite Lundgren-Monin-Novikov chain of equations
\citep{Lundgren67,Friedrich12} or the functional Hopf equation
\citep{Hopf52,McComb90}, will only render every systematic
invariance analysis nearly infeasible. In particular, because the
usage of a nonlocal integral framework, the accompanying existence
of several internal and for wall-bounded flows also additional
geometric boundary conditions, are all obstructive for an analysis
based on Lie-groups, as they all favor the mechanism of symmetry
breaking \citep{Frewer15.1,Frewer16.2,Frewer16}. Thus the {\it
main} possibilities of the Lie-group methodology are already
exhausted at~this~point.

A further serious limit for the methodology of Lie-groups in
turbulence, in particular as standardly used since the key paper
by \cite{Oberlack01}, e.g. as in
\cite{Oberlack03,Khujadze04,Guenther05,Oberlack06,Oberlack10,
Oberlack13.1,Waclawczyk13,Oberlack14} and \cite{Oberlack14.1}, is
in the direction to consistently aim at generating {\it useful}
statistical scaling laws for all higher order moments of a certain
$n$-point correlation function in the inertial range. Even when
applying a Lie-group based invariance analysis for a complex
spatiotemporal (spatially nonlocal and temporally chaotic) system
correctly, the result is standardly only given by a finite
dimensional Lie-group, particularly in elements as scaling or
translation of flow fields, i.e.~the result is in general (up to
some exceptions) repeatedly only given by invariant
transformations of global nature, and predominantly is due to the
non-integrability of these systems. The consequence are global
invariant scaling functions, which are not only incapable of
locating the domain where they should apply, but also in which, in
a na\"ive way, all higher moments are iteratively connected by
constant global group parameters. It is therefore obvious that an
extremely complex and omnipresent spatiotemporal property as that
of intermittency cannot be captured in this restrictive manner. An
invariantly constructed statistical set of {\it global} scaling
laws simply cannot account for the complicated intermittent
behavior which is constantly observed even in the inertial range
of constant energy flux.

Exceptions can be found of course in lower dimensional systems, as
e.g.~in the Burgers equation.~However, the nonlinear Burgers
equation is in itself a very specific and special case as it's
connected to the linear heat equation via the Cole-Hopf
transformation (see e.g.~\cite{Kevorkian96}).~It therefore not
only allows for an infinite dimensional Lie-group due to the
linear superposition principle, but also to {\it formally} express
the general solution of the Burgers equation in a closed form,
which straightforwardly then also extends to any statistical
description of the Burgers equation, as e.g.~to the statistical
functional Hopf approach --- a result somehow or other not
recognized by \cite{Waclawczyk13} as several formal particular
solutions of the Burgers-Hopf equation get calculated although the
{\it general} solution has been already formulated in
\cite{Hosokawa70}. For more details, we refer to our comment
\cite{Frewer16.2} and to its reaction \cite{Frewer.X3}.

Although the origin of intermittency is still object of research,
debates and conjectures there is nevertheless strong evidence that
intermittent behavior in a complex system as found in all
turbulent flows is closely related to the continuous dynamical
breaking of symmetries induced by the system itself
\citep{Tsinober13,Saint13,Brading03,Kurths95,Frisch85}. This
breaking mechanism however acts such that when the system finally
reaches a statistically fully developed state the broken
symmetries are not {\it globally} restored in a statistical sense
\citep{Frisch95,Biferale03}. For example in the inertial range of
turbulence the results clearly show that the flow cannot be
globally invariant under scaling, neither in a deterministic nor
in a statistical sense. Inertial range intermittency when measured
with the longitudinal multi-point structure functions show a clear
lack of global statistical self-similarity. This breakdown of
global scale invariance in the structure functions is expressed by
an anomalous or multifractal scaling behavior. The general rule
is, the higher the order of the structure functions, the larger is
the departure from global statistical self-similarity, that is,
inertial-range intermittency becomes more important for
higher-order moments and becomes even more pronounced the larger
the Reynolds number gets \citep{Falkovich06}.

But in this regime intermittency not only breaks global
statistical self-similarity, also global statistical isotropy may
break, however then in a more weaker sense. The reason is that a
noncompact group as that of scale invariance is more prone to be
broken than a compact group as that of rotation invariance
\citep{Frisch83}. A statistical anisotropy is measured by
comparing the scaling exponents between the longitudinal and the
transversal structure functions, which turn out to be different
even for an initially prepared homogeneous isotropic flow
\citep{Biferale05,Biferale08,Benzi10}. A recent theoretical
investigation could show that this difference does not seem to
depend on the Reynolds number, i.e. that the weak but existent
breaking of isotropy in the inertial range is essentially not
based on a finite Reynolds number effect \citep{Grauer12}. Hence
there is a strong indication that next to global self-similarity
also global isotropy may not be statistically restored in the
limit of infinite Reynolds numbers, as was originally and
contrarily postulated in both the K41-theory (return to global
self-similarity {\it and} isotropy, Kolmogorov 1941) and in the
later refined K62-theory (only return to global isotropy,
Kolmogorov 1962).

It is important to mention here that the dynamical process of
symmetry-breaking inside such an intermittent-evolving system does
not remove the property of scaling per se. Quite the contrary, it
rather allows for fundamentally new and more complex scaling laws,
which all unfold on a {\it local} level without being necessarily
connected to an underlying {\it global} symmetry or {\it global}
invariance principle
\citep{Frisch91,Benzi93,Sreenivasan91,Lvov00,Fujisaka01,Chakraborty10}.

With this knowledge at hand, it is thus highly questionable if
reasonable results can be gained at all when applying the method
of Lie-groups onto any set of multi-point equations in the sense
of Oberlack et al., as it standardly only leads to finite
dimensional transformation groups involving constant (globally
valid) group parameters.~A result, which, with reasonable
certainty, is unable to produce the correct scaling behavior of
the turbulent velocity structure functions, all the more so, as
with each increase in their order the intermittent behavior
consecutively becomes more pronounced. Anyhow, ignoring this
insight by continuously using these finite dimensional (global)
scaling Lie-groups would in fact only give the same wrong results
as a direct dimensional analysis would do.

\appendix
\titleformat{\section}
{\large\bfseries}{Appendix \thetitle.}{0.5em}{}
\numberwithin{equation}{section}

\section{Complete list of all technical errors in
Oberlack (2001)\label{SA}}

This section will show all other mistakes that can be found in
\cite{Oberlack01}. They will be listed, discussed and corrected in
the order as they appear in the text. To note is that all mistakes
shown are independent of each other and, interestingly, also
independent of the main mistake, which was discussed in Section
\ref{S2.1} and which for completeness will be repeated here again
briefly.

\emph{\textbf{(1):}} The first technical mistake appearing in
\cite{Oberlack01} is also the key mistake done in that study. As
was elaborately discussed in Section \ref{S2.1}, the considered
set of\linebreak Eqs.$\,$(3.12a)-(3.12c) are incorrect in the way
that they are too restrictive in not providing the full and
complete set of Lie-point symmetries which the underlying system
of differential equations Eqs.$\,$(3.6a)-(3.6c) can admit. In
order to obtain this complete set, the incorrect
Eqs.$\,$(3.12a)-(3.12c) have to replaced by the correct ones
\eqref{131228:1950}-\eqref{131228:1952}. In solving these
equations, we additionally could prove that Eq.$\,$(3.12c) is
redundant to the information already provided by
Eqs.$\,$(3.12a)-(3.12b), in particular not having the ability to
break any symmetries as it is incorrectly claimed in
\cite{Oberlack01}. Our proof to this redundancy statement is given
from three different perspectives: The proof \eqref{160905:1938}
by only considering the determining equations
\eqref{131228:1950}-\eqref{131228:1952}, then the analytic proof
in Appendix \ref{SB} by also including the underlying differential
equations \eqref{131228:1918}-\eqref{131228:1928}, and finally the
proof in Appendix \ref{SC} via a full systematic symmetry analysis
assisted by a computer algebra system (CAS).

The negative consequence of correcting the symmetry analysis in
\cite{Oberlack01} according to the above mentioned criteria is the
complete arbitrariness when constructing invariant turbulent
scaling laws, as was demonstrated at the end of Section
\ref{S2.1}. Ultimately this just reflects the closure problem of
turbulence, a problem that also the method of Lie-group symmetry
analysis cannot solve or bypass.

\emph{\textbf{(2):}} In the first line of p.$\,$308 it is said,
referring to the result obtained in Eq.$\,$(3.15), that ``the last
two terms do not contribute to the constraints for the
infinitesimals since $\mathscr{N}_iu_j+\mathscr{N}_ju_i$ may be
factored out". This statement we cannot confirm. A correct
analysis rather shows the opposite: Not the last two but rather
the first two terms do not contribute. In fact, the last two terms
just reduce to the already given constraints Eq.$\,$(3.12b). This
finding we will now explicitly demonstrate for the diagonal term
$i=j=1$ which, of course, can be easily transferred to the two
other diagonal and then extended to all off-diagonal terms.

Note that for the following proof we only make use of the
information and notation as given in~\cite{Oberlack01}. Starting
point is Eq.$\,$(3.15), which itself is derived from
Eq.$\,$(3.12c) by carrying out the product rule of differentiation
and thus then to be notated correctly as
\begin{equation}
\Big(\mathscr{N}_i Xu_j+\mathscr{N}_j Xu_i+u_jX\mathscr{N}_i+u_iX
\mathscr{N}_j\Big)\bigg\vert_{(\mathscr{N}_iu_j+\mathscr{N}_ju_i)=0}=0.
\label{160905:2216}
\end{equation}
We will see that it is important not to suppress the preassigned
evaluation constraint of this equation, as it was unfortunately
done in \cite{Oberlack01} with the consequence that important
information got lost. For example, for the diagonal term $i=j=1$
the above constraint reduces to
\begin{equation}
\Big(\mathscr{N}_1
Xu_1+u_1X\mathscr{N}_1\Big)\bigg\vert_{(\mathscr{N}_1u_1)=0}=0,
\label{160905:2137}
\end{equation}
where we divided the equation and its evaluation constraint by 2.
Looking now at the evaluation constraint $\mathscr{N}_1u_1=0$ more
carefully, we see that we can only conclude that
$\mathscr{N}_1=0$, since obviously $u_1\neq 0$. Hence, equation
\eqref{160905:2137} can be equivalently written and thus evaluated
as
\begin{align}
0 &\;\; =\, \Big(\mathscr{N}_1
Xu_1+u_1X\mathscr{N}_1\Big)\bigg\vert_{\mathscr{N}_1u_1=0}
\nonumber\\[0.5em]
& \underset{u_1\neq 0}{=}\, \Big(\mathscr{N}_1
Xu_1+u_1X\mathscr{N}_1\Big)\bigg\vert_{\mathscr{N}_1=0}=
 \Big(\mathscr{N}_1
Xu_1\Big)\bigg\vert_{\mathscr{N}_1=0}+\Big(u_1X\mathscr{N}_1\Big)\bigg\vert_{\mathscr{N}_1=0}
\nonumber\\[0.5em]
&\;\; =\, \underbrace{\mystrut{3.0ex}\mathscr{N}_1
\bigg\vert_{\mathscr{N}_1=0}}_{=0}\cdot\:
\Big(Xu_1\Big)\bigg\vert_{\mathscr{N}_1=0}+u_1\cdot
\Big(X\mathscr{N}_1\Big)\bigg\vert_{\mathscr{N}_1=0}
\;\;\;\;\underset{u_1\neq 0}{=}\;\;
\Big(X\mathscr{N}_1\Big)\bigg\vert_{\mathscr{N}_1=0},\label{160905:2214}
\end{align}
leading to the result that the constraint for the ``velocity
product equation" \eqref{160905:2137} reduces to the given
constraint $(X\mathscr{N}_1)\vert_{\mathscr{N}_1=0}=0$ of
Eq.$\,$(3.12b). Hence, since the above conclusion easily transfers
to all other diagonal and then also naturally to all off-diagonal
terms, this procedure shows again that the praised constraints of
the ``velocity product equations" Eq.$\,$(3.12c) are just
redundant to the constraints Eq.$\,$(3.12b) already given by the
momentum equations and thus not symmetry-breaking as incorrectly
claimed through the (wrongly) induced constraint equations
Eq.$\,$(3.18).

Moreover, the evaluation \eqref{160905:2214} also demonstrates
that it is the last two terms in~\eqref{160905:2216} that give a
contribution and not the first two terms, which, oppositely as
claimed in \cite{Oberlack01}, evaluate to zero. Also the statement
that ``$\mathscr{N}_iu_j+\mathscr{N}_ju_i$ may be factored out"
[p.$\,$308] is not visible at all.

\emph{\textbf{(3):}} Let's assume that Eq.$\,$(3.16) is a
constraint equation as claimed in \cite{Oberlack01} and not an
identity equation as proven before in \emph{\textbf{(2)}}. Then,
nevertheless, it is not justified to deduce Eq.$\,$(3.17) from
Eq.$\,$(3.16) when evaluating it for
$\mathscr{N}_iu_j+\mathscr{N}_ju_i=0$ as by Eq.$\,$(2.12). For
simplicity, we will show this again only for the diagonal term
$i=j=1$; for all other diagonal and off-diagonal terms the same
argument applies:
\begin{align}
0 \,\underset{(3.16)}{=}\, \mathscr{N}_1 \eta_{u_1} &
\,\underset{(3.14)}{=}\,
\Big[a_1(\nu)-a_4(\nu)\Big]\mathscr{N}_1u_1+\mathscr{N}_1\Big[a_2(\nu)u_3+\frac{d
f_1}{d t}-g_1(x_2,\bar{u}_1,\bar{p},\nu)\Big]\nonumber\\[0.5em]
& \!\!\underset{\mathscr{N}_1 u_1=0}{=}\,\,
\mathscr{N}_1\Big[a_2(\nu)u_3+\frac{d f_1}{d
t}-g_1(x_2,\bar{u}_1,\bar{p},\nu)\Big],\label{160906:0803}
\end{align}
where it is now {\it not} justified to conclude that the sum of
the terms in the square bracket has to be zero, as incorrectly
proposed by \cite{Oberlack01} in putting forward Eq.$\,$(3.17) ---
note that the second unjustified constraint in Eq.$\,$(3.17) is
obtained when considering the diagonal term $i=j=3$. The reason
why this choice is not justified is again simply because
$\mathscr{N}_1$ itself is zero when using the evaluation
constraint $\mathscr{N}_1 u_1=0$, which is equivalent to
$\mathscr{N}_1=0$ since $u_1\neq 0$. In fact, as already said and
proven in \emph{\textbf{(2)}}, Eq.$\,$(3.16) and thus
\eqref{160906:0803} is not a constraining equation but only a zero
identity.

\emph{\textbf{(4):}} The statement that ``only the equations for
$R_{22}$ in (B1), $\overline{pu_2}$ in (B3) and $\overline{u_2p}$
in (B4) need to be examined, because these equations decouple from
the other components in the tensor equations" [p.$\,$325] is only
true for the non-rotating case. For the general rotating case
$\Omega_k\neq 0$ in (B1)-(B4) the equations do {\it not} decouple
from the other components as claimed. Hence, the result obtained
in (B8) is thus only valid for the case $\Omega_k=0$.

\emph{\textbf{(5):}} The result (B8), although correctly obtained
from a group classification of the equations $R_{22}$ in (B1),
$\overline{pu_2}$ in (B3) and $\overline{u_2p}$ in (B4), and only
valid in the case for $\Omega_k=0$ (see comment
\emph{\textbf{(4)}} above), is {\it not} uniquely connected to the
proposed symmetry (B7), as the conclusion in Sec.$\,$B.1
[pp.$\,$325-327] in \cite{Oberlack01} misleadingly tries to
suggest. Because, as can be readily seen from the following
(stationary and non-rotating inviscid parallel flow) equations for
$R_{22}$ in (B1),
\begin{align}
0= -\big[\bar{u}_1(x_2+r_2)-\bar{u}_1(x_2)\big]\frac{\partial
R_{22}}{\partial
r_1}&-\left[\frac{\partial\overline{pu_2}}{\partial
x_2}-\frac{\partial\overline{pu_2}}{\partial
r_2}+\frac{\partial\overline{u_2p}}{\partial
r_2}\right]\nonumber\\[0.5em]
& -\frac{\partial R_{(22)2}}{\partial x_2}+ \frac{\partial
R_{(2k)2}}{\partial r_k} - \frac{\partial R_{2(2k)}}{\partial
r_k},
\end{align}
for $\overline{pu_2}$ in (B3)
\begin{align}
\frac{\partial^2 \overline{pu_2}}{\partial x^2_2}
-2\frac{\partial^2 \overline{pu_2}}{\partial x_2\partial r_2}
+\frac{\partial^2 \overline{pu_2}}{\partial r_k\partial r_k} =\, &
\: 2\frac{d\bar{u}_1}{dx_2}\frac{\partial R_{22}}{\partial r_1}
-\left[\frac{\partial^2 R_{(22)2}}{\partial
x^2_2}-2\frac{\partial^2 R_{(2k)2}}{\partial x_2\partial
r_k}+\frac{\partial^2 R_{(kl)2}}{\partial r_k\partial r_l}
\right],
\end{align}
and for $\overline{u_2p}$ in (B4)
\begin{equation}
\frac{\partial^2\overline{u_2p}}{\partial r_k\partial r_k}=-2
\frac{d\bar{u}_1(x_2+r_2)}{dx_2}\frac{\partial R_{22}}{\partial
r_1}-\frac{\partial^2 R_{2(kl)}}{\partial r_k\partial r_l},
\end{equation}
the particularly considered Lie-point symmetry (B7) can be
straightforwardly generalized, for example,~to
\begin{equation}
\left.
\begin{aligned}
\xi_{r_1}=q_1r_1+q_2,\quad \xi_{r_2}=q_1r_2,\quad
\xi_{r_3}=q_1r_3+q_3,\quad \xi_{x_2}=q_1x_2+q_4,\\[0.5em]
\eta_{R_{22}}=q_5R_{22}+\Gamma_{22}(x_2,r_2,r_3),\quad
\eta_{\overline{pu_2}}=q_6\overline{pu_2}-\Lambda^{\!{\scriptscriptstyle
(1)}}_2(x_2,r_1,r_2,r_3),\\[0.5em]
\eta_{\overline{u_2p}}=q_6\overline{u_2p}-\Lambda^{\!{\scriptscriptstyle
(2)}}_2(x_2,r_1,r_2,r_3),\\[0.5em]
\eta_{R_{(kl)2}}=q_6R_{(kl)2}+\delta_{kl}\Lambda^{\!{\scriptscriptstyle
(1)}}_2(x_2,r_1,r_2,r_3),\\[0.5em]
\eta_{R_{2(kl)}}=q_6R_{2(kl)}+\delta_{kl}\Lambda^{\!{\scriptscriptstyle
(2)}}_2(x_2,r_1,r_2,r_3),
\end{aligned}
~~~\right\}
\end{equation}
without effecting or changing the constraint equation (B8) for the
mean flow $\bar{u}_1$. The problem with the generalized symmetry
now is that we are faced with complete arbitrariness due to the
unknown functions $\Gamma_{22}$, $\Lambda^{\!{\scriptscriptstyle
(1)}}_2$ and $\Lambda^{\!{\scriptscriptstyle (2)}}_2$: For
example, for the two-point velocity correlation $R_{22}$, due to
the presence of the unknown and thus arbitrary function
$\Gamma_{22}$, {\it any} arbitrary scaling law or scaling
dependency, in particular in the inhomogeneous direction~$x_2$,
can be generated now by also showing full compatibility to the
log-law ($q_5=q_6$) or to a power-law ($q_5\neq q_6$) for the mean
flow field $\bar{u}_1$ as induced by the constraint equation (B8),
and not only solely by the particular scaling dependency as
misleadingly proposed in (B11) and (B13).

In other words, the Lie-group symmetry method cannot provide an
answer as how, for example, the higher-order moment $R_{22}$
(being the diagonal Reynolds stress $\tau_{22}$ in the one-point
limit~$\vr\to\boldsymbol{0}$) should scale if we assume a log-law
or a power-law for the mean flow field according to the constraint
(B8), or to its integrated form (B9). For that, modelling
procedures and exogenous information from numerical simulations or
physical experiments are necessarily needed to get further
insights. Of course, this problem of arbitrariness in invariant
scaling we not only face for~$R_{22}$, but also for all other
higher-order correlations as $\overline{u_2p}$, $R_{2(ij)}$, etc.
Ultimately this just reflects the closure problem of turbulence,
which, as we have clearly demonstrated in this comment, cannot be
solved or bypassed by the method of Lie-groups alone, as
misleadingly proclaimed in \cite{Oberlack01} and then used to
justify all later results, e.g., in
\cite{Oberlack03,Khujadze04,Oberlack06}. The even later studies,
e.g., in \cite{Oberlack10,Oberlack13.1,Oberlack14.1} and
\cite{Oberlack14} also suffer from the additional problem that new
unphysical symmetries are generated, which in turn violate the
classical principle of cause and effect. For more details, we
refer to our other comments and reviews,\linebreak
\cite{Frewer14.1,Frewer15.0,Frewer15.1,Frewer16.1,Frewer16.2,Frewer16,Frewer16.3},
and to our reactions in \cite{Frewer.X1,Frewer.X2,Frewer.X3} and
\cite{Frewer.X4}.

\section[Explicit symmetry validation without CAS]
{Explicit symmetry\footnote[2]{Note that since the considered
system of equations (\ref{131228:1918})-(\ref{131228:1919}) is
underdetermined (unclosed) {\it all} admitted invariances only act
as equivalence transformations and {\it not} as symmetry
transformations
\citep{Meleshko96,Ibragimov04,Bila11,Chirkunov12,Frewer14.1}.
Although in this section as well as in the next section we will
designate these invariant transformations as symmetries, we have
to keep in mind that in a strict sense they are not symmetries but
only equivalences.} validation without CAS\label{SB}}

This section consists of two parts, where the result of the second
part will depend on the result of the first part. In the first
part we will demonstrate that the key transformation for the mean
velocity field $\bar{u}_1$ (which is taken as a sub-group from the
Lie-group transformation on the right-hand side of Table
\ref{tab1}) is definitely admitted as a symmetry transformation of
the equations (\ref{131228:1918})-(\ref{131228:1921}) and that
it's compatible to all restrictions (\ref{131228:1919}).~Then, in
the second part, we will easily see that at this stage the
remaining ``velocity product equations" (\ref{131228:1928}) are
just redundant to the already transformed momentum equations
(\ref{131228:1921}) and thus not symmetry breaking as contrarily
claimed in \cite{Oberlack01}. Both parts will be performed without
using a computer algebra system (CAS).

\vspace{1em}\noindent {\bf Part I.} The key sub-group
transformation we want to consider is obtained if we put all group
functions to zero (on the right-hand side of Table~\ref{tab1}),
except $F\neq 0$, which, for simplicity, we want to restrict only
to a pure $x_2$-dependence. In non-infinitesimal form the
transformation is thus given as the following arbitrary
$x_2$-dependent translation of the mean and fluctuating streamwise
velocity fields
\begin{equation}
\left. \begin{aligned} \mathsf{T}: & \;\;\; \tilde{t}=t,\;\;\;
\tilde{x}_i=x_i,\;\;\; \tilde{\nu}=\nu,\;\;\;
\tilde{u}^\prime_i=u_i^\prime-\delta_{i1} F(x_2),\;\;\;
\tilde{p}^\prime=p^\prime,\\ & \;\;\; \tilde{\bar{u}}_1=\bar{u}_1+
F(x_2),\;\;\;\tilde{\bar{p}}^*=\bar{p}^*.
\end{aligned}
~~~\right \} \label{141113:2112}
\end{equation}
Now, when inserting this transformation into the continuity
equation (\ref{131228:1918}) we obtain the invariant result
\begin{equation}
0=\mathscr{C}=\frac{\partial u_k^\prime}{\partial x_k}=
\frac{\partial \tilde{u}_k^\prime}{\partial
\tilde{x}_k}+\delta_{k1}\delta_{k2}\frac{d
F(\tilde{x}_2)}{d\tilde{x}_2}=\frac{\partial
\tilde{u}_k^\prime}{\partial \tilde{x}_k}=\tilde{\mathscr{C}},
\label{141114:0846}
\end{equation}
which is also the case for the momentum equations
(\ref{131228:1921})
\begin{align*}
0=\mathscr{N}_i =&\;\frac{\partial u^\prime_i}{\partial
t}+\bar{u}_1\frac{\partial u^\prime_i}{\partial x_1}+\delta_{i1}
u_2^\prime\frac{d\bar{u}_1}{dx_2}-\delta_{i1}\left(K+\nu\frac{d^2\bar{u}_1}{d
x^2_2}\right)\nonumber\\
&\; +\;\delta_{i2}\frac{d\bar{p}^*}{d x_2}+\frac{\partial
u^\prime_i u^\prime_k}{\partial x_k}+\frac{\partial
p^\prime}{\partial x_i}-\nu\frac{\partial^2 u^\prime_i}{\partial
x_k^2}\hspace{6.15cm}
\end{align*}
\begin{align}
\qquad\; = &\; \frac{\partial \tilde{u}^\prime_i}{\partial
\tilde{t}}+\tilde{\bar{u}}_1\frac{\partial
\tilde{u}^\prime_i}{\partial
\tilde{x}_1}-F(\tilde{x}_2)\frac{\partial
\tilde{u}^\prime_i}{\partial \tilde{x}_1}+\delta_{i1}\left(
\tilde{u}_2^\prime\frac{d\tilde{\bar{u}}_1}{d\tilde{x}_2}
-\tilde{u}_2^\prime\frac{dF(\tilde{x}_2)}{d\tilde{x}_2}
-K-\tilde{\nu}\frac{d^2\tilde{\bar{u}}_1}{d
\tilde{x}^2_2}+\tilde{\nu}\frac{d^2F(\tilde{x}_2)}{d\tilde{x}^2_2}\right)\nonumber\\
&\; +\;\delta_{i2}\frac{d\tilde{\bar{p}}^*}{d
\tilde{x}_2}+\frac{\partial \tilde{u}^\prime_i
\tilde{u}^\prime_k}{\partial
\tilde{x}_k}+F(\tilde{x}_2)\frac{\partial\tilde{u}^\prime_i}{\partial\tilde{x}_1}
+\delta_{i1}\tilde{u}^\prime_2\frac{dF(\tilde{x}_2)}{d\tilde{x}_2}
+\frac{\partial \tilde{p}^\prime}{\partial
\tilde{x}_i}-\tilde{\nu}\frac{\partial^2
\tilde{u}^\prime_i}{\partial
\tilde{x}_k^2}-\delta_{i1}\tilde{\nu}\frac{d^2
F(\tilde{x}_2)}{d\tilde{x}_2^2}\nonumber\\[0.75em]
= &\; \frac{\partial \tilde{u}^\prime_i}{\partial
\tilde{t}}+\tilde{\bar{u}}_1\frac{\partial
\tilde{u}^\prime_i}{\partial \tilde{x}_1}+\delta_{i1}
\tilde{u}_2^\prime\frac{d\tilde{\bar{u}}_1}{d\tilde{x}_2}
-\delta_{i1}\left(K+\tilde{\nu}\frac{d^2\tilde{\bar{u}}_1}{d
\tilde{x}^2_2}\right)\nonumber\\
&\; +\;\delta_{i2}\frac{d\tilde{\bar{p}}^*}{d
\tilde{x}_2}+\frac{\partial \tilde{u}^\prime_i
\tilde{u}^\prime_k}{\partial \tilde{x}_k} +\frac{\partial
\tilde{p}^\prime}{\partial
\tilde{x}_i}-\tilde{\nu}\frac{\partial^2
\tilde{u}^\prime_i}{\partial \tilde{x}_k^2}=\tilde{\mathscr{N}}_i,
\hspace{4.51cm}\; \label{141114:0847}
\end{align}

\noindent and for the restriction equations (\ref{131228:1919}),
which, after employing transformation (\ref{141113:2112}),~stay
unchanged as well
\begin{equation}
\frac{\partial \tilde{\bar{u}}_1}{\partial
\tilde{t}}=\frac{\partial \tilde{\bar{u}}_1}{\partial
\tilde{x}_1}= \frac{\partial \tilde{\bar{u}}_1}{\partial
\tilde{x}_3}= \frac{\partial \tilde{\bar{p}}^*}{\partial
\tilde{t}}=\frac{\partial \tilde{\bar{p}}^*}{\partial
\tilde{x}_1}=\frac{\partial \tilde{\bar{p}}^*}{\partial
\tilde{x}_3}=\tilde{\bar{u}}_2=\tilde{\bar{u}}_3=0.
\label{141113:2307}
\end{equation}
Hence, the (\ref{131228:1919})-restricted equations $\mathscr{C}$
(\ref{131228:1918}) and $\mathscr{N}_i$ (\ref{131228:1921}) admit
transformation $\mathsf{T}$ (\ref{141113:2112}) as a symmetry.

\vspace{1em}\noindent {\bf Part II.} In order to complete the
validation procedure of Part I, we also have to check if the
``velocity product equations" $\mathscr{P}_{ij}$
(\ref{131228:1928}) admit $\mathsf{T}$ (\ref{141113:2112}) as a
symmetry transformation. But these ``equations" are redundant to
the momentum equations $\mathscr{N}_i$ (\ref{131228:1921}). They
all show a degenerate behavior when trying to independently
transform them from the underlying momentum equations
$\mathscr{N}_i$. Consider e.g. the first diagonal component
\begin{equation}
0=\mathscr{P}_{11}=\mathscr{N}_1 u^\prime_1+ \mathscr{N}_1
u^\prime_1=2\mathscr{N}_1 u^\prime_1.\label{141114:0005}
\end{equation}
If we would treat (\ref{141114:0005}) as an own and from
$\mathscr{N}_1$ independent equation, we have to globally conclude
that either $u^\prime_1$ must be zero or that $\mathscr{N}_1$ must
be zero in order to satisfy equation (\ref{141114:0005}). But,
since the former choice $u^\prime_1=0$ has to be excluded,
obviously, equation (\ref{141114:0005}) can only be equivalent
(necessarily and sufficiently) to the latter choice
\begin{equation}
0=\mathscr{P}_{11}=2\mathscr{N}_1 u^\prime_1,\;\text{with}\;\;
u^\prime_1\neq 0 \quad \Leftrightarrow\quad \mathscr{N}_1=0,
\end{equation}
i.e. the ``velocity product equation" $\mathscr{P}_{11}$ is
redundant and thus mathematically equivalent to the momentum
equation $\mathscr{N}_1$. Therefore, since the latter equation
stays invariant under transformation $\mathsf{T}$
(\ref{141113:2112}), the former equation will stay invariant too.
To be explicit, let's apply transformation $\mathsf{T}$
(\ref{141113:2112}), with the already obtained result
$\mathscr{N}_1=\tilde{\mathscr{N}}_1$ \eqref{141114:0847}, to
equation (\ref{141114:0005})
\begin{equation}
0=\mathscr{P}_{11}=2\mathscr{N}_1
u^\prime_1=2\tilde{\mathscr{N}}_1\left[\tilde{u}^\prime_1+F(\tilde{x}_2)\right].
\label{141114:0109}
\end{equation}
Now, since $\tilde{\mathscr{N}}_1=0$ (as shown in
(\ref{141114:0847})), and since by construction $F\neq 0$ (the
initial assumption), we obtain the equivalent relation
\begin{equation}
\tilde{\mathscr{N}}_1=0 \quad \Leftrightarrow\quad
\tilde{\mathscr{N}}_1 F(\tilde{x}_2) =0,\;\text{with}\;\; F\neq 0,
\end{equation}
which therefore will turn equation (\ref{141114:0109}) into the
(redundant and necessary) invariant form
\begin{equation}
0=\mathscr{P}_{11}=2\mathscr{N}_1
u^\prime_1=2\tilde{\mathscr{N}}_1\tilde{u}^\prime_1
+2\tilde{\mathscr{N}}_1F(\tilde{x}_2)=2\tilde{\mathscr{N}}_1\tilde{u}^\prime_1
=\tilde{\mathscr{P}}_{11}. \label{141114:1313}
\end{equation}
The conclusion that $\mathscr{P}_{11}$ admits $\mathsf{T}$
(\ref{141113:2112}) as a symmetry transformation due to that
$\mathscr{N}_1$ admits it (and vice versa), can also be readily
verified when using any symmetry-determining computer algebra
package (see Appendix \ref{SC}): The symmetry results stay
completely unaffected when either including or excluding the
``velocity product equation" $\mathscr{P}_{11}$
(\ref{131228:1928}) in a corresponding symmetry analysis next to
the continuity equation $\mathscr{C}$ (\ref{131228:1918}), the
momentum equations $\mathscr{N}_i$ (\ref{131228:1921}) and all
constraint equations (\ref{131228:1919}).

Hence, relation (\ref{141114:1313}) is a fully redundant
invariance of the actual system
(\ref{141114:0846})-(\ref{141113:2307}) and thus clearly not
symmetry breaking. This conclusion, of course, also applies to all
other components of $\mathscr{P}_{ij}$ (\ref{131228:1928}) --- the
procedure is exactly the same as given just before for
$\mathscr{P}_{11}$.

\section{Explicit symmetry validation with CAS\label{SC}}

\vspace{-0.1em}\noindent By using the Maple-based
symmetry-packages GeM \citep{Cheviakov07}, SADE \citep{Filho11}
and DESOLV-II \citep{Vu12}, we will demonstrate that irrespective
of whether the ``velocity product equations" $\mathscr{P}_{ij}$
(\ref{131228:1928}) are excluded from or included into the
analysis, the final result will always stay unchanged in all three
symmetry-determining algorithms. In the case of GeM and SADE this
result is given by (\ref{131228:1839}), while for DESOLV-II it is
further restricted by (\ref{131228:1955}).

Hence, all three CAS results explicitly show that the ``velocity
product equations" $\mathscr{P}_{ij}$ (\ref{131228:1928}) are not
only redundant from the perspective of the fluctuating
Navier-Stokes equations themselves
(\ref{131228:1918})-(\ref{131228:1921}), but also redundant from
the perspective of a symmetry analysis performed upon them
(\ref{131228:1950})-(\ref{131228:1951}). The symmetry breaking
mechanism as claimed in \cite{Oberlack01}, which should solely
arise from $\mathscr{P}_{ij}$ (\ref{131228:1928}), is thus not
supported.

In the following we will list the corresponding codes for all
three packages.~In each of these three sections we begin to first
list the procedure {\it without} the $\mathscr{P}_{ij}$-equations
(\ref{131228:1928}) in order to first compare to the central
result (\ref{131228:1839}).~Then, in a second part, we will list
the procedure {\it with} the $\mathscr{P}_{ij}$-equations in order
to then easily compare and to see that the central symmetry result
(\ref{131228:1839}) stays entirely unchanged.~Finally note again
that the DESOLV-II al\-go\-rithm implicitly restricts the result
(\ref{131228:1839}) consistently to (\ref{131228:1955}) in both
cases, i.e., with or without the $\mathscr{P}_{ij}$-equations, the
corresponding symmetry result simply stays unchanged~too.

\vspace{0.9em}\noindent {\bf Ia. GeM-Package
\textbf{\emph{excluding}} the $\pmb{\mathscr{P}_{ij}}$-equations
(\ref{131228:1928}):}
\vspace{-0.1em}
\begin{maplegroup}
\begin{flushleft}
{\large Header:}
\end{flushleft}

\end{maplegroup}
\begin{maplegroup}
\begin{mapleinput}
\mapleinline{active}{1d}{restart: with(linalg): with(PDEtools): with(DEtools): with(GeM):}{%
}
\end{mapleinput}

\mapleresult

\begin{maplelatex}
\mapleinline{inert}{2d}{`GEM v.021.b3 Copyright (C) Alexei F. Cheviakov, 2004-2006`;}{%
\[
\mathit{GEM\ v.021.b3\ Copyright\ (C)\ Alexei\ F.\ Cheviakov,\
2004-2006}
\]
}
\end{maplelatex}

\end{maplegroup}
\vspace{-0.4em}
\begin{maplegroup}
\begin{flushleft}
{\large Definitions:}
\end{flushleft}

\end{maplegroup}
\begin{maplegroup}
\begin{mapleinput}
\mapleinline{active}{1d}{X:=(t,x,y,z,nu):}{%
}
\end{mapleinput}

\end{maplegroup}
\begin{maplegroup}
\begin{mapleinput}
\mapleinline{active}{1d}{C:=diff(u(X),x)+diff(v(X),y)+diff(w(X),z);}{%
}
\end{mapleinput}

\mapleresult
\begin{maplelatex}
\mapleinline{inert}{2d}{C :=
diff(u(t,x,y,z,nu),x)+diff(v(t,x,y,z,nu),y)+diff(w(t,x,y,z,nu),z);}{%
\[
C := ({\frac {\partial }{\partial x}}\,\mathrm{u}(t, \,x, \,y, \,
z, \,\nu )) + ({\frac {\partial }{\partial y}}\,\mathrm{v}(t, \,x
, \,y, \,z, \,\nu )) + ({\frac {\partial }{\partial z}}\,\mathrm{
w}(t, \,x, \,y, \,z, \,\nu ))
\]
}
\end{maplelatex}

\end{maplegroup}
\begin{maplegroup}
\begin{mapleinput}
\mapleinline{active}{1d}{N1:=diff(u(X),t)+U(X)*diff(u(X),x)+v(X)*diff(U(X),y)-(K+nu*diff(U(X),
y,y))+diff(u(X)*u(X),x)+diff(u(X)*v(X),y)+diff(u(X)*w(X),z)+diff(p(X),
x)-nu*(diff(u(X),x,x)+diff(u(X),y,y)+diff(u(X),z,z));}{%
}
\end{mapleinput}

\mapleresult
\begin{maplelatex}
\mapleinline{inert}{2d}{N1 :=
diff(u(t,x,y,z,nu),t)+U(t,x,y,z,nu)*diff(u(t,x,y,z,nu),x)+v(t,x,y,z,nu
)*diff(U(t,x,y,z,nu),y)-K-nu*diff(U(t,x,y,z,nu),`$`(y,2))+2*u(t,x,y,z,
nu)*diff(u(t,x,y,z,nu),x)+diff(u(t,x,y,z,nu),y)*v(t,x,y,z,nu)+u(t,x,y,
z,nu)*diff(v(t,x,y,z,nu),y)+diff(u(t,x,y,z,nu),z)*w(t,x,y,z,nu)+u(t,x,
y,z,nu)*diff(w(t,x,y,z,nu),z)+diff(p(t,x,y,z,nu),x)-nu*(diff(u(t,x,y,z
,nu),`$`(x,2))+diff(u(t,x,y,z,nu),`$`(y,2))+diff(u(t,x,y,z,nu),`$`(z,2
)));}{%
\maplemultiline{ \mathit{N1} := ({\frac {\partial }{\partial
t}}\,\mathrm{u}(t, \, x, \,y, \,z, \,\nu )) + \mathrm{U}(t, \,x,
\,y, \,z, \,\nu )\,( {\frac {\partial }{\partial
x}}\,\mathrm{u}(t, \,x, \,y, \,z, \,
\nu )) \\
\mbox{} + \mathrm{v}(t, \,x, \,y, \,z, \,\nu )\,({\frac {
\partial }{\partial y}}\,\mathrm{U}(t, \,x, \,y, \,z, \,\nu )) -
K - \nu \,({\frac {\partial ^{2}}{\partial y^{2}}}\,\mathrm{U}(t
, \,x, \,y, \,z, \,\nu )) \\
\mbox{} + 2\,\mathrm{u}(t, \,x, \,y, \,z, \,\nu )\,({\frac {
\partial }{\partial x}}\,\mathrm{u}(t, \,x, \,y, \,z, \,\nu )) +
({\frac {\partial }{\partial y}}\,\mathrm{u}(t, \,x, \,y, \,z, \,
\nu ))\,\mathrm{v}(t, \,x, \,y, \,z, \,\nu ) \\
\mbox{} + \mathrm{u}(t, \,x, \,y, \,z, \,\nu )\,({\frac {
\partial }{\partial y}}\,\mathrm{v}(t, \,x, \,y, \,z, \,\nu )) +
({\frac {\partial }{\partial z}}\,\mathrm{u}(t, \,x, \,y, \,z, \,
\nu ))\,\mathrm{w}(t, \,x, \,y, \,z, \,\nu ) \\
\mbox{} + \mathrm{u}(t, \,x, \,y, \,z, \,\nu )\,({\frac {
\partial }{\partial z}}\,\mathrm{w}(t, \,x, \,y, \,z, \,\nu )) +
({\frac {\partial }{\partial x}}\,\mathrm{p}(t, \,x, \,y, \,z, \,
\nu )) \\
\mbox{} - \nu \,(({\frac {\partial ^{2}}{\partial x^{2}}}\,
\mathrm{u}(t, \,x, \,y, \,z, \,\nu )) + ({\frac {\partial ^{2}}{
\partial y^{2}}}\,\mathrm{u}(t, \,x, \,y, \,z, \,\nu )) + (
{\frac {\partial ^{2}}{\partial z^{2}}}\,\mathrm{u}(t, \,x, \,y,
\,z, \,\nu ))) }
}
\end{maplelatex}

\end{maplegroup}
\begin{maplegroup}
\begin{mapleinput}
\mapleinline{active}{1d}{N2:=diff(v(X),t)+U(X)*diff(v(X),x)+diff(P(X),y)+diff(v(X)*u(X),x)+dif
f(v(X)*v(X),y)+diff(v(X)*w(X),z)+diff(p(X),y)-nu*(diff(v(X),x,x)+diff(
v(X),y,y)+diff(v(X),z,z));}{%
}
\end{mapleinput}

\mapleresult
\begin{maplelatex}
\mapleinline{inert}{2d}{N2 :=
diff(v(t,x,y,z,nu),t)+U(t,x,y,z,nu)*diff(v(t,x,y,z,nu),x)+diff(P(t,x,y
,z,nu),y)+diff(u(t,x,y,z,nu),x)*v(t,x,y,z,nu)+u(t,x,y,z,nu)*diff(v(t,x
,y,z,nu),x)+2*v(t,x,y,z,nu)*diff(v(t,x,y,z,nu),y)+diff(v(t,x,y,z,nu),z
)*w(t,x,y,z,nu)+v(t,x,y,z,nu)*diff(w(t,x,y,z,nu),z)+diff(p(t,x,y,z,nu)
,y)-nu*(diff(v(t,x,y,z,nu),`$`(x,2))+diff(v(t,x,y,z,nu),`$`(y,2))+diff
(v(t,x,y,z,nu),`$`(z,2)));}{%
\maplemultiline{
\mathit{N2} := ({\frac {\partial }{\partial t}}\,\mathrm{v}(t, \,
x, \,y, \,z, \,\nu )) + \mathrm{U}(t, \,x, \,y, \,z, \,\nu )\,(
{\frac {\partial }{\partial x}}\,\mathrm{v}(t, \,x, \,y, \,z, \,
\nu )) + ({\frac {\partial }{\partial y}}\,\mathrm{P}(t, \,x, \,y
, \,z, \,\nu )) \\
\mbox{} + ({\frac {\partial }{\partial x}}\,\mathrm{u}(t, \,x, \,
y, \,z, \,\nu ))\,\mathrm{v}(t, \,x, \,y, \,z, \,\nu ) + \mathrm{
u}(t, \,x, \,y, \,z, \,\nu )\,({\frac {\partial }{\partial x}}\,
\mathrm{v}(t, \,x, \,y, \,z, \,\nu )) \\
\mbox{} + 2\,\mathrm{v}(t, \,x, \,y, \,z, \,\nu )\,({\frac {
\partial }{\partial y}}\,\mathrm{v}(t, \,x, \,y, \,z, \,\nu )) +
({\frac {\partial }{\partial z}}\,\mathrm{v}(t, \,x, \,y, \,z, \,
\nu ))\,\mathrm{w}(t, \,x, \,y, \,z, \,\nu ) \\
\mbox{} + \mathrm{v}(t, \,x, \,y, \,z, \,\nu )\,({\frac {
\partial }{\partial z}}\,\mathrm{w}(t, \,x, \,y, \,z, \,\nu )) +
({\frac {\partial }{\partial y}}\,\mathrm{p}(t, \,x, \,y, \,z, \,
\nu )) \\
\mbox{} - \nu \,(({\frac {\partial ^{2}}{\partial x^{2}}}\,
\mathrm{v}(t, \,x, \,y, \,z, \,\nu )) + ({\frac {\partial ^{2}}{
\partial y^{2}}}\,\mathrm{v}(t, \,x, \,y, \,z, \,\nu )) + (
{\frac {\partial ^{2}}{\partial z^{2}}}\,\mathrm{v}(t, \,x, \,y,
\,z, \,\nu ))) }
}
\end{maplelatex}

\end{maplegroup}
\begin{maplegroup}
\begin{mapleinput}
\mapleinline{active}{1d}{N3:=diff(w(X),t)+U(X)*diff(w(X),x)+diff(w(X)*u(X),x)+diff(w(X)*v(X),y
)+diff(w(X)*w(X),z)+diff(p(X),z)-nu*(diff(w(X),x,x)+diff(w(X),y,y)+dif
f(w(X),z,z));}{%
}
\end{mapleinput}

\mapleresult
\begin{maplelatex}
\mapleinline{inert}{2d}{N3 :=
diff(w(t,x,y,z,nu),t)+U(t,x,y,z,nu)*diff(w(t,x,y,z,nu),x)+diff(u(t,x,y
,z,nu),x)*w(t,x,y,z,nu)+u(t,x,y,z,nu)*diff(w(t,x,y,z,nu),x)+diff(v(t,x
,y,z,nu),y)*w(t,x,y,z,nu)+v(t,x,y,z,nu)*diff(w(t,x,y,z,nu),y)+2*w(t,x,
y,z,nu)*diff(w(t,x,y,z,nu),z)+diff(p(t,x,y,z,nu),z)-nu*(diff(w(t,x,y,z
,nu),`$`(x,2))+diff(w(t,x,y,z,nu),`$`(y,2))+diff(w(t,x,y,z,nu),`$`(z,2
)));}{%
\maplemultiline{
\mathit{N3} := ({\frac {\partial }{\partial t}}\,\mathrm{w}(t, \,
x, \,y, \,z, \,\nu )) + \mathrm{U}(t, \,x, \,y, \,z, \,\nu )\,(
{\frac {\partial }{\partial x}}\,\mathrm{w}(t, \,x, \,y, \,z, \,
\nu )) \\
\mbox{} + ({\frac {\partial }{\partial x}}\,\mathrm{u}(t, \,x, \,
y, \,z, \,\nu ))\,\mathrm{w}(t, \,x, \,y, \,z, \,\nu ) + \mathrm{
u}(t, \,x, \,y, \,z, \,\nu )\,({\frac {\partial }{\partial x}}\,
\mathrm{w}(t, \,x, \,y, \,z, \,\nu )) \\
\mbox{} + ({\frac {\partial }{\partial y}}\,\mathrm{v}(t, \,x, \,
y, \,z, \,\nu ))\,\mathrm{w}(t, \,x, \,y, \,z, \,\nu ) + \mathrm{
v}(t, \,x, \,y, \,z, \,\nu )\,({\frac {\partial }{\partial y}}\,
\mathrm{w}(t, \,x, \,y, \,z, \,\nu )) \\
\mbox{} + 2\,\mathrm{w}(t, \,x, \,y, \,z, \,\nu )\,({\frac {
\partial }{\partial z}}\,\mathrm{w}(t, \,x, \,y, \,z, \,\nu )) +
({\frac {\partial }{\partial z}}\,\mathrm{p}(t, \,x, \,y, \,z, \,
\nu )) \\
\mbox{} - \nu \,(({\frac {\partial ^{2}}{\partial x^{2}}}\,
\mathrm{w}(t, \,x, \,y, \,z, \,\nu )) + ({\frac {\partial ^{2}}{
\partial y^{2}}}\,\mathrm{w}(t, \,x, \,y, \,z, \,\nu )) + (
{\frac {\partial ^{2}}{\partial z^{2}}}\,\mathrm{w}(t, \,x, \,y,
\,z, \,\nu ))) }
}
\end{maplelatex}

\end{maplegroup}
\begin{maplegroup}
\begin{flushleft}
{\large Equations (2.1), (2.2) \& (2.4):}
\end{flushleft}

\end{maplegroup}
\begin{maplegroup}
\begin{mapleinput}
\mapleinline{active}{1d}{eqnC:=C=0: eqnN1:=N1=0: eqnN2:=N2=0: eqnN3:=N3=0:
eqnR1:=diff(U(X),t)=0: eqnR2:=diff(U(X),x)=0: eqnR3:=diff(U(X),z)=0:
eqnR4:=diff(P(X),t)=0: eqnR5:=diff(P(X),x)=0:
eqnR6:=diff(P(X),z)=0:}{%
}
\end{mapleinput}

\end{maplegroup}
\begin{maplegroup}
\begin{flushleft}
{\large Symmetry Algorithm:}
\end{flushleft}

\end{maplegroup}
\begin{maplegroup}
\begin{mapleinput}
\mapleinline{active}{1d}{gem_init_defs([X],[u(X),v(X),w(X),p(X),U(X),P(X)],[],[K],0,[eqnC,eqnN
1,eqnN2,eqnN3,eqnR1,eqnR2,eqnR3,eqnR4,eqnR5,eqnR6],[diff(u(X),x),diff(
u(X),t),diff(v(X),t),diff(w(X),t),diff(U(X),t),diff(U(X),x),diff(U(X),
z),diff(P(X),t),diff(P(X),x),diff(P(X),z)]);}{%
}
\end{mapleinput}

\mapleresult
\begin{maplelatex}
\mapleinline{inert}{2d}{`-- Starting definitions for Point Symmetry or Conservation Law
computation... --`;}{%
\[
\mathit{--\ Starting\ definitions\ for\ Point\ Symmetry\ or\
Conservation\ Law\ computation...\ --}
\]
}
\end{maplelatex}

\begin{maplelatex}
\mapleinline{inert}{2d}{`Independent variables: `, nu, t, x, y, z;}{%
\[
\mathit{Independent\ variables:\ }, \,\nu , \,t, \,x, \,y, \,z
\]
}
\end{maplelatex}

\begin{maplelatex}
\mapleinline{inert}{2d}{`Dependent variables :`, P, U, p, u, v, w;}{%
\[
\mathit{Dependent\ variables\ :}, \,P, \,U, \,p, \,u, \,v, \,w
\]
}
\end{maplelatex}

\begin{maplelatex}
\mapleinline{inert}{2d}{`Free functions: `, vector([]);}{%
\[
\mathit{Free\ functions:\ }, \,[]
\]
}
\end{maplelatex}

\begin{maplelatex}
\mapleinline{inert}{2d}{`Free constants: `, [K];}{%
\[
\mathit{Free\ constants:\ }, \,[K]
\]
}
\end{maplelatex}

\begin{maplelatex}
\mapleinline{inert}{2d}{` - Variable definition successful -`;}{%
\[
\mathit{\ -\ Variable\ definition\ successful\ -}
\]
}
\end{maplelatex}

\begin{maplelatex}
\mapleinline{inert}{2d}{`10 equation(s) defined.`;}{%
\[
\mathit{10\ equation(s)\ defined.}
\]
}
\end{maplelatex}

\begin{maplelatex}
\mapleinline{inert}{2d}{`Computing necessary differential consequences...`;}{%
\[
\mathit{Computing\ necessary\ differential\ consequences...}
\]
}
\end{maplelatex}

\begin{maplelatex}
\mapleinline{inert}{2d}{`29 differential consequence(s) computed.`;}{%
\[
\mathit{29\ differential\ consequence(s)\ computed.}
\]
}
\end{maplelatex}

\begin{maplelatex}
\mapleinline{inert}{2d}{`-- GeM initial definitions for Symmetry/Conservation Law analysis
successful. --`;}{%
\[
\mathit{--\ GeM\ initial\ definitions\ for\ Symmetry/Conservation
\ Law\ analysis\ successful.\ --}
\]
}
\end{maplelatex}

\end{maplegroup}
\begin{maplegroup}
\begin{mapleinput}
\mapleinline{active}{1d}{gem_solvedfor_get_():
overdet_sys:=gem_get_split_sys([X,u(X),v(X),w(X),p(X),U(X),P(X)],0):}{
}
\end{mapleinput}

\mapleresult

\begin{maplelatex}
\mapleinline{inert}{2d}{`Tangent Vector Field Coordinates defined successfuly: `, \{eta_P,
eta_U, eta_p, eta_u, eta_v, eta_w, xi_nu, xi_t, xi_x, xi_y, xi_z\};}{%
\maplemultiline{
\mathit{Tangent\ Vector\ Field\ Coordinates\ defined\
successfuly:\ },  \\
\{\mathit{eta\_P}, \,\mathit{eta\_U}, \,\mathit{eta\_p}, \,
\mathit{eta\_u}, \,\mathit{eta\_v}, \,\mathit{eta\_w}, \,\mathit{
xi\_nu}, \,\mathit{xi\_t}, \,\mathit{xi\_x}, \,\mathit{xi\_y}, \,
\mathit{xi\_z}\} }
}
\end{maplelatex}

\begin{maplelatex}
\mapleinline{inert}{2d}{`Generating t.v.f. coordinates of derivatives...`;}{%
\[
\mathit{Generating\ t.v.f.\ coordinates\ of\ derivatives...}
\]
}
\end{maplelatex}

\begin{maplelatex}
\mapleinline{inert}{2d}{`Done. Generating determining equations...`;}{%
\[
\mathit{Done.\ Generating\ determining\ equations...}
\]
}
\end{maplelatex}

\begin{maplelatex}
\mapleinline{inert}{2d}{`Done. Splitting...`;}{%
\[
\mathit{Done.\ Splitting...}
\]
}
\end{maplelatex}

\begin{maplelatex}
\mapleinline{inert}{2d}{`Split successful. The split system returned. Number of equations: `,
1034;}{%
\[
\mathit{Split\ successful.\ The\ split\ system\ returned.\ Number
\ of\ equations:\ }, \,1034
\]
}
\end{maplelatex}

\end{maplegroup}
\begin{maplegroup}
\begin{mapleinput}
\mapleinline{active}{1d}{twf_coords:=gem_tvf_coords_get_():
simplified_system:=rifsimp(overdet_sys,twf_coords):
sym_sol:=pdsolve(simplified_system[Solved],twf_coords);}{%
}
\end{mapleinput}

\mapleresult
\begin{maplelatex}
\mapleinline{inert}{2d}{sym_sol := \{eta_P =
-_F12(y,nu,U,P)-_F14(nu,U,P)+2*(-_F9(nu)+_F6(nu))*P-_F16(nu,U)+_F18(nu
), eta_U = -_F9(nu)*U+U*_F6(nu)-_F11(y,nu,U,P), eta_p =
-z*_F5[t,t]-x*_F8[t,t]+_F12(y,nu,U,P)+_F14(nu,U,P)+_F16(nu,U)+_F17(t,n
u)+(2*p-K*x)*_F6(nu)+(-2*p+2*K*x)*_F9(nu)-z*_F3(nu)*K, eta_u =
-_F3(nu)*w+(-_F9(nu)+_F6(nu))*u+_F8[t]+_F11(y,nu,U,P), eta_v =
v*(-_F9(nu)+_F6(nu)), eta_w =
_F5[t]+(U+u)*_F3(nu)-_F9(nu)*w+w*_F6(nu), xi_nu =
nu*(2*_F6(nu)-_F9(nu)), xi_t = _F9(nu)*t+_F10(nu), xi_x =
_F6(nu)*x-_F3(nu)*z+_F8(t,nu), xi_y = _F6(nu)*y+_F7(nu), xi_z =
_F3(nu)*x+_F6(nu)*z+_F5(t,nu)\};}{%
\maplemultiline{
\mathit{sym\_sol} := \{\mathit{eta\_P}= - \mathrm{\_F12}(y, \,\nu
 , \,U, \,P) - \mathrm{\_F14}(\nu , \,U, \,P) + 2\,( - \mathrm{
\_F9}(\nu ) + \mathrm{\_F6}(\nu ))\,P \\
\mbox{} - \mathrm{\_F16}(\nu , \,U) + \mathrm{\_F18}(\nu ), \,
\mathit{eta\_U}= - \mathrm{\_F9}(\nu )\,U + U\,\mathrm{\_F6}(\nu
) - \mathrm{\_F11}(y, \,\nu , \,U, \,P),  \\
\mathit{eta\_p}= - z\,{\mathit{\_F5}_{t, \,t}} - x\,{\mathit{\_F8
}_{t, \,t}} + \mathrm{\_F12}(y, \,\nu , \,U, \,P) + \mathrm{\_F14
}(\nu , \,U, \,P) + \mathrm{\_F16}(\nu , \,U) \\
\mbox{} + \mathrm{\_F17}(t, \,\nu ) + (2\,p - K\,x)\,\mathrm{\_F6
}(\nu ) + ( - 2\,p + 2\,K\,x)\,\mathrm{\_F9}(\nu ) - z\,\mathrm{
\_F3}(\nu )\,K,  \\
\mathit{eta\_u}= - \mathrm{\_F3}(\nu )\,w + ( - \mathrm{\_F9}(\nu
 ) + \mathrm{\_F6}(\nu ))\,u + {\mathit{\_F8}_{t}} + \mathrm{
\_F11}(y, \,\nu , \,U, \,P),  \\
\mathit{eta\_v}=v\,( - \mathrm{\_F9}(\nu ) + \mathrm{\_F6}(\nu ))
,  \\
\mathit{eta\_w}={\mathit{\_F5}_{t}} + (U + u)\,\mathrm{\_F3}(\nu
) - \mathrm{\_F9}(\nu )\,w + w\,\mathrm{\_F6}(\nu ),  \\
\mathit{xi\_nu}=\nu \,(2\,\mathrm{\_F6}(\nu ) - \mathrm{\_F9}(\nu
 )), \,\mathit{xi\_t}=\mathrm{\_F9}(\nu )\,t + \mathrm{\_F10}(\nu
 ),  \\
\mathit{xi\_x}=\mathrm{\_F6}(\nu )\,x - \mathrm{\_F3}(\nu )\,z +
\mathrm{\_F8}(t, \,\nu ), \,\mathit{xi\_y}=\mathrm{\_F6}(\nu )\,y
 + \mathrm{\_F7}(\nu ),  \\
\mathit{xi\_z}=\mathrm{\_F3}(\nu )\,x + \mathrm{\_F6}(\nu )\,z +
\mathrm{\_F5}(t, \,\nu )\} }
}
\end{maplelatex}

\end{maplegroup}
\begin{maplegroup}
\begin{flushleft}
{\large Redefinition of group functions as used in (2.10):}
\end{flushleft}

\end{maplegroup}
\begin{maplegroup}
\begin{mapleinput}
\mapleinline{active}{1d}{_F6(nu):=a1(nu); _F3(nu):=-a2(nu); _F8(t,nu):=f1(t,nu);
_F7(nu):=a3(nu); _F5(t,nu):=f2(t,nu); _F9(nu):=a4(nu);
_F10(nu):=a5(nu); _F11(y,nu,U,P):=-g1(y,nu,U,P);
_F17(t,nu):=f3(t,nu)-_F18(nu);
_F12(y,nu,U,P):=-g2(y,nu,U,P)-_F14(nu,U,P)-_F16(nu,U)+_F18(nu);}{%
}
\end{mapleinput}

\mapleresult
\begin{maplelatex}
\mapleinline{inert}{2d}{_F6(nu) := a1(nu);}{%
\[
\mathrm{\_F6}(\nu ) := \mathrm{a1}(\nu )
\]
}
\end{maplelatex}

\begin{maplelatex}
\mapleinline{inert}{2d}{_F3(nu) := -a2(nu);}{%
\[
\mathrm{\_F3}(\nu ) :=  - \mathrm{a2}(\nu )
\]
}
\end{maplelatex}

\begin{maplelatex}
\mapleinline{inert}{2d}{_F8(t,nu) := f1(t,nu);}{%
\[
\mathrm{\_F8}(t, \,\nu ) := \mathrm{f1}(t, \,\nu )
\]
}
\end{maplelatex}

\begin{maplelatex}
\mapleinline{inert}{2d}{_F7(nu) := a3(nu);}{%
\[
\mathrm{\_F7}(\nu ) := \mathrm{a3}(\nu )
\]
}
\end{maplelatex}

\begin{maplelatex}
\mapleinline{inert}{2d}{_F5(t,nu) := f2(t,nu);}{%
\[
\mathrm{\_F5}(t, \,\nu ) := \mathrm{f2}(t, \,\nu )
\]
}
\end{maplelatex}

\begin{maplelatex}
\mapleinline{inert}{2d}{_F9(nu) := a4(nu);}{%
\[
\mathrm{\_F9}(\nu ) := \mathrm{a4}(\nu )
\]
}
\end{maplelatex}

\begin{maplelatex}
\mapleinline{inert}{2d}{_F10(nu) := a5(nu);}{%
\[
\mathrm{\_F10}(\nu ) := \mathrm{a5}(\nu )
\]
}
\end{maplelatex}

\begin{maplelatex}
\mapleinline{inert}{2d}{_F11(y,nu,U,P) := -g1(y,nu,U,P);}{%
\[
\mathrm{\_F11}(y, \,\nu , \,U, \,P) :=  - \mathrm{g1}(y, \,\nu ,
\,U, \,P)
\]
}
\end{maplelatex}

\begin{maplelatex}
\mapleinline{inert}{2d}{_F17(t,nu) := f3(t,nu)-_F18(nu);}{%
\[
\mathrm{\_F17}(t, \,\nu ) := \mathrm{f3}(t, \,\nu ) - \mathrm{
\_F18}(\nu )
\]
}
\end{maplelatex}

\begin{maplelatex}
\mapleinline{inert}{2d}{_F12(y,nu,U,P) := -g2(y,nu,U,P)-_F14(nu,U,P)-_F16(nu,U)+_F18(nu);}{%
\[
\mathrm{\_F12}(y, \,\nu , \,U, \,P) :=  - \mathrm{g2}(y, \,\nu ,
\,U, \,P) - \mathrm{\_F14}(\nu , \,U, \,P) - \mathrm{\_F16}(\nu
, \,U) + \mathrm{\_F18}(\nu )
\]
}
\end{maplelatex}

\end{maplegroup}
\begin{maplegroup}
\begin{flushleft}
{\large Final Result (identical to result (2.10)):}
\end{flushleft}

\end{maplegroup}
\begin{maplegroup}
\begin{mapleinput}
\mapleinline{active}{1d}{sym_sol[9]; sym_sol[10]; sym_sol[11]; sym_sol[8]; sym_sol[7];
sym_sol[4]; sym_sol[5]; sym_sol[6]; sym_sol[3]; sym_sol[2];
sym_sol[1];}{%
}
\end{mapleinput}

\mapleresult
\begin{maplelatex}
\mapleinline{inert}{2d}{xi_x = a1(nu)*x+a2(nu)*z+f1(t,nu);}{%
\[
\mathit{xi\_x}=\mathrm{a1}(\nu )\,x + \mathrm{a2}(\nu )\,z +
\mathrm{f1}(t, \,\nu )
\]
}
\end{maplelatex}

\begin{maplelatex}
\mapleinline{inert}{2d}{xi_y = a1(nu)*y+a3(nu);}{%
\[
\mathit{xi\_y}=\mathrm{a1}(\nu )\,y + \mathrm{a3}(\nu )
\]
}
\end{maplelatex}

\begin{maplelatex}
\mapleinline{inert}{2d}{xi_z = -a2(nu)*x+a1(nu)*z+f2(t,nu);}{%
\[
\mathit{xi\_z}= - \mathrm{a2}(\nu )\,x + \mathrm{a1}(\nu )\,z +
\mathrm{f2}(t, \,\nu )
\]
}
\end{maplelatex}

\begin{maplelatex}
\mapleinline{inert}{2d}{xi_t = a4(nu)*t+a5(nu);}{%
\[
\mathit{xi\_t}=\mathrm{a4}(\nu )\,t + \mathrm{a5}(\nu )
\]
}
\end{maplelatex}

\begin{maplelatex}
\mapleinline{inert}{2d}{xi_nu = nu*(2*a1(nu)-a4(nu));}{%
\[
\mathit{xi\_nu}=\nu \,(2\,\mathrm{a1}(\nu ) - \mathrm{a4}(\nu ))
\]
}
\end{maplelatex}

\begin{maplelatex}
\mapleinline{inert}{2d}{eta_u = a2(nu)*w+(-a4(nu)+a1(nu))*u+f1[t]-g1(y,nu,U,P);}{%
\[
\mathit{eta\_u}=\mathrm{a2}(\nu )\,w + ( - \mathrm{a4}(\nu ) +
\mathrm{a1}(\nu ))\,u + {\mathit{f1}_{t}} - \mathrm{g1}(y, \,\nu
, \,U, \,P)
\]
}
\end{maplelatex}

\begin{maplelatex}
\mapleinline{inert}{2d}{eta_v = v*(-a4(nu)+a1(nu));}{%
\[
\mathit{eta\_v}=v\,( - \mathrm{a4}(\nu ) + \mathrm{a1}(\nu ))
\]
}
\end{maplelatex}

\begin{maplelatex}
\mapleinline{inert}{2d}{eta_w = f2[t]-(U+u)*a2(nu)-a4(nu)*w+w*a1(nu);}{%
\[
\mathit{eta\_w}={\mathit{f2}_{t}} - (U + u)\,\mathrm{a2}(\nu ) -
\mathrm{a4}(\nu )\,w + w\,\mathrm{a1}(\nu )
\]
}
\end{maplelatex}

\vspace{-0.25em}
\begin{maplelatex}
\mapleinline{inert}{2d}{eta_p =
-z*f2[t,t]-x*f1[t,t]-g2(y,nu,U,P)+f3(t,nu)+(2*p-K*x)*a1(nu)+(-2*p+2*K*
x)*a4(nu)+z*a2(nu)*K;}{%
\maplemultiline{
\mathit{eta\_p}= - z\,{\mathit{f2}_{t, \,t}} - x\,{\mathit{f1}_{t
, \,t}} - \mathrm{g2}(y, \,\nu , \,U, \,P) + \mathrm{f3}(t, \,\nu
 ) \\ + (2\,p - K\,x)\,\mathrm{a1}(\nu )  + ( - 2\,p + 2\,K\,x)\,
\mathrm{a4}(\nu ) + z\,\mathrm{a2}(\nu )\,K }
}
\end{maplelatex}

\vspace{0.25em}
\begin{maplelatex}
\mapleinline{inert}{2d}{eta_U = -a4(nu)*U+U*a1(nu)+g1(y,nu,U,P);}{%
\[
\mathit{eta\_U}= - \mathrm{a4}(\nu )\,U + U\,\mathrm{a1}(\nu ) +
\mathrm{g1}(y, \,\nu , \,U, \,P)
\]
}
\end{maplelatex}

\vspace{0.25em}
\begin{maplelatex}
\mapleinline{inert}{2d}{eta_P = g2(y,nu,U,P)+2*(-a4(nu)+a1(nu))*P;}{%
\[
\mathit{eta\_P}=\mathrm{g2}(y, \,\nu , \,U, \,P) + 2\,( -
\mathrm{a4}(\nu ) + \mathrm{a1}(\nu ))\,P
\]
}
\end{maplelatex}

\end{maplegroup}
\begin{maplegroup}
\begin{mapleinput}
\end{mapleinput}

\end{maplegroup}

\vspace{1em}\noindent {\bf Ib. GeM-Package
\textbf{\emph{including}} the $\pmb{\mathscr{P}_{ij}}$-equations
(\ref{131228:1928}):}
\begin{maplegroup}
\begin{flushleft}
{\large Header:}
\end{flushleft}

\end{maplegroup}
\begin{maplegroup}
\begin{mapleinput}
\mapleinline{active}{1d}{restart: with(linalg): with(PDEtools): with(DEtools): with(GeM):}{%
}
\end{mapleinput}

\mapleresult

\begin{maplelatex}
\mapleinline{inert}{2d}{`GEM v.021.b3 Copyright (C) Alexei F. Cheviakov, 2004-2006`;}{%
\[
\mathit{GEM\ v.021.b3\ Copyright\ (C)\ Alexei\ F.\ Cheviakov,\
2004-2006}
\]
}
\end{maplelatex}

\end{maplegroup}
\begin{maplegroup}
\begin{flushleft}
{\large Definitions:}
\end{flushleft}

\end{maplegroup}
\begin{maplegroup}
\begin{mapleinput}
\mapleinline{active}{1d}{X:=(t,x,y,z,nu):}{%
}
\end{mapleinput}

\end{maplegroup}
\begin{maplegroup}
\begin{mapleinput}
\mapleinline{active}{1d}{C:=diff(u(X),x)+diff(v(X),y)+diff(w(X),z);}{%
}
\end{mapleinput}

\mapleresult
\begin{maplelatex}
\mapleinline{inert}{2d}{C :=
diff(u(t,x,y,z,nu),x)+diff(v(t,x,y,z,nu),y)+diff(w(t,x,y,z,nu),z);}{%
\[
C := ({\frac {\partial }{\partial x}}\,\mathrm{u}(t, \,x, \,y, \,
z, \,\nu )) + ({\frac {\partial }{\partial y}}\,\mathrm{v}(t, \,x
, \,y, \,z, \,\nu )) + ({\frac {\partial }{\partial z}}\,\mathrm{
w}(t, \,x, \,y, \,z, \,\nu ))
\]
}
\end{maplelatex}

\end{maplegroup}
\begin{maplegroup}
\begin{mapleinput}
\mapleinline{active}{1d}{N1:=diff(u(X),t)+U(X)*diff(u(X),x)+v(X)*diff(U(X),y)-(K+nu*diff(U(X),
y,y))+diff(u(X)*u(X),x)+diff(u(X)*v(X),y)+diff(u(X)*w(X),z)+diff(p(X),
x)-nu*(diff(u(X),x,x)+diff(u(X),y,y)+diff(u(X),z,z));}{%
}
\end{mapleinput}

\mapleresult
\begin{maplelatex}
\mapleinline{inert}{2d}{N1 :=
diff(u(t,x,y,z,nu),t)+U(t,x,y,z,nu)*diff(u(t,x,y,z,nu),x)+v(t,x,y,z,nu
)*diff(U(t,x,y,z,nu),y)-K-nu*diff(U(t,x,y,z,nu),`$`(y,2))+2*u(t,x,y,z,
nu)*diff(u(t,x,y,z,nu),x)+diff(u(t,x,y,z,nu),y)*v(t,x,y,z,nu)+u(t,x,y,
z,nu)*diff(v(t,x,y,z,nu),y)+diff(u(t,x,y,z,nu),z)*w(t,x,y,z,nu)+u(t,x,
y,z,nu)*diff(w(t,x,y,z,nu),z)+diff(p(t,x,y,z,nu),x)-nu*(diff(u(t,x,y,z
,nu),`$`(x,2))+diff(u(t,x,y,z,nu),`$`(y,2))+diff(u(t,x,y,z,nu),`$`(z,2
)));}{%
\maplemultiline{
\mathit{N1} := ({\frac {\partial }{\partial t}}\,\mathrm{u}(t, \,
x, \,y, \,z, \,\nu )) + \mathrm{U}(t, \,x, \,y, \,z, \,\nu )\,(
{\frac {\partial }{\partial x}}\,\mathrm{u}(t, \,x, \,y, \,z, \,
\nu )) \\
\mbox{} + \mathrm{v}(t, \,x, \,y, \,z, \,\nu )\,({\frac {
\partial }{\partial y}}\,\mathrm{U}(t, \,x, \,y, \,z, \,\nu )) -
K - \nu \,({\frac {\partial ^{2}}{\partial y^{2}}}\,\mathrm{U}(t
, \,x, \,y, \,z, \,\nu )) \\
\mbox{} + 2\,\mathrm{u}(t, \,x, \,y, \,z, \,\nu )\,({\frac {
\partial }{\partial x}}\,\mathrm{u}(t, \,x, \,y, \,z, \,\nu )) +
({\frac {\partial }{\partial y}}\,\mathrm{u}(t, \,x, \,y, \,z, \,
\nu ))\,\mathrm{v}(t, \,x, \,y, \,z, \,\nu ) \\
\mbox{} + \mathrm{u}(t, \,x, \,y, \,z, \,\nu )\,({\frac {
\partial }{\partial y}}\,\mathrm{v}(t, \,x, \,y, \,z, \,\nu )) +
({\frac {\partial }{\partial z}}\,\mathrm{u}(t, \,x, \,y, \,z, \,
\nu ))\,\mathrm{w}(t, \,x, \,y, \,z, \,\nu ) \\
\mbox{} + \mathrm{u}(t, \,x, \,y, \,z, \,\nu )\,({\frac {
\partial }{\partial z}}\,\mathrm{w}(t, \,x, \,y, \,z, \,\nu )) +
({\frac {\partial }{\partial x}}\,\mathrm{p}(t, \,x, \,y, \,z, \,
\nu )) \\
\mbox{} - \nu \,(({\frac {\partial ^{2}}{\partial x^{2}}}\,
\mathrm{u}(t, \,x, \,y, \,z, \,\nu )) + ({\frac {\partial ^{2}}{
\partial y^{2}}}\,\mathrm{u}(t, \,x, \,y, \,z, \,\nu )) + (
{\frac {\partial ^{2}}{\partial z^{2}}}\,\mathrm{u}(t, \,x, \,y,
\,z, \,\nu ))) }
}
\end{maplelatex}

\end{maplegroup}
\begin{maplegroup}
\begin{mapleinput}
\mapleinline{active}{1d}{N2:=diff(v(X),t)+U(X)*diff(v(X),x)+diff(P(X),y)+diff(v(X)*u(X),x)+dif
f(v(X)*v(X),y)+diff(v(X)*w(X),z)+diff(p(X),y)-nu*(diff(v(X),x,x)+diff(
v(X),y,y)+diff(v(X),z,z));}{%
}
\end{mapleinput}

\mapleresult
\begin{maplelatex}
\mapleinline{inert}{2d}{N2 :=
diff(v(t,x,y,z,nu),t)+U(t,x,y,z,nu)*diff(v(t,x,y,z,nu),x)+diff(P(t,x,y
,z,nu),y)+diff(u(t,x,y,z,nu),x)*v(t,x,y,z,nu)+u(t,x,y,z,nu)*diff(v(t,x
,y,z,nu),x)+2*v(t,x,y,z,nu)*diff(v(t,x,y,z,nu),y)+diff(v(t,x,y,z,nu),z
)*w(t,x,y,z,nu)+v(t,x,y,z,nu)*diff(w(t,x,y,z,nu),z)+diff(p(t,x,y,z,nu)
,y)-nu*(diff(v(t,x,y,z,nu),`$`(x,2))+diff(v(t,x,y,z,nu),`$`(y,2))+diff
(v(t,x,y,z,nu),`$`(z,2)));}{%
\maplemultiline{
\mathit{N2} := ({\frac {\partial }{\partial t}}\,\mathrm{v}(t, \,
x, \,y, \,z, \,\nu )) + \mathrm{U}(t, \,x, \,y, \,z, \,\nu )\,(
{\frac {\partial }{\partial x}}\,\mathrm{v}(t, \,x, \,y, \,z, \,
\nu )) + ({\frac {\partial }{\partial y}}\,\mathrm{P}(t, \,x, \,y
, \,z, \,\nu )) \\
\mbox{} + ({\frac {\partial }{\partial x}}\,\mathrm{u}(t, \,x, \,
y, \,z, \,\nu ))\,\mathrm{v}(t, \,x, \,y, \,z, \,\nu ) + \mathrm{
u}(t, \,x, \,y, \,z, \,\nu )\,({\frac {\partial }{\partial x}}\,
\mathrm{v}(t, \,x, \,y, \,z, \,\nu )) \\
\mbox{} + 2\,\mathrm{v}(t, \,x, \,y, \,z, \,\nu )\,({\frac {
\partial }{\partial y}}\,\mathrm{v}(t, \,x, \,y, \,z, \,\nu )) +
({\frac {\partial }{\partial z}}\,\mathrm{v}(t, \,x, \,y, \,z, \,
\nu ))\,\mathrm{w}(t, \,x, \,y, \,z, \,\nu ) \\
\mbox{} + \mathrm{v}(t, \,x, \,y, \,z, \,\nu )\,({\frac {
\partial }{\partial z}}\,\mathrm{w}(t, \,x, \,y, \,z, \,\nu )) +
({\frac {\partial }{\partial y}}\,\mathrm{p}(t, \,x, \,y, \,z, \,
\nu )) \\
\mbox{} - \nu \,(({\frac {\partial ^{2}}{\partial x^{2}}}\,
\mathrm{v}(t, \,x, \,y, \,z, \,\nu )) + ({\frac {\partial ^{2}}{
\partial y^{2}}}\,\mathrm{v}(t, \,x, \,y, \,z, \,\nu )) + (
{\frac {\partial ^{2}}{\partial z^{2}}}\,\mathrm{v}(t, \,x, \,y,
\,z, \,\nu ))) }
}
\end{maplelatex}

\end{maplegroup}
\begin{maplegroup}
\begin{mapleinput}
\mapleinline{active}{1d}{N3:=diff(w(X),t)+U(X)*diff(w(X),x)+diff(w(X)*u(X),x)+diff(w(X)*v(X),y
)+diff(w(X)*w(X),z)+diff(p(X),z)-nu*(diff(w(X),x,x)+diff(w(X),y,y)+dif
f(w(X),z,z));}{%
}
\end{mapleinput}

\mapleresult
\begin{maplelatex}
\mapleinline{inert}{2d}{N3 :=
diff(w(t,x,y,z,nu),t)+U(t,x,y,z,nu)*diff(w(t,x,y,z,nu),x)+diff(u(t,x,y
,z,nu),x)*w(t,x,y,z,nu)+u(t,x,y,z,nu)*diff(w(t,x,y,z,nu),x)+diff(v(t,x
,y,z,nu),y)*w(t,x,y,z,nu)+v(t,x,y,z,nu)*diff(w(t,x,y,z,nu),y)+2*w(t,x,
y,z,nu)*diff(w(t,x,y,z,nu),z)+diff(p(t,x,y,z,nu),z)-nu*(diff(w(t,x,y,z
,nu),`$`(x,2))+diff(w(t,x,y,z,nu),`$`(y,2))+diff(w(t,x,y,z,nu),`$`(z,2
)));}{%
\maplemultiline{
\mathit{N3} := ({\frac {\partial }{\partial t}}\,\mathrm{w}(t, \,
x, \,y, \,z, \,\nu )) + \mathrm{U}(t, \,x, \,y, \,z, \,\nu )\,(
{\frac {\partial }{\partial x}}\,\mathrm{w}(t, \,x, \,y, \,z, \,
\nu )) \\
\mbox{} + ({\frac {\partial }{\partial x}}\,\mathrm{u}(t, \,x, \,
y, \,z, \,\nu ))\,\mathrm{w}(t, \,x, \,y, \,z, \,\nu ) + \mathrm{
u}(t, \,x, \,y, \,z, \,\nu )\,({\frac {\partial }{\partial x}}\,
\mathrm{w}(t, \,x, \,y, \,z, \,\nu )) \\
\mbox{} + ({\frac {\partial }{\partial y}}\,\mathrm{v}(t, \,x, \,
y, \,z, \,\nu ))\,\mathrm{w}(t, \,x, \,y, \,z, \,\nu ) + \mathrm{
v}(t, \,x, \,y, \,z, \,\nu )\,({\frac {\partial }{\partial y}}\,
\mathrm{w}(t, \,x, \,y, \,z, \,\nu )) \\
\mbox{} + 2\,\mathrm{w}(t, \,x, \,y, \,z, \,\nu )\,({\frac {
\partial }{\partial z}}\,\mathrm{w}(t, \,x, \,y, \,z, \,\nu )) +
({\frac {\partial }{\partial z}}\,\mathrm{p}(t, \,x, \,y, \,z, \,
\nu )) \\
\mbox{} - \nu \,(({\frac {\partial ^{2}}{\partial x^{2}}}\,
\mathrm{w}(t, \,x, \,y, \,z, \,\nu )) + ({\frac {\partial ^{2}}{
\partial y^{2}}}\,\mathrm{w}(t, \,x, \,y, \,z, \,\nu )) + (
{\frac {\partial ^{2}}{\partial z^{2}}}\,\mathrm{w}(t, \,x, \,y,
\,z, \,\nu ))) }
}
\end{maplelatex}

\end{maplegroup}
\begin{maplegroup}
\begin{flushleft}
{\large Equations (2.1), (2.2) \& (2.4) including the "velocity
product equations" (2.3):}
\end{flushleft}

\end{maplegroup}
\begin{maplegroup}
\begin{mapleinput}
\mapleinline{active}{1d}{eqnC:=C=0: eqnN1:=N1=0: eqnN2:=N2=0: eqnN3:=N3=0:
eqnP11:=2*N1*u(X)=0: eqnP22:=2*N2*v(X)=0: eqnP33:=2*N3*w(X)=0:
eqnP12:=N1*v(X)+N2*u(X)=0: eqnP13:=N1*w(X)+N3*u(X)=0:
eqnP23:=N2*w(X)+N3*v(X)=0:
eqnR1:=diff(U(X),t)=0: eqnR2:=diff(U(X),x)=0: eqnR3:=diff(U(X),z)=0:
eqnR4:=diff(P(X),t)=0: eqnR5:=diff(P(X),x)=0:
eqnR6:=diff(P(X),z)=0:}{%
}
\end{mapleinput}

\vspace{2em}
\end{maplegroup}
\begin{maplegroup}
\begin{flushleft}
{\large Symmetry Algorithm:}
\end{flushleft}

\end{maplegroup}
\begin{maplegroup}
\begin{mapleinput}
\mapleinline{active}{1d}{gem_init_defs([X],[u(X),v(X),w(X),p(X),U(X),P(X)],[],[K],0,[eqnC,eqnN
1,eqnN2,eqnN3,eqnP11,eqnP22,eqnP33,eqnP12,eqnP13,eqnP23,eqnR1,eqnR2,eq
nR3,eqnR4,eqnR5,eqnR6],[diff(u(X),x),diff(u(X),t),diff(v(X),t),diff(w(
X),t),diff(U(X),y,y),diff(P(X),y),diff(w(X),z,z),diff(v(X),x,x),diff(u
(X),z,z),diff(p(X),z),diff(U(X),t),diff(U(X),x),diff(U(X),z),diff(P(X)
,t),diff(P(X),x),diff(P(X),z)]);}{%
}
\end{mapleinput}

\mapleresult
\begin{maplelatex}
\mapleinline{inert}{2d}{`-- Starting definitions for Point Symmetry or Conservation Law
computation... --`;}{%
\[
\mathit{--\ Starting\ definitions\ for\ Point\ Symmetry\ or\
Conservation\ Law\ computation...\ --}
\]
}
\end{maplelatex}

\begin{maplelatex}
\mapleinline{inert}{2d}{`Independent variables: `, nu, t, x, y, z;}{%
\[
\mathit{Independent\ variables:\ }, \,\nu , \,t, \,x, \,y, \,z
\]
}
\end{maplelatex}

\begin{maplelatex}
\mapleinline{inert}{2d}{`Dependent variables :`, P, U, p, u, v, w;}{%
\[
\mathit{Dependent\ variables\ :}, \,P, \,U, \,p, \,u, \,v, \,w
\]
}
\end{maplelatex}

\begin{maplelatex}
\mapleinline{inert}{2d}{`Free functions: `, vector([]);}{%
\[
\mathit{Free\ functions:\ }, \,[]
\]
}
\end{maplelatex}

\begin{maplelatex}
\mapleinline{inert}{2d}{`Free constants: `, [K];}{%
\[
\mathit{Free\ constants:\ }, \,[K]
\]
}
\end{maplelatex}

\begin{maplelatex}
\mapleinline{inert}{2d}{` - Variable definition successful -`;}{%
\[
\mathit{\ -\ Variable\ definition\ successful\ -}
\]
}
\end{maplelatex}

\begin{maplelatex}
\mapleinline{inert}{2d}{`16 equation(s) defined.`;}{%
\[
\mathit{16\ equation(s)\ defined.}
\]
}
\end{maplelatex}

\begin{maplelatex}
\mapleinline{inert}{2d}{`Computing necessary differential consequences...`;}{%
\[
\mathit{Computing\ necessary\ differential\ consequences...}
\]
}
\end{maplelatex}

\begin{maplelatex}
\mapleinline{inert}{2d}{`29 differential consequence(s) computed.`;}{%
\[
\mathit{29\ differential\ consequence(s)\ computed.}
\]
}
\end{maplelatex}

\begin{maplelatex}
\mapleinline{inert}{2d}{`-- GeM initial definitions for Symmetry/Conservation Law analysis
successful. --`;}{%
\[
\mathit{--\ GeM\ initial\ definitions\ for\ Symmetry/Conservation
\ Law\ analysis\ successful.\ --}
\]
}
\end{maplelatex}

\end{maplegroup}
\begin{maplegroup}
\begin{mapleinput}
\mapleinline{active}{1d}{gem_solvedfor_get_():
overdet_sys:=gem_get_split_sys([X,u(X),v(X),w(X),p(X),U(X),P(X)],0):}{
}
\end{mapleinput}

\mapleresult

\begin{maplelatex}
\mapleinline{inert}{2d}{`Tangent Vector Field Coordinates defined successfuly: `, \{eta_P,
eta_U, eta_p, eta_u, eta_v, eta_w, xi_nu, xi_t, xi_x, xi_y, xi_z\};}{%
\maplemultiline{
\mathit{Tangent\ Vector\ Field\ Coordinates\ defined\
successfuly:\ },  \\
\{\mathit{eta\_P}, \,\mathit{eta\_U}, \,\mathit{eta\_p}, \,
\mathit{eta\_u}, \,\mathit{eta\_v}, \,\mathit{eta\_w}, \,\mathit{
xi\_nu}, \,\mathit{xi\_t}, \,\mathit{xi\_x}, \,\mathit{xi\_y}, \,
\mathit{xi\_z}\} }
}
\end{maplelatex}

\begin{maplelatex}
\mapleinline{inert}{2d}{`Generating t.v.f. coordinates of derivatives...`;}{%
\[
\mathit{Generating\ t.v.f.\ coordinates\ of\ derivatives...}
\]
}
\end{maplelatex}

\begin{maplelatex}
\mapleinline{inert}{2d}{`Done. Generating determining equations...`;}{%
\[
\mathit{Done.\ Generating\ determining\ equations...}
\]
}
\end{maplelatex}

\begin{maplelatex}
\mapleinline{inert}{2d}{`Done. Splitting...`;}{%
\[
\mathit{Done.\ Splitting...}
\]
}
\end{maplelatex}

\begin{maplelatex}
\mapleinline{inert}{2d}{`Split successful. The split system returned. Number of equations: `,
4865;}{%
\[
\mathit{Split\ successful.\ The\ split\ system\ returned.\ Number
\ of\ equations:\ }, \,4865
\]
}
\end{maplelatex}

\end{maplegroup}
\begin{maplegroup}
\begin{mapleinput}
\mapleinline{active}{1d}{twf_coords:=gem_tvf_coords_get_():
simplified_system:=rifsimp(overdet_sys,twf_coords):
sym_sol:=pdsolve(simplified_system[Solved],twf_coords);}{%
}
\end{mapleinput}

\mapleresult
\begin{maplelatex}
\mapleinline{inert}{2d}{sym_sol := \{eta_P =
-_F12(y,nu,U,P)-_F14(nu,U,P)+2*(-_F9(nu)+_F6(nu))*P-_F16(nu,U)+_F18(nu
), eta_U = -_F11(y,nu,U,P)-_F9(nu)*U+_F6(nu)*U, eta_p =
-z*_F5[t,t]-x*_F8[t,t]+_F12(y,nu,U,P)+_F14(nu,U,P)+_F16(nu,U)+_F17(t,n
u)+(2*p-x*K)*_F6(nu)+(-2*p+2*x*K)*_F9(nu)-z*_F3(nu)*K, eta_u =
-_F3(nu)*w+(-_F9(nu)+_F6(nu))*u+_F8[t]+_F11(y,nu,U,P), eta_v =
v*(-_F9(nu)+_F6(nu)), eta_w =
_F5[t]+(u+U)*_F3(nu)-_F9(nu)*w+w*_F6(nu), xi_nu =
nu*(2*_F6(nu)-_F9(nu)), xi_t = _F9(nu)*t+_F10(nu), xi_x =
_F6(nu)*x-_F3(nu)*z+_F8(t,nu), xi_y = _F6(nu)*y+_F7(nu), xi_z =
_F3(nu)*x+_F6(nu)*z+_F5(t,nu)\};}{%
\maplemultiline{
\mathit{sym\_sol} := \{\mathit{eta\_P}= - \mathrm{\_F12}(y, \,\nu
 , \,U, \,P) - \mathrm{\_F14}(\nu , \,U, \,P) + 2\,( - \mathrm{
\_F9}(\nu ) + \mathrm{\_F6}(\nu ))\,P \\
\mbox{} - \mathrm{\_F16}(\nu , \,U) + \mathrm{\_F18}(\nu ), \,
\mathit{eta\_U}= - \mathrm{\_F11}(y, \,\nu , \,U, \,P) - \mathrm{
\_F9}(\nu )\,U + \mathrm{\_F6}(\nu )\,U,  \\
\mathit{eta\_p}= - z\,{\mathit{\_F5}_{t, \,t}} - x\,{\mathit{\_F8
}_{t, \,t}} + \mathrm{\_F12}(y, \,\nu , \,U, \,P) + \mathrm{\_F14
}(\nu , \,U, \,P) + \mathrm{\_F16}(\nu , \,U) \\
\mbox{} + \mathrm{\_F17}(t, \,\nu ) + (2\,p - x\,K)\,\mathrm{\_F6
}(\nu ) + ( - 2\,p + 2\,x\,K)\,\mathrm{\_F9}(\nu ) - z\,\mathrm{
\_F3}(\nu )\,K,  \\
\mathit{eta\_u}= - \mathrm{\_F3}(\nu )\,w + ( - \mathrm{\_F9}(\nu
 ) + \mathrm{\_F6}(\nu ))\,u + {\mathit{\_F8}_{t}} + \mathrm{
\_F11}(y, \,\nu , \,U, \,P),  \\
\mathit{eta\_v}=v\,( - \mathrm{\_F9}(\nu ) + \mathrm{\_F6}(\nu ))
,  \\
\mathit{eta\_w}={\mathit{\_F5}_{t}} + (u + U)\,\mathrm{\_F3}(\nu
) - \mathrm{\_F9}(\nu )\,w + w\,\mathrm{\_F6}(\nu ),  \\
\mathit{xi\_nu}=\nu \,(2\,\mathrm{\_F6}(\nu ) - \mathrm{\_F9}(\nu
 )), \,\mathit{xi\_t}=\mathrm{\_F9}(\nu )\,t + \mathrm{\_F10}(\nu
 ),  \\
\mathit{xi\_x}=\mathrm{\_F6}(\nu )\,x - \mathrm{\_F3}(\nu )\,z +
\mathrm{\_F8}(t, \,\nu ), \,\mathit{xi\_y}=\mathrm{\_F6}(\nu )\,y
 + \mathrm{\_F7}(\nu ),  \\
\mathit{xi\_z}=\mathrm{\_F3}(\nu )\,x + \mathrm{\_F6}(\nu )\,z +
\mathrm{\_F5}(t, \,\nu )\} }
}
\end{maplelatex}

\end{maplegroup}
\begin{maplegroup}
\begin{flushleft}
{\large Redefinition of group functions as used in (2.10):}
\end{flushleft}

\end{maplegroup}
\begin{maplegroup}
\begin{mapleinput}
\mapleinline{active}{1d}{_F6(nu):=a1(nu); _F3(nu):=-a2(nu); _F8(t,nu):=f1(t,nu);
_F7(nu):=a3(nu); _F5(t,nu):=f2(t,nu); _F9(nu):=a4(nu);
_F10(nu):=a5(nu); _F11(y,nu,U,P):=-g1(y,nu,U,P);
_F17(t,nu):=f3(t,nu)-_F18(nu);
_F12(y,nu,U,P):=-g2(y,nu,U,P)-_F14(nu,U,P)-_F16(nu,U)+_F18(nu);}{%
}
\end{mapleinput}

\newpage
\mapleresult
\begin{maplelatex}
\mapleinline{inert}{2d}{_F6(nu) := a1(nu);}{%
\[
\mathrm{\_F6}(\nu ) := \mathrm{a1}(\nu )
\]
}
\end{maplelatex}

\begin{maplelatex}
\mapleinline{inert}{2d}{_F3(nu) := -a2(nu);}{%
\[
\mathrm{\_F3}(\nu ) :=  - \mathrm{a2}(\nu )
\]
}
\end{maplelatex}

\begin{maplelatex}
\mapleinline{inert}{2d}{_F8(t,nu) := f1(t,nu);}{%
\[
\mathrm{\_F8}(t, \,\nu ) := \mathrm{f1}(t, \,\nu )
\]
}
\end{maplelatex}

\begin{maplelatex}
\mapleinline{inert}{2d}{_F7(nu) := a3(nu);}{%
\[
\mathrm{\_F7}(\nu ) := \mathrm{a3}(\nu )
\]
}
\end{maplelatex}

\begin{maplelatex}
\mapleinline{inert}{2d}{_F5(t,nu) := f2(t,nu);}{%
\[
\mathrm{\_F5}(t, \,\nu ) := \mathrm{f2}(t, \,\nu )
\]
}
\end{maplelatex}

\begin{maplelatex}
\mapleinline{inert}{2d}{_F9(nu) := a4(nu);}{%
\[
\mathrm{\_F9}(\nu ) := \mathrm{a4}(\nu )
\]
}
\end{maplelatex}

\begin{maplelatex}
\mapleinline{inert}{2d}{_F10(nu) := a5(nu);}{%
\[
\mathrm{\_F10}(\nu ) := \mathrm{a5}(\nu )
\]
}
\end{maplelatex}

\begin{maplelatex}
\mapleinline{inert}{2d}{_F11(y,nu,U,P) := -g1(y,nu,U,P);}{%
\[
\mathrm{\_F11}(y, \,\nu , \,U, \,P) :=  - \mathrm{g1}(y, \,\nu ,
\,U, \,P)
\]
}
\end{maplelatex}

\begin{maplelatex}
\mapleinline{inert}{2d}{_F17(t,nu) := f3(t,nu)-_F18(nu);}{%
\[
\mathrm{\_F17}(t, \,\nu ) := \mathrm{f3}(t, \,\nu ) - \mathrm{
\_F18}(\nu )
\]
}
\end{maplelatex}

\begin{maplelatex}
\mapleinline{inert}{2d}{_F12(y,nu,U,P) := -g2(y,nu,U,P)-_F14(nu,U,P)-_F16(nu,U)+_F18(nu);}{%
\[
\mathrm{\_F12}(y, \,\nu , \,U, \,P) :=  - \mathrm{g2}(y, \,\nu ,
\,U, \,P) - \mathrm{\_F14}(\nu , \,U, \,P) - \mathrm{\_F16}(\nu
, \,U) + \mathrm{\_F18}(\nu )
\]
}
\end{maplelatex}

\end{maplegroup}
\begin{maplegroup}
\begin{flushleft}
{\large Final Result (still identical to result (2.10)):}
\end{flushleft}

\end{maplegroup}
\begin{maplegroup}
\begin{mapleinput}
\mapleinline{active}{1d}{sym_sol[9]; sym_sol[10]; sym_sol[11]; sym_sol[8]; sym_sol[7];
sym_sol[4]; sym_sol[5]; sym_sol[6]; sym_sol[3]; sym_sol[2];
sym_sol[1];}{%
}
\end{mapleinput}

\mapleresult
\begin{maplelatex}
\mapleinline{inert}{2d}{xi_x = a1(nu)*x+a2(nu)*z+f1(t,nu);}{%
\[
\mathit{xi\_x}=\mathrm{a1}(\nu )\,x + \mathrm{a2}(\nu )\,z +
\mathrm{f1}(t, \,\nu )
\]
}
\end{maplelatex}

\begin{maplelatex}
\mapleinline{inert}{2d}{xi_y = a1(nu)*y+a3(nu);}{%
\[
\mathit{xi\_y}=\mathrm{a1}(\nu )\,y + \mathrm{a3}(\nu )
\]
}
\end{maplelatex}

\begin{maplelatex}
\mapleinline{inert}{2d}{xi_z = -a2(nu)*x+a1(nu)*z+f2(t,nu);}{%
\[
\mathit{xi\_z}= - \mathrm{a2}(\nu )\,x + \mathrm{a1}(\nu )\,z +
\mathrm{f2}(t, \,\nu )
\]
}
\end{maplelatex}

\begin{maplelatex}
\mapleinline{inert}{2d}{xi_t = a4(nu)*t+a5(nu);}{%
\[
\mathit{xi\_t}=\mathrm{a4}(\nu )\,t + \mathrm{a5}(\nu )
\]
}
\end{maplelatex}

\begin{maplelatex}
\mapleinline{inert}{2d}{xi_nu = nu*(2*a1(nu)-a4(nu));}{%
\[
\mathit{xi\_nu}=\nu \,(2\,\mathrm{a1}(\nu ) - \mathrm{a4}(\nu ))
\]
}
\end{maplelatex}

\begin{maplelatex}
\mapleinline{inert}{2d}{eta_u = a2(nu)*w+(-a4(nu)+a1(nu))*u+f1[t]-g1(y,nu,U,P);}{%
\[
\mathit{eta\_u}=\mathrm{a2}(\nu )\,w + ( - \mathrm{a4}(\nu ) +
\mathrm{a1}(\nu ))\,u + {\mathit{f1}_{t}} - \mathrm{g1}(y, \,\nu
, \,U, \,P)
\]
}
\end{maplelatex}

\begin{maplelatex}
\mapleinline{inert}{2d}{eta_v = v*(-a4(nu)+a1(nu));}{%
\[
\mathit{eta\_v}=v\,( - \mathrm{a4}(\nu ) + \mathrm{a1}(\nu ))
\]
}
\end{maplelatex}

\begin{maplelatex}
\mapleinline{inert}{2d}{eta_w = f2[t]-(u+U)*a2(nu)-a4(nu)*w+w*a1(nu);}{%
\[
\mathit{eta\_w}={\mathit{f2}_{t}} - (u + U)\,\mathrm{a2}(\nu ) -
\mathrm{a4}(\nu )\,w + w\,\mathrm{a1}(\nu )
\]
}
\end{maplelatex}

\vspace{-0.25em}
\begin{maplelatex}
\mapleinline{inert}{2d}{eta_p =
-z*f2[t,t]-x*f1[t,t]-g2(y,nu,U,P)+f3(t,nu)+(2*p-x*K)*a1(nu)+(-2*p+2*x*
K)*a4(nu)+z*a2(nu)*K;}{%
\maplemultiline{
\mathit{eta\_p}= - z\,{\mathit{f2}_{t, \,t}} - x\,{\mathit{f1}_{t
, \,t}} - \mathrm{g2}(y, \,\nu , \,U, \,P) + \mathrm{f3}(t, \,\nu
 )\\ + (2\,p - x\,K)\,\mathrm{a1}(\nu ) + ( - 2\,p + 2\,x\,K)\,
\mathrm{a4}(\nu ) + z\,\mathrm{a2}(\nu )\,K }
}
\end{maplelatex}

\vspace{0.25em}
\begin{maplelatex}
\mapleinline{inert}{2d}{eta_U = g1(y,nu,U,P)-a4(nu)*U+a1(nu)*U;}{%
\[
\mathit{eta\_U}=\mathrm{g1}(y, \,\nu , \,U, \,P) - \mathrm{a4}(
\nu )\,U + \mathrm{a1}(\nu )\,U
\]
}
\end{maplelatex}

\vspace{0.25em}
\begin{maplelatex}
\mapleinline{inert}{2d}{eta_P = g2(y,nu,U,P)+2*(-a4(nu)+a1(nu))*P;}{%
\[
\mathit{eta\_P}=\mathrm{g2}(y, \,\nu , \,U, \,P) + 2\,( -
\mathrm{a4}(\nu ) + \mathrm{a1}(\nu ))\,P
\]
}
\end{maplelatex}

\end{maplegroup}
\begin{maplegroup}
\begin{mapleinput}
\end{mapleinput}

\end{maplegroup}

\vspace{1em}\noindent {\bf IIa. SADE-Package
\textbf{\emph{excluding}} the $\pmb{\mathscr{P}_{ij}}$-equations
(\ref{131228:1928}):}
\begin{maplegroup}
\begin{flushleft}
{\large Header:}
\end{flushleft}

\end{maplegroup}
\begin{maplegroup}
\begin{mapleinput}
\mapleinline{active}{1d}{restart: with(sade):}{%
}
\end{mapleinput}

\mapleresult
\begin{maplelatex}
\mapleinline{inert}{2d}{`Symmetry Analysis of Differential Equations`;}{%
\[
\mathit{Symmetry\ Analysis\ of\ Differential\ Equations}
\]
}
\end{maplelatex}

\begin{maplelatex}
\mapleinline{inert}{2d}{`By Tarcisio M. Rocha Filho - Annibal Figueiredo - 2010`;}{%
\[
\mathit{By\ Tarcisio\ M.\ Rocha\ Filho\ -\ Annibal\ Figueiredo\ -
\ 2010}
\]
}
\end{maplelatex}

\end{maplegroup}
\begin{maplegroup}
\begin{flushleft}
{\large Definitions:}
\end{flushleft}

\end{maplegroup}
\begin{maplegroup}
\begin{mapleinput}
\mapleinline{active}{1d}{alias(sigma=(t,x,y,z,nu,u,v,w,p,U,P)): X:=(t,x,y,z,nu): }{%
}
\end{mapleinput}

\end{maplegroup}
\begin{maplegroup}
\begin{mapleinput}
\mapleinline{active}{1d}{C:=diff(u(X),x)+diff(v(X),y)+diff(w(X),z);}{%
}
\end{mapleinput}

\mapleresult
\begin{maplelatex}
\mapleinline{inert}{2d}{C :=
diff(u(t,x,y,z,nu),x)+diff(v(t,x,y,z,nu),y)+diff(w(t,x,y,z,nu),z);}{%
\[
C := ({\frac {\partial }{\partial x}}\,\mathrm{u}(t, \,x, \,y, \,
z, \,\nu )) + ({\frac {\partial }{\partial y}}\,\mathrm{v}(t, \,x
, \,y, \,z, \,\nu )) + ({\frac {\partial }{\partial z}}\,\mathrm{
w}(t, \,x, \,y, \,z, \,\nu ))
\]
}
\end{maplelatex}

\end{maplegroup}
\begin{maplegroup}
\begin{mapleinput}
\mapleinline{active}{1d}{N1:=diff(u(X),t)+U(X)*diff(u(X),x)+v(X)*diff(U(X),y)-(K+nu*diff(U(X),
y,y))+diff(u(X)*u(X),x)+diff(u(X)*v(X),y)+diff(u(X)*w(X),z)+diff(p(X),
x)-nu*(diff(u(X),x,x)+diff(u(X),y,y)+diff(u(X),z,z));}{%
}
\end{mapleinput}

\mapleresult
\begin{maplelatex}
\mapleinline{inert}{2d}{N1 :=
diff(u(t,x,y,z,nu),t)+U(t,x,y,z,nu)*diff(u(t,x,y,z,nu),x)+v(t,x,y,z,nu
)*diff(U(t,x,y,z,nu),y)-K-nu*diff(U(t,x,y,z,nu),`$`(y,2))+2*u(t,x,y,z,
nu)*diff(u(t,x,y,z,nu),x)+diff(u(t,x,y,z,nu),y)*v(t,x,y,z,nu)+u(t,x,y,
z,nu)*diff(v(t,x,y,z,nu),y)+diff(u(t,x,y,z,nu),z)*w(t,x,y,z,nu)+u(t,x,
y,z,nu)*diff(w(t,x,y,z,nu),z)+diff(p(t,x,y,z,nu),x)-nu*(diff(u(t,x,y,z
,nu),`$`(x,2))+diff(u(t,x,y,z,nu),`$`(y,2))+diff(u(t,x,y,z,nu),`$`(z,2
)));}{%
\maplemultiline{
\mathit{N1} := ({\frac {\partial }{\partial t}}\,\mathrm{u}(t, \,
x, \,y, \,z, \,\nu )) + \mathrm{U}(t, \,x, \,y, \,z, \,\nu )\,(
{\frac {\partial }{\partial x}}\,\mathrm{u}(t, \,x, \,y, \,z, \,
\nu )) \\
\mbox{} + \mathrm{v}(t, \,x, \,y, \,z, \,\nu )\,({\frac {
\partial }{\partial y}}\,\mathrm{U}(t, \,x, \,y, \,z, \,\nu )) -
K - \nu \,({\frac {\partial ^{2}}{\partial y^{2}}}\,\mathrm{U}(t
, \,x, \,y, \,z, \,\nu )) \\
\mbox{} + 2\,\mathrm{u}(t, \,x, \,y, \,z, \,\nu )\,({\frac {
\partial }{\partial x}}\,\mathrm{u}(t, \,x, \,y, \,z, \,\nu )) +
({\frac {\partial }{\partial y}}\,\mathrm{u}(t, \,x, \,y, \,z, \,
\nu ))\,\mathrm{v}(t, \,x, \,y, \,z, \,\nu ) \\
\mbox{} + \mathrm{u}(t, \,x, \,y, \,z, \,\nu )\,({\frac {
\partial }{\partial y}}\,\mathrm{v}(t, \,x, \,y, \,z, \,\nu )) +
({\frac {\partial }{\partial z}}\,\mathrm{u}(t, \,x, \,y, \,z, \,
\nu ))\,\mathrm{w}(t, \,x, \,y, \,z, \,\nu ) \\
\mbox{} + \mathrm{u}(t, \,x, \,y, \,z, \,\nu )\,({\frac {
\partial }{\partial z}}\,\mathrm{w}(t, \,x, \,y, \,z, \,\nu )) +
({\frac {\partial }{\partial x}}\,\mathrm{p}(t, \,x, \,y, \,z, \,
\nu )) \\
\mbox{} - \nu \,(({\frac {\partial ^{2}}{\partial x^{2}}}\,
\mathrm{u}(t, \,x, \,y, \,z, \,\nu )) + ({\frac {\partial ^{2}}{
\partial y^{2}}}\,\mathrm{u}(t, \,x, \,y, \,z, \,\nu )) + (
{\frac {\partial ^{2}}{\partial z^{2}}}\,\mathrm{u}(t, \,x, \,y,
\,z, \,\nu ))) }
}
\end{maplelatex}

\end{maplegroup}
\begin{maplegroup}
\begin{mapleinput}
\mapleinline{active}{1d}{N2:=diff(v(X),t)+U(X)*diff(v(X),x)+diff(P(X),y)+diff(v(X)*u(X),x)+dif
f(v(X)*v(X),y)+diff(v(X)*w(X),z)+diff(p(X),y)-nu*(diff(v(X),x,x)+diff(
v(X),y,y)+diff(v(X),z,z));}{%
}
\end{mapleinput}

\mapleresult
\begin{maplelatex}
\mapleinline{inert}{2d}{N2 :=
diff(v(t,x,y,z,nu),t)+U(t,x,y,z,nu)*diff(v(t,x,y,z,nu),x)+diff(P(t,x,y
,z,nu),y)+diff(u(t,x,y,z,nu),x)*v(t,x,y,z,nu)+u(t,x,y,z,nu)*diff(v(t,x
,y,z,nu),x)+2*v(t,x,y,z,nu)*diff(v(t,x,y,z,nu),y)+diff(v(t,x,y,z,nu),z
)*w(t,x,y,z,nu)+v(t,x,y,z,nu)*diff(w(t,x,y,z,nu),z)+diff(p(t,x,y,z,nu)
,y)-nu*(diff(v(t,x,y,z,nu),`$`(x,2))+diff(v(t,x,y,z,nu),`$`(y,2))+diff
(v(t,x,y,z,nu),`$`(z,2)));}{%
\maplemultiline{
\mathit{N2} := ({\frac {\partial }{\partial t}}\,\mathrm{v}(t, \,
x, \,y, \,z, \,\nu )) + \mathrm{U}(t, \,x, \,y, \,z, \,\nu )\,(
{\frac {\partial }{\partial x}}\,\mathrm{v}(t, \,x, \,y, \,z, \,
\nu )) + ({\frac {\partial }{\partial y}}\,\mathrm{P}(t, \,x, \,y
, \,z, \,\nu )) \\
\mbox{} + ({\frac {\partial }{\partial x}}\,\mathrm{u}(t, \,x, \,
y, \,z, \,\nu ))\,\mathrm{v}(t, \,x, \,y, \,z, \,\nu ) + \mathrm{
u}(t, \,x, \,y, \,z, \,\nu )\,({\frac {\partial }{\partial x}}\,
\mathrm{v}(t, \,x, \,y, \,z, \,\nu )) \\
\mbox{} + 2\,\mathrm{v}(t, \,x, \,y, \,z, \,\nu )\,({\frac {
\partial }{\partial y}}\,\mathrm{v}(t, \,x, \,y, \,z, \,\nu )) +
({\frac {\partial }{\partial z}}\,\mathrm{v}(t, \,x, \,y, \,z, \,
\nu ))\,\mathrm{w}(t, \,x, \,y, \,z, \,\nu ) \\
\mbox{} + \mathrm{v}(t, \,x, \,y, \,z, \,\nu )\,({\frac {
\partial }{\partial z}}\,\mathrm{w}(t, \,x, \,y, \,z, \,\nu )) +
({\frac {\partial }{\partial y}}\,\mathrm{p}(t, \,x, \,y, \,z, \,
\nu )) \\
\mbox{} - \nu \,(({\frac {\partial ^{2}}{\partial x^{2}}}\,
\mathrm{v}(t, \,x, \,y, \,z, \,\nu )) + ({\frac {\partial ^{2}}{
\partial y^{2}}}\,\mathrm{v}(t, \,x, \,y, \,z, \,\nu )) + (
{\frac {\partial ^{2}}{\partial z^{2}}}\,\mathrm{v}(t, \,x, \,y,
\,z, \,\nu ))) }
}
\end{maplelatex}

\end{maplegroup}
\begin{maplegroup}
\begin{mapleinput}
\mapleinline{active}{1d}{N3:=diff(w(X),t)+U(X)*diff(w(X),x)+diff(w(X)*u(X),x)+diff(w(X)*v(X),y
)+diff(w(X)*w(X),z)+diff(p(X),z)-nu*(diff(w(X),x,x)+diff(w(X),y,y)+dif
f(w(X),z,z));}{%
}
\end{mapleinput}

\mapleresult
\begin{maplelatex}
\mapleinline{inert}{2d}{N3 :=
diff(w(t,x,y,z,nu),t)+U(t,x,y,z,nu)*diff(w(t,x,y,z,nu),x)+diff(u(t,x,y
,z,nu),x)*w(t,x,y,z,nu)+u(t,x,y,z,nu)*diff(w(t,x,y,z,nu),x)+diff(v(t,x
,y,z,nu),y)*w(t,x,y,z,nu)+v(t,x,y,z,nu)*diff(w(t,x,y,z,nu),y)+2*w(t,x,
y,z,nu)*diff(w(t,x,y,z,nu),z)+diff(p(t,x,y,z,nu),z)-nu*(diff(w(t,x,y,z
,nu),`$`(x,2))+diff(w(t,x,y,z,nu),`$`(y,2))+diff(w(t,x,y,z,nu),`$`(z,2
)));}{%
\maplemultiline{
\mathit{N3} := ({\frac {\partial }{\partial t}}\,\mathrm{w}(t, \,
x, \,y, \,z, \,\nu )) + \mathrm{U}(t, \,x, \,y, \,z, \,\nu )\,(
{\frac {\partial }{\partial x}}\,\mathrm{w}(t, \,x, \,y, \,z, \,
\nu )) \\
\mbox{} + ({\frac {\partial }{\partial x}}\,\mathrm{u}(t, \,x, \,
y, \,z, \,\nu ))\,\mathrm{w}(t, \,x, \,y, \,z, \,\nu ) + \mathrm{
u}(t, \,x, \,y, \,z, \,\nu )\,({\frac {\partial }{\partial x}}\,
\mathrm{w}(t, \,x, \,y, \,z, \,\nu )) \\
\mbox{} + ({\frac {\partial }{\partial y}}\,\mathrm{v}(t, \,x, \,
y, \,z, \,\nu ))\,\mathrm{w}(t, \,x, \,y, \,z, \,\nu ) + \mathrm{
v}(t, \,x, \,y, \,z, \,\nu )\,({\frac {\partial }{\partial y}}\,
\mathrm{w}(t, \,x, \,y, \,z, \,\nu )) \\
\mbox{} + 2\,\mathrm{w}(t, \,x, \,y, \,z, \,\nu )\,({\frac {
\partial }{\partial z}}\,\mathrm{w}(t, \,x, \,y, \,z, \,\nu )) +
({\frac {\partial }{\partial z}}\,\mathrm{p}(t, \,x, \,y, \,z, \,
\nu )) \\
\mbox{} - \nu \,(({\frac {\partial ^{2}}{\partial x^{2}}}\,
\mathrm{w}(t, \,x, \,y, \,z, \,\nu )) + ({\frac {\partial ^{2}}{
\partial y^{2}}}\,\mathrm{w}(t, \,x, \,y, \,z, \,\nu )) + (
{\frac {\partial ^{2}}{\partial z^{2}}}\,\mathrm{w}(t, \,x, \,y,
\,z, \,\nu ))) }
}
\end{maplelatex}

\end{maplegroup}
\begin{maplegroup}
\begin{flushleft}
{\large Equations (2.1), (2.2) \& (2.4):}
\end{flushleft}

\end{maplegroup}
\begin{maplegroup}
\begin{mapleinput}
\mapleinline{active}{1d}{eqnC:=C=0: eqnN1:=N1=0: eqnN2:=N2=0: eqnN3:=N3=0:
eqnR1:=diff(U(X),t)=0: eqnR2:=diff(U(X),x)=0: eqnR3:=diff(U(X),z)=0:
eqnR4:=diff(P(X),t)=0: eqnR5:=diff(P(X),x)=0:
eqnR6:=diff(P(X),z)=0:
}{%
}
\end{mapleinput}

\end{maplegroup}
\begin{maplegroup}
\begin{mapleinput}
\mapleinline{active}{1d}{eqns:=[eqnC,eqnN1,eqnN2,eqnN3,eqnR1,eqnR2,eqnR3,eqnR4,eqnR5,eqnR6]:}{
}
\end{mapleinput}

\end{maplegroup}
\begin{maplegroup}
\begin{flushleft}
{\large Symmetry Algorithm:}
\end{flushleft}

\end{maplegroup}
\begin{maplegroup}
\begin{flushleft}
\textit{{\large a) Size of the determining system:}}
\end{flushleft}

\end{maplegroup}
\begin{maplegroup}
\begin{mapleinput}
\mapleinline{active}{1d}{detsys:=liesymmetries(eqns,[u(X),v(X),w(X),p(X),U(X),P(X)],determining):
nops(detsys[1]);}{%
}
\end{mapleinput}

\mapleresult
\begin{maplelatex}
\mapleinline{inert}{2d}{202;}{%
\[
202
\]
}
\end{maplelatex}

\end{maplegroup}
\begin{maplegroup}
\begin{flushleft}
\textit{{\large b) Solving the determining system: }}
\end{flushleft}

\end{maplegroup}
\begin{maplegroup}
\begin{mapleinput}
\mapleinline{active}{1d}{symsol:=liesymmetries(eqns,[u(X),v(X),w(X),p(X),U(X),P(X)]);}{%
}
\end{mapleinput}

\mapleresult
\begin{maplelatex}
\mapleinline{inert}{2d}{symsol := [\{_F5(nu)*D[y], _F7(nu)*D[t], _F10(nu,t)*D[p],
-_F1(U,P,nu,y)*D[u]+_F1(U,P,nu,y)*D[U],
-_F2(U,P,nu,y)*D[p]+_F2(U,P,nu,y)*D[P],
diff(_F4(nu,t),t)*D[w]-z*diff(_F4(nu,t),`$`(t,2))*D[p]+_F4(nu,t)*D[z],
diff(_F6(nu,t),t)*D[u]-x*diff(_F6(nu,t),`$`(t,2))*D[p]+_F6(nu,t)*D[x],
z*_F8(nu)*K*D[p]+_F8(nu)*w*D[u]+_F8(nu)*(-u-U)*D[w]+_F8(nu)*z*D[x]-_F8
(nu)*x*D[z],
-_F9(nu)*nu*D[nu]+_F9(nu)*(-2*P-2*p+2*x*K)*D[p]+t*_F9(nu)*D[t]+_F9(nu)
*(-u-U)*D[u]-_F9(nu)*v*D[v]-_F9(nu)*w*D[w],
x*_F3(nu)*K*D[p]+_F3(nu)*nu*D[nu]+t*_F3(nu)*D[t]+_F3(nu)*x*D[x]+_F3(nu
)*y*D[y]+_F3(nu)*z*D[z]\}, \{\}];}{%
\maplemultiline{
\mathit{symsol} := [\{\mathrm{\_F5}(\nu )\,{\mathrm{D}_{y}}, \,
\mathrm{\_F7}(\nu )\,{\mathrm{D}_{t}}, \,\mathrm{\_F10}(\nu , \,t
)\,{\mathrm{D}_{p}}, \, - \mathrm{\_F1}(U, \,P, \,\nu , \,y)\,{
\mathrm{D}_{u}} + \mathrm{\_F1}(U, \,P, \,\nu , \,y)\,{\mathrm{D}
_{U}},  \\
 - \mathrm{\_F2}(U, \,P, \,\nu , \,y)\,{\mathrm{D}_{p}} +
\mathrm{\_F2}(U, \,P, \,\nu , \,y)\,{\mathrm{D}_{P}},  \\
({\frac {\partial }{\partial t}}\,\mathrm{\_F4}(\nu , \,t))\,{
\mathrm{D}_{w}} - z\,({\frac {\partial ^{2}}{\partial t^{2}}}\,
\mathrm{\_F4}(\nu , \,t))\,{\mathrm{D}_{p}} + \mathrm{\_F4}(\nu
, \,t)\,{\mathrm{D}_{z}},  \\
({\frac {\partial }{\partial t}}\,\mathrm{\_F6}(\nu , \,t))\,{
\mathrm{D}_{u}} - x\,({\frac {\partial ^{2}}{\partial t^{2}}}\,
\mathrm{\_F6}(\nu , \,t))\,{\mathrm{D}_{p}} + \mathrm{\_F6}(\nu
, \,t)\,{\mathrm{D}_{x}},  \\
z\,\mathrm{\_F8}(\nu )\,K\,{\mathrm{D}_{p}} + \mathrm{\_F8}(\nu )
\,w\,{\mathrm{D}_{u}} + \mathrm{\_F8}(\nu )\,( - u - U)\,{
\mathrm{D}_{w}} + \mathrm{\_F8}(\nu )\,z\,{\mathrm{D}_{x}} -
\mathrm{\_F8}(\nu )\,x\,{\mathrm{D}_{z}},  \\
 - \mathrm{\_F9}(\nu )\,\nu \,{\mathrm{D}_{\nu }} + \mathrm{\_F9}
(\nu )\,( - 2\,P - 2\,p + 2\,x\,K)\,{\mathrm{D}_{p}} + t\,
\mathrm{\_F9}(\nu )\,{\mathrm{D}_{t}} \\
\mbox{} + \mathrm{\_F9}(\nu )\,( - u - U)\,{\mathrm{D}_{u}} -
\mathrm{\_F9}(\nu )\,v\,{\mathrm{D}_{v}} - \mathrm{\_F9}(\nu )\,w
\,{\mathrm{D}_{w}}, x\,\mathrm{\_F3}(\nu )\,K\,{\mathrm{D}_{p}}
 + \mathrm{\_F3}(\nu )\,\nu \,{\mathrm{D}_{\nu }} \\
\mbox{} + t\,\mathrm{\_F3}(\nu )\,{\mathrm{D}_{t}} + \mathrm{\_F3
}(\nu )\,x\,{\mathrm{D}_{x}} + \mathrm{\_F3}(\nu )\,y\,{\mathrm{D
}_{y}} + \mathrm{\_F3}(\nu )\,z\,{\mathrm{D}_{z}}\}, \,\{\}] }
}
\end{maplelatex}

\end{maplegroup}
\begin{maplegroup}
\begin{flushleft}
{\large Redefinition of group functions as used (2.10):}
\end{flushleft}

\end{maplegroup}
\begin{maplegroup}
\begin{mapleinput}
\mapleinline{active}{1d}{_F6(nu,t):=f1(t,nu); _F3(nu):=a1(nu); _F8(nu):=a2(nu);
_F5(nu):=a3(nu); _F4(nu,t):=f2(t,nu); _F9(nu):=-a1(nu)+a4(nu);
_F7(nu):=a5(nu); _F1(U,P,nu,y):=(a1(nu)-a4(nu))*U+g1(y,nu,U,P);
_F2(U,P,nu,y):=2*(a1(nu)-a4(nu))*P+g2(y,nu,U,P);
_F10(nu,t):=f3(t,nu);}{%
}
\end{mapleinput}

\mapleresult
\begin{maplelatex}
\mapleinline{inert}{2d}{_F6(nu,t) := f1(t,nu);}{%
\[
\mathrm{\_F6}(\nu , \,t) := \mathrm{f1}(t, \,\nu )
\]
}
\end{maplelatex}

\begin{maplelatex}
\mapleinline{inert}{2d}{_F3(nu) := a1(nu);}{%
\[
\mathrm{\_F3}(\nu ) := \mathrm{a1}(\nu )
\]
}
\end{maplelatex}

\begin{maplelatex}
\mapleinline{inert}{2d}{_F8(nu) := a2(nu);}{%
\[
\mathrm{\_F8}(\nu ) := \mathrm{a2}(\nu )
\]
}
\end{maplelatex}

\begin{maplelatex}
\mapleinline{inert}{2d}{_F5(nu) := a3(nu);}{%
\[
\mathrm{\_F5}(\nu ) := \mathrm{a3}(\nu )
\]
}
\end{maplelatex}

\begin{maplelatex}
\mapleinline{inert}{2d}{_F4(nu,t) := f2(t,nu);}{%
\[
\mathrm{\_F4}(\nu , \,t) := \mathrm{f2}(t, \,\nu )
\]
}
\end{maplelatex}

\begin{maplelatex}
\mapleinline{inert}{2d}{_F9(nu) := -a1(nu)+a4(nu);}{%
\[
\mathrm{\_F9}(\nu ) :=  - \mathrm{a1}(\nu ) + \mathrm{a4}(\nu )
\]
}
\end{maplelatex}

\begin{maplelatex}
\mapleinline{inert}{2d}{_F7(nu) := a5(nu);}{%
\[
\mathrm{\_F7}(\nu ) := \mathrm{a5}(\nu )
\]
}
\end{maplelatex}

\begin{maplelatex}
\mapleinline{inert}{2d}{_F1(U,P,nu,y) := (a1(nu)-a4(nu))*U+g1(y,nu,U,P);}{%
\[
\mathrm{\_F1}(U, \,P, \,\nu , \,y) := (\mathrm{a1}(\nu ) -
\mathrm{a4}(\nu ))\,U + \mathrm{g1}(y, \,\nu , \,U, \,P)
\]
}
\end{maplelatex}

\begin{maplelatex}
\mapleinline{inert}{2d}{_F2(U,P,nu,y) := 2*(a1(nu)-a4(nu))*P+g2(y,nu,U,P);}{%
\[
\mathrm{\_F2}(U, \,P, \,\nu , \,y) := 2\,(\mathrm{a1}(\nu ) -
\mathrm{a4}(\nu ))\,P + \mathrm{g2}(y, \,\nu , \,U, \,P)
\]
}
\end{maplelatex}

\begin{maplelatex}
\mapleinline{inert}{2d}{_F10(nu,t) := f3(t,nu);}{%
\[
\mathrm{\_F10}(\nu , \,t) := \mathrm{f3}(t, \,\nu )
\]
}
\end{maplelatex}

\end{maplegroup}
\begin{maplegroup}
\begin{flushleft}
{\large Final Result (identical to (2.10)):}
\end{flushleft}

\end{maplegroup}
\begin{maplegroup}
\begin{mapleinput}
\mapleinline{active}{1d}{n:=nops(symsol[1]): infgen:=sum(symsol[1,i],i=1..n):}{%
}
\end{mapleinput}

\end{maplegroup}
\begin{maplegroup}
\begin{mapleinput}
\mapleinline{active}{1d}{xi[x](sigma)=simplify(coeff(infgen,D[x]));
xi[y](sigma)=simplify(coeff(infgen,D[y]));
xi[z](sigma)=simplify(coeff(infgen,D[z]));
xi[t](sigma)=simplify(coeff(infgen,D[t]));
xi[nu](sigma)=simplify(coeff(infgen,D[nu]));
eta[u](sigma)=simplify(coeff(infgen,D[u]));
eta[v](sigma)=simplify(coeff(infgen,D[v]));
eta[w](sigma)=simplify(coeff(infgen,D[w]));
eta[p](sigma)=simplify(coeff(infgen,D[p]));
eta[U](sigma)=simplify(coeff(infgen,D[U]));
eta[P](sigma)=simplify(coeff(infgen,D[P]));}{%
}
\end{mapleinput}

\mapleresult
\begin{maplelatex}
\mapleinline{inert}{2d}{xi[x](sigma) = f1(t,nu)+a2(nu)*z+a1(nu)*x;}{%
\[
{\xi _{x}}(\sigma )=\mathrm{f1}(t, \,\nu ) + \mathrm{a2}(\nu )\,z
 + \mathrm{a1}(\nu )\,x
\]
}
\end{maplelatex}

\begin{maplelatex}
\mapleinline{inert}{2d}{xi[y](sigma) = a3(nu)+a1(nu)*y;}{%
\[
{\xi _{y}}(\sigma )=\mathrm{a3}(\nu ) + \mathrm{a1}(\nu )\,y
\]
}
\end{maplelatex}

\begin{maplelatex}
\mapleinline{inert}{2d}{xi[z](sigma) = f2(t,nu)+a1(nu)*z-a2(nu)*x;}{%
\[
{\xi _{z}}(\sigma )=\mathrm{f2}(t, \,\nu ) + \mathrm{a1}(\nu )\,z
 - \mathrm{a2}(\nu )\,x
\]
}
\end{maplelatex}

\begin{maplelatex}
\mapleinline{inert}{2d}{xi[t](sigma) = a5(nu)+t*a4(nu);}{%
\[
{\xi _{t}}(\sigma )=\mathrm{a5}(\nu ) + t\,\mathrm{a4}(\nu )
\]
}
\end{maplelatex}

\begin{maplelatex}
\mapleinline{inert}{2d}{xi[nu](sigma) = 2*a1(nu)*nu-nu*a4(nu);}{%
\[
{\xi _{\nu }}(\sigma )=2\,\mathrm{a1}(\nu )\,\nu  - \nu \,
\mathrm{a4}(\nu )
\]
}
\end{maplelatex}

\begin{maplelatex}
\mapleinline{inert}{2d}{eta[u](sigma) =
-g1(y,nu,U,P)+diff(f1(t,nu),t)+a1(nu)*u-a4(nu)*u+a2(nu)*w;}{%
\[
{\eta _{u}}(\sigma )= - \mathrm{g1}(y, \,\nu , \,U, \,P) + (
{\frac {\partial }{\partial t}}\,\mathrm{f1}(t, \,\nu )) +
\mathrm{a1}(\nu )\,u - \mathrm{a4}(\nu )\,u + \mathrm{a2}(\nu )\,
w
\]
}
\end{maplelatex}

\begin{maplelatex}
\mapleinline{inert}{2d}{eta[v](sigma) = (a1(nu)-a4(nu))*v;}{%
\[
{\eta _{v}}(\sigma )=(\mathrm{a1}(\nu ) - \mathrm{a4}(\nu ))\,v
\]
}
\end{maplelatex}

\begin{maplelatex}
\mapleinline{inert}{2d}{eta[w](sigma) =
diff(f2(t,nu),t)+w*a1(nu)-w*a4(nu)-a2(nu)*u-a2(nu)*U;}{%
\[
{\eta _{w}}(\sigma )=({\frac {\partial }{\partial t}}\,\mathrm{f2
}(t, \,\nu )) + w\,\mathrm{a1}(\nu ) - w\,\mathrm{a4}(\nu ) -
\mathrm{a2}(\nu )\,u - \mathrm{a2}(\nu )\,U
\]
}
\end{maplelatex}

\begin{maplelatex}
\mapleinline{inert}{2d}{eta[p](sigma) =
z*a2(nu)*K-x*a1(nu)*K+f3(t,nu)-g2(y,nu,U,P)+2*a1(nu)*p-2*a4(nu)*p+2*a4
(nu)*x*K-z*diff(f2(t,nu),`$`(t,2))-x*diff(f1(t,nu),`$`(t,2));}{%
\maplemultiline{
{\eta _{p}}(\sigma )=z\,\mathrm{a2}(\nu )\,K - x\,\mathrm{a1}(\nu
 )\,K + \mathrm{f3}(t, \,\nu ) - \mathrm{g2}(y, \,\nu , \,U, \,P)
 + 2\,\mathrm{a1}(\nu )\,p - 2\,\mathrm{a4}(\nu )\,p \\
\mbox{} + 2\,\mathrm{a4}(\nu )\,x\,K - z\,({\frac {\partial ^{2}
}{\partial t^{2}}}\,\mathrm{f2}(t, \,\nu )) - x\,({\frac {
\partial ^{2}}{\partial t^{2}}}\,\mathrm{f1}(t, \,\nu )) }
}
\end{maplelatex}

\begin{maplelatex}
\mapleinline{inert}{2d}{eta[U](sigma) = U*a1(nu)-U*a4(nu)+g1(y,nu,U,P);}{%
\[
{\eta _{U}}(\sigma )=U\,\mathrm{a1}(\nu ) - U\,\mathrm{a4}(\nu )
 + \mathrm{g1}(y, \,\nu , \,U, \,P)
\]
}
\end{maplelatex}

\begin{maplelatex}
\mapleinline{inert}{2d}{eta[P](sigma) = 2*P*a1(nu)-2*P*a4(nu)+g2(y,nu,U,P);}{%
\[
{\eta _{P}}(\sigma )=2\,P\,\mathrm{a1}(\nu ) - 2\,P\,\mathrm{a4}(
\nu ) + \mathrm{g2}(y, \,\nu , \,U, \,P)
\]
}
\end{maplelatex}

\end{maplegroup}
\begin{maplegroup}
\begin{mapleinput}
\end{mapleinput}

\end{maplegroup}

\vspace{1em}\noindent {\bf IIb. SADE-Package
\textbf{\emph{including}} the $\pmb{\mathscr{P}_{ij}}$-equations
(\ref{131228:1928}):}
\begin{maplegroup}
\begin{flushleft}
{\large Header:}
\end{flushleft}

\end{maplegroup}
\begin{maplegroup}
\begin{mapleinput}
\mapleinline{active}{1d}{restart: with(sade):}{%
}
\end{mapleinput}

\mapleresult
\begin{maplelatex}
\mapleinline{inert}{2d}{`Symmetry Analysis of Differential Equations`;}{%
\[
\mathit{Symmetry\ Analysis\ of\ Differential\ Equations}
\]
}
\end{maplelatex}

\begin{maplelatex}
\mapleinline{inert}{2d}{`By Tarcisio M. Rocha Filho - Annibal Figueiredo - 2010`;}{%
\[
\mathit{By\ Tarcisio\ M.\ Rocha\ Filho\ -\ Annibal\ Figueiredo\ -
\ 2010}
\]
}
\end{maplelatex}

\end{maplegroup}
\begin{maplegroup}
\begin{flushleft}
{\large Definitions:}
\end{flushleft}

\end{maplegroup}
\begin{maplegroup}
\begin{mapleinput}
\mapleinline{active}{1d}{alias(sigma=(t,x,y,z,nu,u,v,w,p,U,P)): X:=(t,x,y,z,nu): }{%
}
\end{mapleinput}

\end{maplegroup}
\begin{maplegroup}
\begin{mapleinput}
\mapleinline{active}{1d}{C:=diff(u(X),x)+diff(v(X),y)+diff(w(X),z);}{%
}
\end{mapleinput}

\mapleresult
\begin{maplelatex}
\mapleinline{inert}{2d}{C :=
diff(u(t,x,y,z,nu),x)+diff(v(t,x,y,z,nu),y)+diff(w(t,x,y,z,nu),z);}{%
\[
C := ({\frac {\partial }{\partial x}}\,\mathrm{u}(t, \,x, \,y, \,
z, \,\nu )) + ({\frac {\partial }{\partial y}}\,\mathrm{v}(t, \,x
, \,y, \,z, \,\nu )) + ({\frac {\partial }{\partial z}}\,\mathrm{
w}(t, \,x, \,y, \,z, \,\nu ))
\]
}
\end{maplelatex}

\end{maplegroup}
\begin{maplegroup}
\begin{mapleinput}
\mapleinline{active}{1d}{N1:=diff(u(X),t)+U(X)*diff(u(X),x)+v(X)*diff(U(X),y)-(K+nu*diff(U(X),
y,y))+diff(u(X)*u(X),x)+diff(u(X)*v(X),y)+diff(u(X)*w(X),z)+diff(p(X),
x)-nu*(diff(u(X),x,x)+diff(u(X),y,y)+diff(u(X),z,z));}{%
}
\end{mapleinput}

\mapleresult
\begin{maplelatex}
\mapleinline{inert}{2d}{N1 :=
diff(u(t,x,y,z,nu),t)+U(t,x,y,z,nu)*diff(u(t,x,y,z,nu),x)+v(t,x,y,z,nu
)*diff(U(t,x,y,z,nu),y)-K-nu*diff(U(t,x,y,z,nu),`$`(y,2))+2*u(t,x,y,z,
nu)*diff(u(t,x,y,z,nu),x)+diff(u(t,x,y,z,nu),y)*v(t,x,y,z,nu)+u(t,x,y,
z,nu)*diff(v(t,x,y,z,nu),y)+diff(u(t,x,y,z,nu),z)*w(t,x,y,z,nu)+u(t,x,
y,z,nu)*diff(w(t,x,y,z,nu),z)+diff(p(t,x,y,z,nu),x)-nu*(diff(u(t,x,y,z
,nu),`$`(x,2))+diff(u(t,x,y,z,nu),`$`(y,2))+diff(u(t,x,y,z,nu),`$`(z,2
)));}{%
\maplemultiline{
\mathit{N1} := ({\frac {\partial }{\partial t}}\,\mathrm{u}(t, \,
x, \,y, \,z, \,\nu )) + \mathrm{U}(t, \,x, \,y, \,z, \,\nu )\,(
{\frac {\partial }{\partial x}}\,\mathrm{u}(t, \,x, \,y, \,z, \,
\nu )) \\
\mbox{} + \mathrm{v}(t, \,x, \,y, \,z, \,\nu )\,({\frac {
\partial }{\partial y}}\,\mathrm{U}(t, \,x, \,y, \,z, \,\nu )) -
K - \nu \,({\frac {\partial ^{2}}{\partial y^{2}}}\,\mathrm{U}(t
, \,x, \,y, \,z, \,\nu )) \\
\mbox{} + 2\,\mathrm{u}(t, \,x, \,y, \,z, \,\nu )\,({\frac {
\partial }{\partial x}}\,\mathrm{u}(t, \,x, \,y, \,z, \,\nu )) +
({\frac {\partial }{\partial y}}\,\mathrm{u}(t, \,x, \,y, \,z, \,
\nu ))\,\mathrm{v}(t, \,x, \,y, \,z, \,\nu ) \\
\mbox{} + \mathrm{u}(t, \,x, \,y, \,z, \,\nu )\,({\frac {
\partial }{\partial y}}\,\mathrm{v}(t, \,x, \,y, \,z, \,\nu )) +
({\frac {\partial }{\partial z}}\,\mathrm{u}(t, \,x, \,y, \,z, \,
\nu ))\,\mathrm{w}(t, \,x, \,y, \,z, \,\nu ) \\
\mbox{} + \mathrm{u}(t, \,x, \,y, \,z, \,\nu )\,({\frac {
\partial }{\partial z}}\,\mathrm{w}(t, \,x, \,y, \,z, \,\nu )) +
({\frac {\partial }{\partial x}}\,\mathrm{p}(t, \,x, \,y, \,z, \,
\nu )) \\
\mbox{} - \nu \,(({\frac {\partial ^{2}}{\partial x^{2}}}\,
\mathrm{u}(t, \,x, \,y, \,z, \,\nu )) + ({\frac {\partial ^{2}}{
\partial y^{2}}}\,\mathrm{u}(t, \,x, \,y, \,z, \,\nu )) + (
{\frac {\partial ^{2}}{\partial z^{2}}}\,\mathrm{u}(t, \,x, \,y,
\,z, \,\nu ))) }
}
\end{maplelatex}

\end{maplegroup}
\begin{maplegroup}
\begin{mapleinput}
\mapleinline{active}{1d}{N2:=diff(v(X),t)+U(X)*diff(v(X),x)+diff(P(X),y)+diff(v(X)*u(X),x)+dif
f(v(X)*v(X),y)+diff(v(X)*w(X),z)+diff(p(X),y)-nu*(diff(v(X),x,x)+diff(
v(X),y,y)+diff(v(X),z,z));}{%
}
\end{mapleinput}

\mapleresult
\begin{maplelatex}
\mapleinline{inert}{2d}{N2 :=
diff(v(t,x,y,z,nu),t)+U(t,x,y,z,nu)*diff(v(t,x,y,z,nu),x)+diff(P(t,x,y
,z,nu),y)+diff(u(t,x,y,z,nu),x)*v(t,x,y,z,nu)+u(t,x,y,z,nu)*diff(v(t,x
,y,z,nu),x)+2*v(t,x,y,z,nu)*diff(v(t,x,y,z,nu),y)+diff(v(t,x,y,z,nu),z
)*w(t,x,y,z,nu)+v(t,x,y,z,nu)*diff(w(t,x,y,z,nu),z)+diff(p(t,x,y,z,nu)
,y)-nu*(diff(v(t,x,y,z,nu),`$`(x,2))+diff(v(t,x,y,z,nu),`$`(y,2))+diff
(v(t,x,y,z,nu),`$`(z,2)));}{%
\maplemultiline{
\mathit{N2} := ({\frac {\partial }{\partial t}}\,\mathrm{v}(t, \,
x, \,y, \,z, \,\nu )) + \mathrm{U}(t, \,x, \,y, \,z, \,\nu )\,(
{\frac {\partial }{\partial x}}\,\mathrm{v}(t, \,x, \,y, \,z, \,
\nu )) + ({\frac {\partial }{\partial y}}\,\mathrm{P}(t, \,x, \,y
, \,z, \,\nu )) \\
\mbox{} + ({\frac {\partial }{\partial x}}\,\mathrm{u}(t, \,x, \,
y, \,z, \,\nu ))\,\mathrm{v}(t, \,x, \,y, \,z, \,\nu ) + \mathrm{
u}(t, \,x, \,y, \,z, \,\nu )\,({\frac {\partial }{\partial x}}\,
\mathrm{v}(t, \,x, \,y, \,z, \,\nu )) \\
\mbox{} + 2\,\mathrm{v}(t, \,x, \,y, \,z, \,\nu )\,({\frac {
\partial }{\partial y}}\,\mathrm{v}(t, \,x, \,y, \,z, \,\nu )) +
({\frac {\partial }{\partial z}}\,\mathrm{v}(t, \,x, \,y, \,z, \,
\nu ))\,\mathrm{w}(t, \,x, \,y, \,z, \,\nu ) \\
\mbox{} + \mathrm{v}(t, \,x, \,y, \,z, \,\nu )\,({\frac {
\partial }{\partial z}}\,\mathrm{w}(t, \,x, \,y, \,z, \,\nu )) +
({\frac {\partial }{\partial y}}\,\mathrm{p}(t, \,x, \,y, \,z, \,
\nu )) \\
\mbox{} - \nu \,(({\frac {\partial ^{2}}{\partial x^{2}}}\,
\mathrm{v}(t, \,x, \,y, \,z, \,\nu )) + ({\frac {\partial ^{2}}{
\partial y^{2}}}\,\mathrm{v}(t, \,x, \,y, \,z, \,\nu )) + (
{\frac {\partial ^{2}}{\partial z^{2}}}\,\mathrm{v}(t, \,x, \,y,
\,z, \,\nu ))) }
}
\end{maplelatex}

\end{maplegroup}
\begin{maplegroup}
\begin{mapleinput}
\mapleinline{active}{1d}{N3:=diff(w(X),t)+U(X)*diff(w(X),x)+diff(w(X)*u(X),x)+diff(w(X)*v(X),y
)+diff(w(X)*w(X),z)+diff(p(X),z)-nu*(diff(w(X),x,x)+diff(w(X),y,y)+dif
f(w(X),z,z));}{%
}
\end{mapleinput}

\mapleresult
\begin{maplelatex}
\mapleinline{inert}{2d}{N3 :=
diff(w(t,x,y,z,nu),t)+U(t,x,y,z,nu)*diff(w(t,x,y,z,nu),x)+diff(u(t,x,y
,z,nu),x)*w(t,x,y,z,nu)+u(t,x,y,z,nu)*diff(w(t,x,y,z,nu),x)+diff(v(t,x
,y,z,nu),y)*w(t,x,y,z,nu)+v(t,x,y,z,nu)*diff(w(t,x,y,z,nu),y)+2*w(t,x,
y,z,nu)*diff(w(t,x,y,z,nu),z)+diff(p(t,x,y,z,nu),z)-nu*(diff(w(t,x,y,z
,nu),`$`(x,2))+diff(w(t,x,y,z,nu),`$`(y,2))+diff(w(t,x,y,z,nu),`$`(z,2
)));}{%
\maplemultiline{
\mathit{N3} := ({\frac {\partial }{\partial t}}\,\mathrm{w}(t, \,
x, \,y, \,z, \,\nu )) + \mathrm{U}(t, \,x, \,y, \,z, \,\nu )\,(
{\frac {\partial }{\partial x}}\,\mathrm{w}(t, \,x, \,y, \,z, \,
\nu )) \\
\mbox{} + ({\frac {\partial }{\partial x}}\,\mathrm{u}(t, \,x, \,
y, \,z, \,\nu ))\,\mathrm{w}(t, \,x, \,y, \,z, \,\nu ) + \mathrm{
u}(t, \,x, \,y, \,z, \,\nu )\,({\frac {\partial }{\partial x}}\,
\mathrm{w}(t, \,x, \,y, \,z, \,\nu )) \\
\mbox{} + ({\frac {\partial }{\partial y}}\,\mathrm{v}(t, \,x, \,
y, \,z, \,\nu ))\,\mathrm{w}(t, \,x, \,y, \,z, \,\nu ) + \mathrm{
v}(t, \,x, \,y, \,z, \,\nu )\,({\frac {\partial }{\partial y}}\,
\mathrm{w}(t, \,x, \,y, \,z, \,\nu )) \\
\mbox{} + 2\,\mathrm{w}(t, \,x, \,y, \,z, \,\nu )\,({\frac {
\partial }{\partial z}}\,\mathrm{w}(t, \,x, \,y, \,z, \,\nu )) +
({\frac {\partial }{\partial z}}\,\mathrm{p}(t, \,x, \,y, \,z, \,
\nu )) \\
\mbox{} - \nu \,(({\frac {\partial ^{2}}{\partial x^{2}}}\,
\mathrm{w}(t, \,x, \,y, \,z, \,\nu )) + ({\frac {\partial ^{2}}{
\partial y^{2}}}\,\mathrm{w}(t, \,x, \,y, \,z, \,\nu )) + (
{\frac {\partial ^{2}}{\partial z^{2}}}\,\mathrm{w}(t, \,x, \,y,
\,z, \,\nu ))) }
}
\end{maplelatex}

\end{maplegroup}
\begin{maplegroup}
\begin{flushleft}
{\large Equations (2.1), (2.2) \& (2.4) including the "velocity
product equations" (2.3):}
\end{flushleft}

\end{maplegroup}
\begin{maplegroup}
\begin{mapleinput}
\mapleinline{active}{1d}{eqnC:=C=0: eqnN1:=N1=0: eqnN2:=N2=0: eqnN3:=N3=0:
eqnP11:=2*N1*u(X)=0: eqnP22:=2*N2*v(X)=0: eqnP33:=2*N3*w(X)=0:
eqnP12:=N1*v(X)+N2*u(X)=0: eqnP13:=N1*w(X)+N3*u(X)=0:
eqnP23:=N2*w(X)+N3*v(X)=0:
eqnR1:=diff(U(X),t)=0: eqnR2:=diff(U(X),x)=0: eqnR3:=diff(U(X),z)=0:
eqnR4:=diff(P(X),t)=0: eqnR5:=diff(P(X),x)=0:
eqnR6:=diff(P(X),z)=0: }{%
}
\end{mapleinput}

\end{maplegroup}
\begin{maplegroup}
\begin{mapleinput}
\mapleinline{active}{1d}{eqns:=[eqnC,eqnN1,eqnN2,eqnN3,eqnP11,eqnP22,eqnP33,eqnP12,eqnP13,eqnP
23,eqnR1,eqnR2,eqnR3,eqnR4,eqnR5,eqnR6]:}{%
}
\end{mapleinput}

\end{maplegroup}
\begin{maplegroup}
\begin{flushleft}
{\large Symmetry Algorithm:}
\end{flushleft}

\end{maplegroup}
\begin{maplegroup}
\begin{flushleft}
\textit{{\large a) Size of the determining system:}}
\end{flushleft}

\end{maplegroup}
\begin{maplegroup}
\begin{mapleinput}
\mapleinline{active}{1d}{detsys:=liesymmetries(eqns,[u(X),v(X),w(X),p(X),U(X),P(X)],determining):
nops(detsys[1]);}{%
}
\end{mapleinput}

\mapleresult
\begin{maplelatex}
\mapleinline{inert}{2d}{522;}{%
\[
522
\]
}
\end{maplelatex}

\end{maplegroup}
\begin{maplegroup}
\begin{flushleft}
\textit{{\large b) Solving the determining system: }}
\end{flushleft}

\end{maplegroup}
\begin{maplegroup}
\begin{mapleinput}
\mapleinline{active}{1d}{symsol:=liesymmetries(eqns,[u(X),v(X),w(X),p(X),U(X),P(X)]);}{%
}
\end{mapleinput}

\mapleresult
\begin{maplelatex}
\mapleinline{inert}{2d}{symsol := [\{_F5(nu)*D[y], _F7(nu)*D[t], _F10(nu,t)*D[p],
-_F1(U,P,nu,y)*D[u]+_F1(U,P,nu,y)*D[U],
-_F2(U,P,nu,y)*D[p]+_F2(U,P,nu,y)*D[P],
diff(_F4(nu,t),t)*D[w]-z*diff(_F4(nu,t),`$`(t,2))*D[p]+_F4(nu,t)*D[z],
diff(_F6(nu,t),t)*D[u]-x*diff(_F6(nu,t),`$`(t,2))*D[p]+_F6(nu,t)*D[x],
z*_F8(nu)*K*D[p]+_F8(nu)*w*D[u]+_F8(nu)*(-u-U)*D[w]+_F8(nu)*z*D[x]-_F8
(nu)*x*D[z],
-_F9(nu)*nu*D[nu]+_F9(nu)*(-2*p-2*P+2*x*K)*D[p]+t*_F9(nu)*D[t]+_F9(nu)
*(-u-U)*D[u]-_F9(nu)*v*D[v]-_F9(nu)*w*D[w],
x*_F3(nu)*K*D[p]+_F3(nu)*nu*D[nu]+t*_F3(nu)*D[t]+_F3(nu)*x*D[x]+_F3(nu
)*y*D[y]+_F3(nu)*z*D[z]\}, \{\}];}{%
\maplemultiline{
\mathit{symsol} := [\{\mathrm{\_F5}(\nu )\,{\mathrm{D}_{y}}, \,
\mathrm{\_F7}(\nu )\,{\mathrm{D}_{t}}, \,\mathrm{\_F10}(\nu , \,t
)\,{\mathrm{D}_{p}}, \, - \mathrm{\_F1}(U, \,P, \,\nu , \,y)\,{
\mathrm{D}_{u}} + \mathrm{\_F1}(U, \,P, \,\nu , \,y)\,{\mathrm{D}
_{U}},  \\
 - \mathrm{\_F2}(U, \,P, \,\nu , \,y)\,{\mathrm{D}_{p}} +
\mathrm{\_F2}(U, \,P, \,\nu , \,y)\,{\mathrm{D}_{P}},  \\
({\frac {\partial }{\partial t}}\,\mathrm{\_F4}(\nu , \,t))\,{
\mathrm{D}_{w}} - z\,({\frac {\partial ^{2}}{\partial t^{2}}}\,
\mathrm{\_F4}(\nu , \,t))\,{\mathrm{D}_{p}} + \mathrm{\_F4}(\nu
, \,t)\,{\mathrm{D}_{z}},  \\
({\frac {\partial }{\partial t}}\,\mathrm{\_F6}(\nu , \,t))\,{
\mathrm{D}_{u}} - x\,({\frac {\partial ^{2}}{\partial t^{2}}}\,
\mathrm{\_F6}(\nu , \,t))\,{\mathrm{D}_{p}} + \mathrm{\_F6}(\nu
, \,t)\,{\mathrm{D}_{x}},  \\
z\,\mathrm{\_F8}(\nu )\,K\,{\mathrm{D}_{p}} + \mathrm{\_F8}(\nu )
\,w\,{\mathrm{D}_{u}} + \mathrm{\_F8}(\nu )\,( - u - U)\,{
\mathrm{D}_{w}} + \mathrm{\_F8}(\nu )\,z\,{\mathrm{D}_{x}} -
\mathrm{\_F8}(\nu )\,x\,{\mathrm{D}_{z}},  \\
 - \mathrm{\_F9}(\nu )\,\nu \,{\mathrm{D}_{\nu }} + \mathrm{\_F9}
(\nu )\,( - 2\,p - 2\,P + 2\,x\,K)\,{\mathrm{D}_{p}} + t\,
\mathrm{\_F9}(\nu )\,{\mathrm{D}_{t}} \\
\mbox{} + \mathrm{\_F9}(\nu )\,( - u - U)\,{\mathrm{D}_{u}} -
\mathrm{\_F9}(\nu )\,v\,{\mathrm{D}_{v}} - \mathrm{\_F9}(\nu )\,w
\,{\mathrm{D}_{w}}, x\,\mathrm{\_F3}(\nu )\,K\,{\mathrm{D}_{p}}
 + \mathrm{\_F3}(\nu )\,\nu \,{\mathrm{D}_{\nu }} \\
\mbox{} + t\,\mathrm{\_F3}(\nu )\,{\mathrm{D}_{t}} + \mathrm{\_F3
}(\nu )\,x\,{\mathrm{D}_{x}} + \mathrm{\_F3}(\nu )\,y\,{\mathrm{D
}_{y}} + \mathrm{\_F3}(\nu )\,z\,{\mathrm{D}_{z}}\}, \,\{\}] }
}
\end{maplelatex}

\end{maplegroup}
\begin{maplegroup}
\begin{flushleft}
{\large Redefinition of group functions as used in (2.10):}
\end{flushleft}

\end{maplegroup}
\begin{maplegroup}
\begin{mapleinput}
\mapleinline{active}{1d}{_F6(nu,t):=f1(t,nu); _F3(nu):=a1(nu); _F8(nu):=a2(nu);
_F5(nu):=a3(nu); _F4(nu,t):=f2(t,nu); _F9(nu):=-a1(nu)+a4(nu);
_F7(nu):=a5(nu); _F1(U,P,nu,y):=(a1(nu)-a4(nu))*U+g1(y,nu,U,P);
_F2(U,P,nu,y):=2*(a1(nu)-a4(nu))*P+g2(y,nu,U,P);
_F10(nu,t):=f3(t,nu);}{%
}
\end{mapleinput}

\mapleresult
\begin{maplelatex}
\mapleinline{inert}{2d}{_F6(nu,t) := f1(t,nu);}{%
\[
\mathrm{\_F6}(\nu , \,t) := \mathrm{f1}(t, \,\nu )
\]
}
\end{maplelatex}

\begin{maplelatex}
\mapleinline{inert}{2d}{_F3(nu) := a1(nu);}{%
\[
\mathrm{\_F3}(\nu ) := \mathrm{a1}(\nu )
\]
}
\end{maplelatex}

\begin{maplelatex}
\mapleinline{inert}{2d}{_F8(nu) := a2(nu);}{%
\[
\mathrm{\_F8}(\nu ) := \mathrm{a2}(\nu )
\]
}
\end{maplelatex}

\begin{maplelatex}
\mapleinline{inert}{2d}{_F5(nu) := a3(nu);}{%
\[
\mathrm{\_F5}(\nu ) := \mathrm{a3}(\nu )
\]
}
\end{maplelatex}

\begin{maplelatex}
\mapleinline{inert}{2d}{_F4(nu,t) := f2(t,nu);}{%
\[
\mathrm{\_F4}(\nu , \,t) := \mathrm{f2}(t, \,\nu )
\]
}
\end{maplelatex}

\begin{maplelatex}
\mapleinline{inert}{2d}{_F9(nu) := -a1(nu)+a4(nu);}{%
\[
\mathrm{\_F9}(\nu ) :=  - \mathrm{a1}(\nu ) + \mathrm{a4}(\nu )
\]
}
\end{maplelatex}

\begin{maplelatex}
\mapleinline{inert}{2d}{_F7(nu) := a5(nu);}{%
\[
\mathrm{\_F7}(\nu ) := \mathrm{a5}(\nu )
\]
}
\end{maplelatex}

\begin{maplelatex}
\mapleinline{inert}{2d}{_F1(U,P,nu,y) := (a1(nu)-a4(nu))*U+g1(y,nu,U,P);}{%
\[
\mathrm{\_F1}(U, \,P, \,\nu , \,y) := (\mathrm{a1}(\nu ) -
\mathrm{a4}(\nu ))\,U + \mathrm{g1}(y, \,\nu , \,U, \,P)
\]
}
\end{maplelatex}

\begin{maplelatex}
\mapleinline{inert}{2d}{_F2(U,P,nu,y) := 2*(a1(nu)-a4(nu))*P+g2(y,nu,U,P);}{%
\[
\mathrm{\_F2}(U, \,P, \,\nu , \,y) := 2\,(\mathrm{a1}(\nu ) -
\mathrm{a4}(\nu ))\,P + \mathrm{g2}(y, \,\nu , \,U, \,P)
\]
}
\end{maplelatex}

\begin{maplelatex}
\mapleinline{inert}{2d}{_F10(nu,t) := f3(t,nu);}{%
\[
\mathrm{\_F10}(\nu , \,t) := \mathrm{f3}(t, \,\nu )
\]
}
\end{maplelatex}

\end{maplegroup}
\begin{maplegroup}
\begin{flushleft}
{\large Final Result (still identical to (2.10)):}
\end{flushleft}

\end{maplegroup}
\begin{maplegroup}
\begin{mapleinput}
\mapleinline{active}{1d}{n:=nops(symsol[1]): infgen:=sum(symsol[1,i],i=1..n):}{%
}
\end{mapleinput}

\end{maplegroup}
\begin{maplegroup}
\begin{mapleinput}
\mapleinline{active}{1d}{xi[x](sigma)=simplify(coeff(infgen,D[x]));
xi[y](sigma)=simplify(coeff(infgen,D[y]));
xi[z](sigma)=simplify(coeff(infgen,D[z]));
xi[t](sigma)=simplify(coeff(infgen,D[t]));
xi[nu](sigma)=simplify(coeff(infgen,D[nu]));
eta[u](sigma)=simplify(coeff(infgen,D[u]));
eta[v](sigma)=simplify(coeff(infgen,D[v]));
eta[w](sigma)=simplify(coeff(infgen,D[w]));
eta[p](sigma)=simplify(coeff(infgen,D[p]));
eta[U](sigma)=simplify(coeff(infgen,D[U]));
eta[P](sigma)=simplify(coeff(infgen,D[P]));}{%
}
\end{mapleinput}

\mapleresult
\begin{maplelatex}
\mapleinline{inert}{2d}{xi[x](sigma) = f1(t,nu)+a2(nu)*z+a1(nu)*x;}{%
\[
{\xi _{x}}(\sigma )=\mathrm{f1}(t, \,\nu ) + \mathrm{a2}(\nu )\,z
 + \mathrm{a1}(\nu )\,x
\]
}
\end{maplelatex}

\begin{maplelatex}
\mapleinline{inert}{2d}{xi[y](sigma) = a3(nu)+a1(nu)*y;}{%
\[
{\xi _{y}}(\sigma )=\mathrm{a3}(\nu ) + \mathrm{a1}(\nu )\,y
\]
}
\end{maplelatex}

\begin{maplelatex}
\mapleinline{inert}{2d}{xi[z](sigma) = -a2(nu)*x+a1(nu)*z+f2(t,nu);}{%
\[
{\xi _{z}}(\sigma )= - \mathrm{a2}(\nu )\,x + \mathrm{a1}(\nu )\,
z + \mathrm{f2}(t, \,\nu )
\]
}
\end{maplelatex}

\begin{maplelatex}
\mapleinline{inert}{2d}{xi[t](sigma) = a5(nu)+t*a4(nu);}{%
\[
{\xi _{t}}(\sigma )=\mathrm{a5}(\nu ) + t\,\mathrm{a4}(\nu )
\]
}
\end{maplelatex}

\begin{maplelatex}
\mapleinline{inert}{2d}{xi[nu](sigma) = 2*a1(nu)*nu-nu*a4(nu);}{%
\[
{\xi _{\nu }}(\sigma )=2\,\mathrm{a1}(\nu )\,\nu  - \nu \,
\mathrm{a4}(\nu )
\]
}
\end{maplelatex}

\begin{maplelatex}
\mapleinline{inert}{2d}{eta[u](sigma) =
-g1(y,nu,U,P)+diff(f1(t,nu),t)+a2(nu)*w+a1(nu)*u-a4(nu)*u;}{%
\[
{\eta _{u}}(\sigma )= - \mathrm{g1}(y, \,\nu , \,U, \,P) + (
{\frac {\partial }{\partial t}}\,\mathrm{f1}(t, \,\nu )) +
\mathrm{a2}(\nu )\,w + \mathrm{a1}(\nu )\,u - \mathrm{a4}(\nu )\,
u
\]
}
\end{maplelatex}

\begin{maplelatex}
\mapleinline{inert}{2d}{eta[v](sigma) = -(-a1(nu)+a4(nu))*v;}{%
\[
{\eta _{v}}(\sigma )= - ( - \mathrm{a1}(\nu ) + \mathrm{a4}(\nu )
)\,v
\]
}
\end{maplelatex}

\begin{maplelatex}
\mapleinline{inert}{2d}{eta[w](sigma) =
-a2(nu)*u-a2(nu)*U+diff(f2(t,nu),t)+w*a1(nu)-w*a4(nu);}{%
\[
{\eta _{w}}(\sigma )= - \mathrm{a2}(\nu )\,u - \mathrm{a2}(\nu )
\,U + ({\frac {\partial }{\partial t}}\,\mathrm{f2}(t, \,\nu ))
 + w\,\mathrm{a1}(\nu ) - w\,\mathrm{a4}(\nu )
\]
}
\end{maplelatex}

\begin{maplelatex}
\mapleinline{inert}{2d}{eta[p](sigma) =
z*a2(nu)*K-x*a1(nu)*K-z*diff(f2(t,nu),`$`(t,2))-x*diff(f1(t,nu),`$`(t,
2))+2*a1(nu)*p-2*a4(nu)*p+2*a4(nu)*x*K+f3(t,nu)-g2(y,nu,U,P);}{%
\maplemultiline{
{\eta _{p}}(\sigma )=z\,\mathrm{a2}(\nu )\,K - x\,\mathrm{a1}(\nu
 )\,K - z\,({\frac {\partial ^{2}}{\partial t^{2}}}\,\mathrm{f2}(
t, \,\nu )) - x\,({\frac {\partial ^{2}}{\partial t^{2}}}\,
\mathrm{f1}(t, \,\nu )) + 2\,\mathrm{a1}(\nu )\,p - 2\,\mathrm{a4
}(\nu )\,p \\
\mbox{} + 2\,\mathrm{a4}(\nu )\,x\,K + \mathrm{f3}(t, \,\nu ) -
\mathrm{g2}(y, \,\nu , \,U, \,P) }
}
\end{maplelatex}

\begin{maplelatex}
\mapleinline{inert}{2d}{eta[U](sigma) = U*a1(nu)-U*a4(nu)+g1(y,nu,U,P);}{%
\[
{\eta _{U}}(\sigma )=U\,\mathrm{a1}(\nu ) - U\,\mathrm{a4}(\nu )
 + \mathrm{g1}(y, \,\nu , \,U, \,P)
\]
}
\end{maplelatex}

\begin{maplelatex}
\mapleinline{inert}{2d}{eta[P](sigma) = 2*a1(nu)*P-2*a4(nu)*P+g2(y,nu,U,P);}{%
\[
{\eta _{P}}(\sigma )=2\,\mathrm{a1}(\nu )\,P - 2\,\mathrm{a4}(\nu
 )\,P + \mathrm{g2}(y, \,\nu , \,U, \,P)
\]
}
\end{maplelatex}

\end{maplegroup}
\begin{maplegroup}
\begin{mapleinput}
\end{mapleinput}

\end{maplegroup}

\vspace{1em}\noindent {\bf IIIa. DESOLV-II-Package
\textbf{\emph{excluding}} the $\pmb{\mathscr{P}_{ij}}$-equations
(\ref{131228:1928}):}
\begin{maplegroup}
\begin{flushleft}
{\large Header:}
\end{flushleft}

\end{maplegroup}
\begin{maplegroup}
\begin{mapleinput}
\mapleinline{active}{1d}{restart: read "Desolv-V5R5.mpl": with(desolv):}{%
}
\end{mapleinput}

\mapleresult
\begin{maplelatex}
\mapleinline{inert}{2d}{`DESOLVII_V5R5 (March-2011)(c) by Dr. K.
T. Vu, Dr. J. Carminati and
Miss. G. Jefferson`;}{%
\maplemultiline{ \mathit{\phantom{xxxxxxxx} DESOLVII\_V5R5\ (March-2011)(c)} \\
\mathit{by\ Dr.\ K.\ T.\ Vu,\ Dr.\ J.\ Carminati\ and\ Miss.\ G.\
Jefferson} }
}
\end{maplelatex}

\end{maplegroup}
\begin{maplegroup}
\begin{flushleft}
{\large Definitions:}
\end{flushleft}

\end{maplegroup}
\begin{maplegroup}
\begin{mapleinput}
\mapleinline{active}{1d}{alias(sigma=(t,x,y,z,nu,u,v,w,p,U,P)): X:=(t,x,y,z,nu): }{%
}
\end{mapleinput}

\end{maplegroup}
\begin{maplegroup}
\begin{mapleinput}
\mapleinline{active}{1d}{C:=diff(u(X),x)+diff(v(X),y)+diff(w(X),z);}{%
}
\end{mapleinput}

\mapleresult
\begin{maplelatex}
\mapleinline{inert}{2d}{C :=
diff(u(t,x,y,z,nu),x)+diff(v(t,x,y,z,nu),y)+diff(w(t,x,y,z,nu),z);}{%
\[
C := ({\frac {\partial }{\partial x}}\,\mathrm{u}(t, \,x, \,y, \,
z, \,\nu )) + ({\frac {\partial }{\partial y}}\,\mathrm{v}(t, \,x
, \,y, \,z, \,\nu )) + ({\frac {\partial }{\partial z}}\,\mathrm{
w}(t, \,x, \,y, \,z, \,\nu ))
\]
}
\end{maplelatex}

\end{maplegroup}
\begin{maplegroup}
\begin{mapleinput}
\mapleinline{active}{1d}{N1:=diff(u(X),t)+U(X)*diff(u(X),x)+v(X)*diff(U(X),y)-(K+nu*diff(U(X),
y,y))+diff(u(X)*u(X),x)+diff(u(X)*v(X),y)+diff(u(X)*w(X),z)+diff(p(X),
x)-nu*(diff(u(X),x,x)+diff(u(X),y,y)+diff(u(X),z,z));}{%
}
\end{mapleinput}

\mapleresult
\begin{maplelatex}
\mapleinline{inert}{2d}{N1 :=
diff(u(t,x,y,z,nu),t)+U(t,x,y,z,nu)*diff(u(t,x,y,z,nu),x)+v(t,x,y,z,nu
)*diff(U(t,x,y,z,nu),y)-K-nu*diff(U(t,x,y,z,nu),`$`(y,2))+2*u(t,x,y,z,
nu)*diff(u(t,x,y,z,nu),x)+diff(u(t,x,y,z,nu),y)*v(t,x,y,z,nu)+u(t,x,y,
z,nu)*diff(v(t,x,y,z,nu),y)+diff(u(t,x,y,z,nu),z)*w(t,x,y,z,nu)+u(t,x,
y,z,nu)*diff(w(t,x,y,z,nu),z)+diff(p(t,x,y,z,nu),x)-nu*(diff(u(t,x,y,z
,nu),`$`(x,2))+diff(u(t,x,y,z,nu),`$`(y,2))+diff(u(t,x,y,z,nu),`$`(z,2
)));}{%
\maplemultiline{
\mathit{N1} := ({\frac {\partial }{\partial t}}\,\mathrm{u}(t, \,
x, \,y, \,z, \,\nu )) + \mathrm{U}(t, \,x, \,y, \,z, \,\nu )\,(
{\frac {\partial }{\partial x}}\,\mathrm{u}(t, \,x, \,y, \,z, \,
\nu )) \\
\mbox{} + \mathrm{v}(t, \,x, \,y, \,z, \,\nu )\,({\frac {
\partial }{\partial y}}\,\mathrm{U}(t, \,x, \,y, \,z, \,\nu )) -
K - \nu \,({\frac {\partial ^{2}}{\partial y^{2}}}\,\mathrm{U}(t
, \,x, \,y, \,z, \,\nu )) \\
\mbox{} + 2\,\mathrm{u}(t, \,x, \,y, \,z, \,\nu )\,({\frac {
\partial }{\partial x}}\,\mathrm{u}(t, \,x, \,y, \,z, \,\nu )) +
({\frac {\partial }{\partial y}}\,\mathrm{u}(t, \,x, \,y, \,z, \,
\nu ))\,\mathrm{v}(t, \,x, \,y, \,z, \,\nu ) \\
\mbox{} + \mathrm{u}(t, \,x, \,y, \,z, \,\nu )\,({\frac {
\partial }{\partial y}}\,\mathrm{v}(t, \,x, \,y, \,z, \,\nu )) +
({\frac {\partial }{\partial z}}\,\mathrm{u}(t, \,x, \,y, \,z, \,
\nu ))\,\mathrm{w}(t, \,x, \,y, \,z, \,\nu ) \\
\mbox{} + \mathrm{u}(t, \,x, \,y, \,z, \,\nu )\,({\frac {
\partial }{\partial z}}\,\mathrm{w}(t, \,x, \,y, \,z, \,\nu )) +
({\frac {\partial }{\partial x}}\,\mathrm{p}(t, \,x, \,y, \,z, \,
\nu )) \\
\mbox{} - \nu \,(({\frac {\partial ^{2}}{\partial x^{2}}}\,
\mathrm{u}(t, \,x, \,y, \,z, \,\nu )) + ({\frac {\partial ^{2}}{
\partial y^{2}}}\,\mathrm{u}(t, \,x, \,y, \,z, \,\nu )) + (
{\frac {\partial ^{2}}{\partial z^{2}}}\,\mathrm{u}(t, \,x, \,y,
\,z, \,\nu ))) }
}
\end{maplelatex}

\end{maplegroup}
\begin{maplegroup}
\begin{mapleinput}
\mapleinline{active}{1d}{N2:=diff(v(X),t)+U(X)*diff(v(X),x)+diff(P(X),y)+diff(v(X)*u(X),x)+dif
f(v(X)*v(X),y)+diff(v(X)*w(X),z)+diff(p(X),y)-nu*(diff(v(X),x,x)+diff(
v(X),y,y)+diff(v(X),z,z));}{%
}
\end{mapleinput}

\mapleresult
\begin{maplelatex}
\mapleinline{inert}{2d}{N2 :=
diff(v(t,x,y,z,nu),t)+U(t,x,y,z,nu)*diff(v(t,x,y,z,nu),x)+diff(P(t,x,y
,z,nu),y)+diff(u(t,x,y,z,nu),x)*v(t,x,y,z,nu)+u(t,x,y,z,nu)*diff(v(t,x
,y,z,nu),x)+2*v(t,x,y,z,nu)*diff(v(t,x,y,z,nu),y)+diff(v(t,x,y,z,nu),z
)*w(t,x,y,z,nu)+v(t,x,y,z,nu)*diff(w(t,x,y,z,nu),z)+diff(p(t,x,y,z,nu)
,y)-nu*(diff(v(t,x,y,z,nu),`$`(x,2))+diff(v(t,x,y,z,nu),`$`(y,2))+diff
(v(t,x,y,z,nu),`$`(z,2)));}{%
\maplemultiline{
\mathit{N2} := ({\frac {\partial }{\partial t}}\,\mathrm{v}(t, \,
x, \,y, \,z, \,\nu )) + \mathrm{U}(t, \,x, \,y, \,z, \,\nu )\,(
{\frac {\partial }{\partial x}}\,\mathrm{v}(t, \,x, \,y, \,z, \,
\nu )) + ({\frac {\partial }{\partial y}}\,\mathrm{P}(t, \,x, \,y
, \,z, \,\nu )) \\
\mbox{} + ({\frac {\partial }{\partial x}}\,\mathrm{u}(t, \,x, \,
y, \,z, \,\nu ))\,\mathrm{v}(t, \,x, \,y, \,z, \,\nu ) + \mathrm{
u}(t, \,x, \,y, \,z, \,\nu )\,({\frac {\partial }{\partial x}}\,
\mathrm{v}(t, \,x, \,y, \,z, \,\nu )) \\
\mbox{} + 2\,\mathrm{v}(t, \,x, \,y, \,z, \,\nu )\,({\frac {
\partial }{\partial y}}\,\mathrm{v}(t, \,x, \,y, \,z, \,\nu )) +
({\frac {\partial }{\partial z}}\,\mathrm{v}(t, \,x, \,y, \,z, \,
\nu ))\,\mathrm{w}(t, \,x, \,y, \,z, \,\nu ) \\
\mbox{} + \mathrm{v}(t, \,x, \,y, \,z, \,\nu )\,({\frac {
\partial }{\partial z}}\,\mathrm{w}(t, \,x, \,y, \,z, \,\nu )) +
({\frac {\partial }{\partial y}}\,\mathrm{p}(t, \,x, \,y, \,z, \,
\nu )) \\
\mbox{} - \nu \,(({\frac {\partial ^{2}}{\partial x^{2}}}\,
\mathrm{v}(t, \,x, \,y, \,z, \,\nu )) + ({\frac {\partial ^{2}}{
\partial y^{2}}}\,\mathrm{v}(t, \,x, \,y, \,z, \,\nu )) + (
{\frac {\partial ^{2}}{\partial z^{2}}}\,\mathrm{v}(t, \,x, \,y,
\,z, \,\nu ))) }
}
\end{maplelatex}

\end{maplegroup}
\begin{maplegroup}
\begin{mapleinput}
\mapleinline{active}{1d}{N3:=diff(w(X),t)+U(X)*diff(w(X),x)+diff(w(X)*u(X),x)+diff(w(X)*v(X),y
)+diff(w(X)*w(X),z)+diff(p(X),z)-nu*(diff(w(X),x,x)+diff(w(X),y,y)+dif
f(w(X),z,z));}{%
}
\end{mapleinput}

\mapleresult
\begin{maplelatex}
\mapleinline{inert}{2d}{N3 :=
diff(w(t,x,y,z,nu),t)+U(t,x,y,z,nu)*diff(w(t,x,y,z,nu),x)+diff(u(t,x,y
,z,nu),x)*w(t,x,y,z,nu)+u(t,x,y,z,nu)*diff(w(t,x,y,z,nu),x)+diff(v(t,x
,y,z,nu),y)*w(t,x,y,z,nu)+v(t,x,y,z,nu)*diff(w(t,x,y,z,nu),y)+2*w(t,x,
y,z,nu)*diff(w(t,x,y,z,nu),z)+diff(p(t,x,y,z,nu),z)-nu*(diff(w(t,x,y,z
,nu),`$`(x,2))+diff(w(t,x,y,z,nu),`$`(y,2))+diff(w(t,x,y,z,nu),`$`(z,2
)));}{%
\maplemultiline{
\mathit{N3} := ({\frac {\partial }{\partial t}}\,\mathrm{w}(t, \,
x, \,y, \,z, \,\nu )) + \mathrm{U}(t, \,x, \,y, \,z, \,\nu )\,(
{\frac {\partial }{\partial x}}\,\mathrm{w}(t, \,x, \,y, \,z, \,
\nu )) \\
\mbox{} + ({\frac {\partial }{\partial x}}\,\mathrm{u}(t, \,x, \,
y, \,z, \,\nu ))\,\mathrm{w}(t, \,x, \,y, \,z, \,\nu ) + \mathrm{
u}(t, \,x, \,y, \,z, \,\nu )\,({\frac {\partial }{\partial x}}\,
\mathrm{w}(t, \,x, \,y, \,z, \,\nu )) \\
\mbox{} + ({\frac {\partial }{\partial y}}\,\mathrm{v}(t, \,x, \,
y, \,z, \,\nu ))\,\mathrm{w}(t, \,x, \,y, \,z, \,\nu ) + \mathrm{
v}(t, \,x, \,y, \,z, \,\nu )\,({\frac {\partial }{\partial y}}\,
\mathrm{w}(t, \,x, \,y, \,z, \,\nu )) \\
\mbox{} + 2\,\mathrm{w}(t, \,x, \,y, \,z, \,\nu )\,({\frac {
\partial }{\partial z}}\,\mathrm{w}(t, \,x, \,y, \,z, \,\nu )) +
({\frac {\partial }{\partial z}}\,\mathrm{p}(t, \,x, \,y, \,z, \,
\nu )) \\
\mbox{} - \nu \,(({\frac {\partial ^{2}}{\partial x^{2}}}\,
\mathrm{w}(t, \,x, \,y, \,z, \,\nu )) + ({\frac {\partial ^{2}}{
\partial y^{2}}}\,\mathrm{w}(t, \,x, \,y, \,z, \,\nu )) + (
{\frac {\partial ^{2}}{\partial z^{2}}}\,\mathrm{w}(t, \,x, \,y,
\,z, \,\nu ))) }
}
\end{maplelatex}

\end{maplegroup}
\begin{maplegroup}
\begin{flushleft}
{\large Equations (2.1), (2.2) \& (2.4):}
\end{flushleft}

\end{maplegroup}
\begin{maplegroup}
\begin{mapleinput}
\mapleinline{active}{1d}{eqnC:=C=0: eqnN1:=N1=0: eqnN2:=N2=0: eqnN3:=N3=0:
eqnR1:=diff(U(X),t)=0: eqnR2:=diff(U(X),x)=0: eqnR3:=diff(U(X),z)=0:
eqnR4:=diff(P(X),t)=0: eqnR5:=diff(P(X),x)=0:
eqnR6:=diff(P(X),z)=0:}{%
}
\end{mapleinput}

\end{maplegroup}
\begin{maplegroup}
\begin{mapleinput}
\mapleinline{active}{1d}{eqns:=[eqnC,eqnN1,eqnN2,eqnN3,eqnR1,eqnR2,eqnR3,eqnR4,eqnR5,eqnR6]:}{
}
\end{mapleinput}

\end{maplegroup}
\begin{maplegroup}
\begin{flushleft}
{\large Symmetry Algorithm:}
\end{flushleft}

\end{maplegroup}
\begin{maplegroup}
\begin{flushleft}
\textit{{\large a) Size of the determining system:}}
\end{flushleft}

\end{maplegroup}
\begin{maplegroup}
\begin{mapleinput}
\mapleinline{active}{1d}{detsys:=gendef(eqns,[u,v,w,p,U,P],[X]): nops(detsys[1]);}{%
}
\end{mapleinput}

\mapleresult
\begin{maplelatex}
\mapleinline{inert}{2d}{113;}{%
\[
113
\]
}
\end{maplelatex}

\end{maplegroup}
\begin{maplegroup}
\begin{flushleft}
\textit{{\large b) Solving the determining system:}}
\end{flushleft}

\end{maplegroup}
\begin{maplegroup}
\begin{mapleinput}
\mapleinline{active}{1d}{sym:=pdesolv(op(detsys));}{%
}
\end{mapleinput}

\mapleresult
\begin{maplelatex}
\mapleinline{inert}{2d}{sym := [[], [], [xi[t](sigma) =
1/nu*F_109(nu)+t/nu*F_51(nu)+2*t*F_97(nu), xi[x](sigma) =
F_108(t,nu)+x*F_51(nu)/nu+x*F_97(nu), xi[y](sigma) =
F_110(nu)+y*F_51(nu)/nu+y*F_97(nu), xi[z](sigma) =
-F_107(t,nu)+z*F_51(nu)/nu+z*F_97(nu), xi[nu](sigma) = F_51(nu),
eta[u](sigma) = F_81(y,nu)+diff(F_108(t,nu),t)-u*F_97(nu),
eta[v](sigma) = -v*F_97(nu), eta[w](sigma) =
-diff(F_107(t,nu),t)-w*F_97(nu), eta[p](sigma) =
-2*p*F_97(nu)+F_41(y,nu,U,P)-2*P*F_97(nu)+diff(F_107(t,nu),`$`(t,2))*z
-x*diff(F_108(t,nu),`$`(t,2))+1/nu*F_114(t,nu)+K*x/nu*F_51(nu)+3*K*x*F
_97(nu), eta[U](sigma) = -F_81(y,nu)-U*F_97(nu), eta[P](sigma) =
-F_41(y,nu,U,P)], [F_41(y,nu,U,P), F_114(t,nu), F_81(y,nu),
F_108(t,nu), F_107(t,nu), F_97(nu), F_51(nu), F_110(nu),
F_109(nu)]];}{%
\maplemultiline{
\mathit{sym} := [[], \,[], [{\xi _{t}}(\sigma )={\displaystyle
\frac {\mathrm{F\_109}(\nu )}{\nu }}  + {\displaystyle \frac {t\,
\mathrm{F\_51}(\nu )}{\nu }}  + 2\,t\,\mathrm{F\_97}(\nu ),  \\
{\xi _{x}}(\sigma )=\mathrm{F\_108}(t, \,\nu ) + {\displaystyle
\frac {x\,\mathrm{F\_51}(\nu )}{\nu }}  + x\,\mathrm{F\_97}(\nu )
,  \\
{\xi _{y}}(\sigma )=\mathrm{F\_110}(\nu ) + {\displaystyle
\frac {y\,\mathrm{F\_51}(\nu )}{\nu }}  + y\,\mathrm{F\_97}(\nu )
,  \\
{\xi _{z}}(\sigma )= - \mathrm{F\_107}(t, \,\nu ) +
{\displaystyle \frac {z\,\mathrm{F\_51}(\nu )}{\nu }}  + z\,
\mathrm{F\_97}(\nu ), \,{\xi _{\nu }}(\sigma )=\mathrm{F\_51}(\nu
 ),  \\
{\eta _{u}}(\sigma )=\mathrm{F\_81}(y, \,\nu ) + ({\frac {
\partial }{\partial t}}\,\mathrm{F\_108}(t, \,\nu )) - u\,
\mathrm{F\_97}(\nu ), \,{\eta _{v}}(\sigma )= - v\,\mathrm{F\_97}
(\nu ),  \\
{\eta _{w}}(\sigma )= - ({\frac {\partial }{\partial t}}\,
\mathrm{F\_107}(t, \,\nu )) - w\,\mathrm{F\_97}(\nu ), {\eta _{p}
}(\sigma )= - 2\,p\,\mathrm{F\_97}(\nu ) + \mathrm{F\_41}(y, \,
\nu , \,U, \,P) \\
\mbox{} - 2\,P\,\mathrm{F\_97}(\nu ) + ({\frac {\partial ^{2}}{
\partial t^{2}}}\,\mathrm{F\_107}(t, \,\nu ))\,z - x\,({\frac {
\partial ^{2}}{\partial t^{2}}}\,\mathrm{F\_108}(t, \,\nu )) +
{\displaystyle \frac {\mathrm{F\_114}(t, \,\nu )}{\nu }}  \\
\mbox{} + {\displaystyle \frac {K\,x\,\mathrm{F\_51}(\nu )}{\nu }
}  + 3\,K\,x\,\mathrm{F\_97}(\nu ), \,{\eta _{U}}(\sigma )= -
\mathrm{F\_81}(y, \,\nu ) - U\,\mathrm{F\_97}(\nu ),  \\
{\eta _{P}}(\sigma )= - \mathrm{F\_41}(y, \,\nu , \,U, \,P)], [
\mathrm{F\_41}(y, \,\nu , \,U, \,P), \,\mathrm{F\_114}(t, \,\nu )
, \,\mathrm{F\_81}(y, \,\nu ), \,\mathrm{F\_108}(t, \,\nu ),  \\
\mathrm{F\_107}(t, \,\nu ), \,\mathrm{F\_97}(\nu ), \,\mathrm{
F\_51}(\nu ), \,\mathrm{F\_110}(\nu ), \,\mathrm{F\_109}(\nu )]]
 }
}
\end{maplelatex}

\end{maplegroup}
\begin{maplegroup}
\begin{flushleft}
{\large Redefinition of group functions as used in (2.10) and
(2.11):}
\end{flushleft}

\end{maplegroup}
\begin{maplegroup}
\begin{mapleinput}
\mapleinline{active}{1d}{F_51(nu):=(2*a1(nu)-a4(nu))*nu; F_97(nu):=-(a1(nu)-a4(nu));
F_108(t,nu):=f1(t,nu); F_110(nu):=a3(nu); F_107(t,nu):=-f2(t,nu);
F_109(nu):=nu*a5(nu); F_81(y,nu):=-F(y,nu); F_114(t,nu):=nu*f3(t,nu);
F_41(y,nu,U,P):=-2*(a1(nu)-a4(nu))*P-g2(y,nu,U,P);}{%
}
\end{mapleinput}

\mapleresult
\begin{maplelatex}
\mapleinline{inert}{2d}{F_51(nu) := (2*a1(nu)-a4(nu))*nu;}{%
\[
\mathrm{F\_51}(\nu ) := (2\,\mathrm{a1}(\nu ) - \mathrm{a4}(\nu )
)\,\nu
\]
}
\end{maplelatex}

\begin{maplelatex}
\mapleinline{inert}{2d}{F_97(nu) := -a1(nu)+a4(nu);}{%
\[
\mathrm{F\_97}(\nu ) :=  - \mathrm{a1}(\nu ) + \mathrm{a4}(\nu )
\]
}
\end{maplelatex}

\begin{maplelatex}
\mapleinline{inert}{2d}{F_108(t,nu) := f1(t,nu);}{%
\[
\mathrm{F\_108}(t, \,\nu ) := \mathrm{f1}(t, \,\nu )
\]
}
\end{maplelatex}

\begin{maplelatex}
\mapleinline{inert}{2d}{F_110(nu) := a3(nu);}{%
\[
\mathrm{F\_110}(\nu ) := \mathrm{a3}(\nu )
\]
}
\end{maplelatex}

\begin{maplelatex}
\mapleinline{inert}{2d}{F_107(t,nu) := -f2(t,nu);}{%
\[
\mathrm{F\_107}(t, \,\nu ) :=  - \mathrm{f2}(t, \,\nu )
\]
}
\end{maplelatex}

\begin{maplelatex}
\mapleinline{inert}{2d}{F_109(nu) := nu*a5(nu);}{%
\[
\mathrm{F\_109}(\nu ) := \nu \,\mathrm{a5}(\nu )
\]
}
\end{maplelatex}

\begin{maplelatex}
\mapleinline{inert}{2d}{F_81(y,nu) := -F(y,nu);}{%
\[
\mathrm{F\_81}(y, \,\nu ) :=  - \mathrm{F}(y, \,\nu )
\]
}
\end{maplelatex}

\begin{maplelatex}
\mapleinline{inert}{2d}{F_114(t,nu) := nu*f3(t,nu);}{%
\[
\mathrm{F\_114}(t, \,\nu ) := \nu \,\mathrm{f3}(t, \,\nu )
\]
}
\end{maplelatex}

\begin{maplelatex}
\mapleinline{inert}{2d}{F_41(y,nu,U,P) := -2*(a1(nu)-a4(nu))*P-g2(y,nu,U,P);}{%
\[
\mathrm{F\_41}(y, \,\nu , \,U, \,P) := -2\,(\mathrm{a1}(\nu ) -
\mathrm{a4}(\nu ))\,P - \mathrm{g2}(y, \,\nu , \,U, \,P)
\]
}
\end{maplelatex}

\end{maplegroup}
\begin{maplegroup}
\begin{flushleft}
{\large Final Result (identical to (2.10) and (2.11)):}
\end{flushleft}

\end{maplegroup}
\begin{maplegroup}
\begin{mapleinput}
\mapleinline{active}{1d}{simplify(sym[3,2]); simplify(sym[3,3]); simplify(sym[3,4]);
simplify(sym[3,1]); simplify(sym[3,5]); simplify(sym[3,6]);
simplify(sym[3,7]); simplify(sym[3,8]); simplify(sym[3,9]);
simplify(sym[3,10]); simplify(sym[3,11]);}{%
}
\end{mapleinput}

\mapleresult
\begin{maplelatex}
\mapleinline{inert}{2d}{xi[x](sigma) = f1(t,nu)+x*a1(nu);}{%
\[
{\xi _{x}}(\sigma )=\mathrm{f1}(t, \,\nu ) + x\,\mathrm{a1}(\nu )
\]
}
\end{maplelatex}

\begin{maplelatex}
\mapleinline{inert}{2d}{xi[y](sigma) = a3(nu)+y*a1(nu);}{%
\[
{\xi _{y}}(\sigma )=\mathrm{a3}(\nu ) + y\,\mathrm{a1}(\nu )
\]
}
\end{maplelatex}

\begin{maplelatex}
\mapleinline{inert}{2d}{xi[z](sigma) = f2(t,nu)+z*a1(nu);}{%
\[
{\xi _{z}}(\sigma )=\mathrm{f2}(t, \,\nu ) + z\,\mathrm{a1}(\nu )
\]
}
\end{maplelatex}

\begin{maplelatex}
\mapleinline{inert}{2d}{xi[t](sigma) = a5(nu)+t*a4(nu);}{%
\[
{\xi _{t}}(\sigma )=\mathrm{a5}(\nu ) + t\,\mathrm{a4}(\nu )
\]
}
\end{maplelatex}

\begin{maplelatex}
\mapleinline{inert}{2d}{xi[nu](sigma) = (2*a1(nu)-a4(nu))*nu;}{%
\[
{\xi _{\nu }}(\sigma )=(2\,\mathrm{a1}(\nu ) - \mathrm{a4}(\nu ))
\,\nu
\]
}
\end{maplelatex}

\begin{maplelatex}
\mapleinline{inert}{2d}{eta[u](sigma) = -F(y,nu)+diff(f1(t,nu),t)+u*a1(nu)-u*a4(nu);}{%
\[
{\eta _{u}}(\sigma )= - \mathrm{F}(y, \,\nu ) + ({\frac {
\partial }{\partial t}}\,\mathrm{f1}(t, \,\nu )) + u\,\mathrm{a1}
(\nu ) - u\,\mathrm{a4}(\nu )
\]
}
\end{maplelatex}

\begin{maplelatex}
\mapleinline{inert}{2d}{eta[v](sigma) = v*(a1(nu)-a4(nu));}{%
\[
{\eta _{v}}(\sigma )=v\,(\mathrm{a1}(\nu ) - \mathrm{a4}(\nu ))
\]
}
\end{maplelatex}

\begin{maplelatex}
\mapleinline{inert}{2d}{eta[w](sigma) = diff(f2(t,nu),t)+w*a1(nu)-w*a4(nu);}{%
\[
{\eta _{w}}(\sigma )=({\frac {\partial }{\partial t}}\,\mathrm{f2
}(t, \,\nu )) + w\,\mathrm{a1}(\nu ) - w\,\mathrm{a4}(\nu )
\]
}
\end{maplelatex}

\vspace{-0.5em}
\begin{maplelatex}
\mapleinline{inert}{2d}{eta[p](sigma) =
2*p*a1(nu)-2*p*a4(nu)-g2(y,nu,U,P)-diff(f2(t,nu)
,`$`(t,2))*z-x*diff(f1(t,nu),`$`(t,2))+f3(t,nu)-K*x*a1(nu)+2*K*x*a4(nu
);}{%
\maplemultiline{ {\eta _{p}}(\sigma )=2\,p\,\mathrm{a1}(\nu ) -
2\,p\,\mathrm{a4}( \nu ) - \mathrm{g2}(y, \,\nu , \,U, \,P) -
({\frac {\partial ^{2}}{
\partial t^{2}}}\,\mathrm{f2}(t, \,\nu ))\,z \\
\mbox{} - x\,({\frac {\partial ^{2}}{\partial t^{2}}}\,\mathrm{f1
}(t, \,\nu )) + \mathrm{f3}(t, \,\nu ) - K\,x\,\mathrm{a1}(\nu )
 + 2\,K\,x\,\mathrm{a4}(\nu ) }
}
\end{maplelatex}

\begin{maplelatex}
\mapleinline{inert}{2d}{eta[U](sigma) = F(y,nu)+U*a1(nu)-U*a4(nu);}{%
\[
{\eta _{U}}(\sigma )=\mathrm{F}(y, \,\nu ) + U\,\mathrm{a1}(\nu )
 - U\,\mathrm{a4}(\nu )
\]
}
\end{maplelatex}

\begin{maplelatex}
\mapleinline{inert}{2d}{eta[P](sigma) = 2*P*a1(nu)-2*P*a4(nu)+g2(y,nu,U,P);}{%
\[
{\eta _{P}}(\sigma )=  2\,P\,\mathrm{a1}(\nu ) - 2\,P\,\mathrm{
a4}(\nu ) + \mathrm{g2}(y, \,\nu , \,U, \,P)
\]
}
\end{maplelatex}

\end{maplegroup}
\begin{maplegroup}
\begin{mapleinput}
\end{mapleinput}

\end{maplegroup}

\vspace{1em}\noindent {\bf IIIb. DESOLV-II-Package
\textbf{\emph{including}} the $\pmb{\mathscr{P}_{ij}}$-equations
(\ref{131228:1928}):}
\begin{maplegroup}
\begin{flushleft}
{\large Header:}
\end{flushleft}

\end{maplegroup}
\begin{maplegroup}
\begin{mapleinput}
\mapleinline{active}{1d}{restart: read "Desolv-V5R5.mpl": with(desolv):}{%
}
\end{mapleinput}

\mapleresult
\begin{maplelatex}
\mapleinline{inert}{2d}{`DESOLVII_V5R5 (March-2011)(c) by Dr. K.
T. Vu, Dr. J. Carminati and
Miss. G. Jefferson`;}{%
\maplemultiline{ \mathit{\phantom{xxxxxxxx} DESOLVII\_V5R5\ (March-2011)(c)} \\
\mathit{by\ Dr.\ K.\ T.\ Vu,\ Dr.\ J.\ Carminati\ and\ Miss.\ G.\
Jefferson} }
}
\end{maplelatex}

\end{maplegroup}
\begin{maplegroup}
\begin{flushleft}
{\large Definitions:}
\end{flushleft}

\end{maplegroup}
\begin{maplegroup}
\begin{mapleinput}
\mapleinline{active}{1d}{alias(sigma=(t,x,y,z,nu,u,v,w,p,U,P)): X:=(t,x,y,z,nu): }{%
}
\end{mapleinput}

\end{maplegroup}
\begin{maplegroup}
\begin{mapleinput}
\mapleinline{active}{1d}{C:=diff(u(X),x)+diff(v(X),y)+diff(w(X),z);}{%
}
\end{mapleinput}

\mapleresult
\begin{maplelatex}
\mapleinline{inert}{2d}{C :=
diff(u(t,x,y,z,nu),x)+diff(v(t,x,y,z,nu),y)+diff(w(t,x,y,z,nu),z);}{%
\[
C := ({\frac {\partial }{\partial x}}\,\mathrm{u}(t, \,x, \,y, \,
z, \,\nu )) + ({\frac {\partial }{\partial y}}\,\mathrm{v}(t, \,x
, \,y, \,z, \,\nu )) + ({\frac {\partial }{\partial z}}\,\mathrm{
w}(t, \,x, \,y, \,z, \,\nu ))
\]
}
\end{maplelatex}

\end{maplegroup}
\begin{maplegroup}
\begin{mapleinput}
\mapleinline{active}{1d}{N1:=diff(u(X),t)+U(X)*diff(u(X),x)+v(X)*diff(U(X),y)-(K+nu*diff(U(X),
y,y))+diff(u(X)*u(X),x)+diff(u(X)*v(X),y)+diff(u(X)*w(X),z)+diff(p(X),
x)-nu*(diff(u(X),x,x)+diff(u(X),y,y)+diff(u(X),z,z));}{%
}
\end{mapleinput}

\mapleresult
\begin{maplelatex}
\mapleinline{inert}{2d}{N1 :=
diff(u(t,x,y,z,nu),t)+U(t,x,y,z,nu)*diff(u(t,x,y,z,nu),x)+v(t,x,y,z,nu
)*diff(U(t,x,y,z,nu),y)-K-nu*diff(U(t,x,y,z,nu),`$`(y,2))+2*u(t,x,y,z,
nu)*diff(u(t,x,y,z,nu),x)+diff(u(t,x,y,z,nu),y)*v(t,x,y,z,nu)+u(t,x,y,
z,nu)*diff(v(t,x,y,z,nu),y)+diff(u(t,x,y,z,nu),z)*w(t,x,y,z,nu)+u(t,x,
y,z,nu)*diff(w(t,x,y,z,nu),z)+diff(p(t,x,y,z,nu),x)-nu*(diff(u(t,x,y,z
,nu),`$`(x,2))+diff(u(t,x,y,z,nu),`$`(y,2))+diff(u(t,x,y,z,nu),`$`(z,2
)));}{%
\maplemultiline{
\mathit{N1} := ({\frac {\partial }{\partial t}}\,\mathrm{u}(t, \,
x, \,y, \,z, \,\nu )) + \mathrm{U}(t, \,x, \,y, \,z, \,\nu )\,(
{\frac {\partial }{\partial x}}\,\mathrm{u}(t, \,x, \,y, \,z, \,
\nu )) \\
\mbox{} + \mathrm{v}(t, \,x, \,y, \,z, \,\nu )\,({\frac {
\partial }{\partial y}}\,\mathrm{U}(t, \,x, \,y, \,z, \,\nu )) -
K - \nu \,({\frac {\partial ^{2}}{\partial y^{2}}}\,\mathrm{U}(t
, \,x, \,y, \,z, \,\nu )) \\
\mbox{} + 2\,\mathrm{u}(t, \,x, \,y, \,z, \,\nu )\,({\frac {
\partial }{\partial x}}\,\mathrm{u}(t, \,x, \,y, \,z, \,\nu )) +
({\frac {\partial }{\partial y}}\,\mathrm{u}(t, \,x, \,y, \,z, \,
\nu ))\,\mathrm{v}(t, \,x, \,y, \,z, \,\nu ) \\
\mbox{} + \mathrm{u}(t, \,x, \,y, \,z, \,\nu )\,({\frac {
\partial }{\partial y}}\,\mathrm{v}(t, \,x, \,y, \,z, \,\nu )) +
({\frac {\partial }{\partial z}}\,\mathrm{u}(t, \,x, \,y, \,z, \,
\nu ))\,\mathrm{w}(t, \,x, \,y, \,z, \,\nu ) \\
\mbox{} + \mathrm{u}(t, \,x, \,y, \,z, \,\nu )\,({\frac {
\partial }{\partial z}}\,\mathrm{w}(t, \,x, \,y, \,z, \,\nu )) +
({\frac {\partial }{\partial x}}\,\mathrm{p}(t, \,x, \,y, \,z, \,
\nu )) \\
\mbox{} - \nu \,(({\frac {\partial ^{2}}{\partial x^{2}}}\,
\mathrm{u}(t, \,x, \,y, \,z, \,\nu )) + ({\frac {\partial ^{2}}{
\partial y^{2}}}\,\mathrm{u}(t, \,x, \,y, \,z, \,\nu )) + (
{\frac {\partial ^{2}}{\partial z^{2}}}\,\mathrm{u}(t, \,x, \,y,
\,z, \,\nu ))) }
}
\end{maplelatex}

\end{maplegroup}
\begin{maplegroup}
\begin{mapleinput}
\mapleinline{active}{1d}{N2:=diff(v(X),t)+U(X)*diff(v(X),x)+diff(P(X),y)+diff(v(X)*u(X),x)+dif
f(v(X)*v(X),y)+diff(v(X)*w(X),z)+diff(p(X),y)-nu*(diff(v(X),x,x)+diff(
v(X),y,y)+diff(v(X),z,z));}{%
}
\end{mapleinput}

\mapleresult
\begin{maplelatex}
\mapleinline{inert}{2d}{N2 :=
diff(v(t,x,y,z,nu),t)+U(t,x,y,z,nu)*diff(v(t,x,y,z,nu),x)+diff(P(t,x,y
,z,nu),y)+diff(u(t,x,y,z,nu),x)*v(t,x,y,z,nu)+u(t,x,y,z,nu)*diff(v(t,x
,y,z,nu),x)+2*v(t,x,y,z,nu)*diff(v(t,x,y,z,nu),y)+diff(v(t,x,y,z,nu),z
)*w(t,x,y,z,nu)+v(t,x,y,z,nu)*diff(w(t,x,y,z,nu),z)+diff(p(t,x,y,z,nu)
,y)-nu*(diff(v(t,x,y,z,nu),`$`(x,2))+diff(v(t,x,y,z,nu),`$`(y,2))+diff
(v(t,x,y,z,nu),`$`(z,2)));}{%
\maplemultiline{
\mathit{N2} := ({\frac {\partial }{\partial t}}\,\mathrm{v}(t, \,
x, \,y, \,z, \,\nu )) + \mathrm{U}(t, \,x, \,y, \,z, \,\nu )\,(
{\frac {\partial }{\partial x}}\,\mathrm{v}(t, \,x, \,y, \,z, \,
\nu )) + ({\frac {\partial }{\partial y}}\,\mathrm{P}(t, \,x, \,y
, \,z, \,\nu )) \\
\mbox{} + ({\frac {\partial }{\partial x}}\,\mathrm{u}(t, \,x, \,
y, \,z, \,\nu ))\,\mathrm{v}(t, \,x, \,y, \,z, \,\nu ) + \mathrm{
u}(t, \,x, \,y, \,z, \,\nu )\,({\frac {\partial }{\partial x}}\,
\mathrm{v}(t, \,x, \,y, \,z, \,\nu )) \\
\mbox{} + 2\,\mathrm{v}(t, \,x, \,y, \,z, \,\nu )\,({\frac {
\partial }{\partial y}}\,\mathrm{v}(t, \,x, \,y, \,z, \,\nu )) +
({\frac {\partial }{\partial z}}\,\mathrm{v}(t, \,x, \,y, \,z, \,
\nu ))\,\mathrm{w}(t, \,x, \,y, \,z, \,\nu ) \\
\mbox{} + \mathrm{v}(t, \,x, \,y, \,z, \,\nu )\,({\frac {
\partial }{\partial z}}\,\mathrm{w}(t, \,x, \,y, \,z, \,\nu )) +
({\frac {\partial }{\partial y}}\,\mathrm{p}(t, \,x, \,y, \,z, \,
\nu )) \\
\mbox{} - \nu \,(({\frac {\partial ^{2}}{\partial x^{2}}}\,
\mathrm{v}(t, \,x, \,y, \,z, \,\nu )) + ({\frac {\partial ^{2}}{
\partial y^{2}}}\,\mathrm{v}(t, \,x, \,y, \,z, \,\nu )) + (
{\frac {\partial ^{2}}{\partial z^{2}}}\,\mathrm{v}(t, \,x, \,y,
\,z, \,\nu ))) }
}
\end{maplelatex}

\end{maplegroup}
\begin{maplegroup}
\begin{mapleinput}
\mapleinline{active}{1d}{N3:=diff(w(X),t)+U(X)*diff(w(X),x)+diff(w(X)*u(X),x)+diff(w(X)*v(X),y
)+diff(w(X)*w(X),z)+diff(p(X),z)-nu*(diff(w(X),x,x)+diff(w(X),y,y)+dif
f(w(X),z,z));}{%
}
\end{mapleinput}

\mapleresult
\begin{maplelatex}
\mapleinline{inert}{2d}{N3 :=
diff(w(t,x,y,z,nu),t)+U(t,x,y,z,nu)*diff(w(t,x,y,z,nu),x)+diff(u(t,x,y
,z,nu),x)*w(t,x,y,z,nu)+u(t,x,y,z,nu)*diff(w(t,x,y,z,nu),x)+diff(v(t,x
,y,z,nu),y)*w(t,x,y,z,nu)+v(t,x,y,z,nu)*diff(w(t,x,y,z,nu),y)+2*w(t,x,
y,z,nu)*diff(w(t,x,y,z,nu),z)+diff(p(t,x,y,z,nu),z)-nu*(diff(w(t,x,y,z
,nu),`$`(x,2))+diff(w(t,x,y,z,nu),`$`(y,2))+diff(w(t,x,y,z,nu),`$`(z,2
)));}{%
\maplemultiline{
\mathit{N3} := ({\frac {\partial }{\partial t}}\,\mathrm{w}(t, \,
x, \,y, \,z, \,\nu )) + \mathrm{U}(t, \,x, \,y, \,z, \,\nu )\,(
{\frac {\partial }{\partial x}}\,\mathrm{w}(t, \,x, \,y, \,z, \,
\nu )) \\
\mbox{} + ({\frac {\partial }{\partial x}}\,\mathrm{u}(t, \,x, \,
y, \,z, \,\nu ))\,\mathrm{w}(t, \,x, \,y, \,z, \,\nu ) + \mathrm{
u}(t, \,x, \,y, \,z, \,\nu )\,({\frac {\partial }{\partial x}}\,
\mathrm{w}(t, \,x, \,y, \,z, \,\nu )) \\
\mbox{} + ({\frac {\partial }{\partial y}}\,\mathrm{v}(t, \,x, \,
y, \,z, \,\nu ))\,\mathrm{w}(t, \,x, \,y, \,z, \,\nu ) + \mathrm{
v}(t, \,x, \,y, \,z, \,\nu )\,({\frac {\partial }{\partial y}}\,
\mathrm{w}(t, \,x, \,y, \,z, \,\nu )) \\
\mbox{} + 2\,\mathrm{w}(t, \,x, \,y, \,z, \,\nu )\,({\frac {
\partial }{\partial z}}\,\mathrm{w}(t, \,x, \,y, \,z, \,\nu )) +
({\frac {\partial }{\partial z}}\,\mathrm{p}(t, \,x, \,y, \,z, \,
\nu )) \\
\mbox{} - \nu \,(({\frac {\partial ^{2}}{\partial x^{2}}}\,
\mathrm{w}(t, \,x, \,y, \,z, \,\nu )) + ({\frac {\partial ^{2}}{
\partial y^{2}}}\,\mathrm{w}(t, \,x, \,y, \,z, \,\nu )) + (
{\frac {\partial ^{2}}{\partial z^{2}}}\,\mathrm{w}(t, \,x, \,y,
\,z, \,\nu ))) }
}
\end{maplelatex}

\end{maplegroup}
\begin{maplegroup}
\begin{flushleft}
{\large Equations (2.1), (2.2) \& (2.4) including the "velocity
product equations" (2.3):}
\end{flushleft}

\end{maplegroup}
\begin{maplegroup}
\begin{mapleinput}
\mapleinline{active}{1d}{eqnC:=C=0: eqnN1:=N1=0: eqnN2:=N2=0: eqnN3:=N3=0:
eqnP11:=2*N1*u(X)=0: eqnP22:=2*N2*v(X)=0: eqnP33:=2*N3*w(X)=0:
eqnP12:=N1*v(X)+N2*u(X)=0: eqnP13:=N1*w(X)+N3*u(X)=0:
eqnP23:=N2*w(X)+N3*v(X)=0:
eqnR1:=diff(U(X),t)=0: eqnR2:=diff(U(X),x)=0: eqnR3:=diff(U(X),z)=0:
eqnR4:=diff(P(X),t)=0: eqnR5:=diff(P(X),x)=0:
eqnR6:=diff(P(X),z)=0:  }{%
}
\end{mapleinput}

\end{maplegroup}
\begin{maplegroup}
\begin{mapleinput}
\mapleinline{active}{1d}{eqns:=[eqnC,eqnN1,eqnN2,eqnN3,eqnP11,eqnP22,eqnP33,eqnP12,eqnP13,eqnP
23,eqnR1,eqnR2,eqnR3,eqnR4,eqnR5,eqnR6]:}{%
}
\end{mapleinput}

\end{maplegroup}
\begin{maplegroup}
\begin{flushleft}
{\large Symmetry Algorithm:}
\end{flushleft}

\end{maplegroup}
\begin{maplegroup}
\begin{flushleft}
\textit{{\large a) Size of the determining system:}}
\end{flushleft}

\end{maplegroup}
\begin{maplegroup}
\begin{mapleinput}
\mapleinline{active}{1d}{detsys:=gendef(eqns,[u,v,w,p,U,P],[X]): nops(detsys[1]);}{%
}
\end{mapleinput}

\mapleresult
\begin{maplelatex}
\mapleinline{inert}{2d}{113;}{%
\[
113
\]
}
\end{maplelatex}

\end{maplegroup}

\begin{maplegroup}
\begin{flushleft}
\textit{{\large Note that since by default the}}
{\large\texttt{gendef}}\textit{{\large -routine checks on
redundancy, it automatically returns, as a consequence, the
same number of determining equations as in the previous
non-extended case IIIa.}}
\end{flushleft}

\end{maplegroup}

\vspace{1em}
\begin{maplegroup}
\begin{flushleft}
\textit{{\large b) Solving the determining system:}}
\end{flushleft}

\end{maplegroup}
\begin{maplegroup}
\begin{mapleinput}
\mapleinline{active}{1d}{sym:=pdesolv(op(detsys));}{%
}
\end{mapleinput}

\mapleresult
\begin{maplelatex}
\mapleinline{inert}{2d}{sym := [[], [], [xi[t](sigma) =
1/nu*F_109(nu)+t/nu*F_71(nu)+2*F_97(nu)*t, xi[x](sigma) =
F_108(t,nu)+x*F_71(nu)/nu+x*F_97(nu), xi[y](sigma) =
F_110(nu)+y*F_71(nu)/nu+y*F_97(nu), xi[z](sigma) =
-F_107(t,nu)+z*F_71(nu)/nu+z*F_97(nu), xi[nu](sigma) = F_71(nu),
eta[u](sigma) = F_60(y,nu)+diff(F_108(t,nu),t)-u*F_97(nu),
eta[v](sigma) = -v*F_97(nu), eta[w](sigma) =
-diff(F_107(t,nu),t)-w*F_97(nu), eta[p](sigma) =
-2*p*F_97(nu)-F_51(y,nu,U,P)-2*P*F_97(nu)+diff(F_107(t,nu),`$`(t,2))*z
-x*diff(F_108(t,nu),`$`(t,2))+1/nu*F_114(t,nu)+K*x/nu*F_71(nu)+3*K*x*F
_97(nu), eta[U](sigma) = -F_60(y,nu)-U*F_97(nu), eta[P](sigma) =
F_51(y,nu,U,P)], [F_51(y,nu,U,P), F_114(t,nu), F_60(y,nu),
F_108(t,nu), F_107(t,nu), F_97(nu), F_71(nu), F_110(nu),
F_109(nu)]];}{%
\maplemultiline{
\mathit{sym} := [[], \,[], [{\xi _{t}}(\sigma )={\displaystyle
\frac {\mathrm{F\_109}(\nu )}{\nu }}  + {\displaystyle \frac {t\,
\mathrm{F\_71}(\nu )}{\nu }}  + 2\,\mathrm{F\_97}(\nu )\,t,  \\
{\xi _{x}}(\sigma )=\mathrm{F\_108}(t, \,\nu ) + {\displaystyle
\frac {x\,\mathrm{F\_71}(\nu )}{\nu }}  + x\,\mathrm{F\_97}(\nu )
,  \\
{\xi _{y}}(\sigma )=\mathrm{F\_110}(\nu ) + {\displaystyle
\frac {y\,\mathrm{F\_71}(\nu )}{\nu }}  + y\,\mathrm{F\_97}(\nu )
,  \\
{\xi _{z}}(\sigma )= - \mathrm{F\_107}(t, \,\nu ) +
{\displaystyle \frac {z\,\mathrm{F\_71}(\nu )}{\nu }}  + z\,
\mathrm{F\_97}(\nu ), \,{\xi _{\nu }}(\sigma )=\mathrm{F\_71}(\nu
 ),  \\
{\eta _{u}}(\sigma )=\mathrm{F\_60}(y, \,\nu ) + ({\frac {
\partial }{\partial t}}\,\mathrm{F\_108}(t, \,\nu )) - u\,
\mathrm{F\_97}(\nu ), \,{\eta _{v}}(\sigma )= - v\,\mathrm{F\_97}
(\nu ),  \\
{\eta _{w}}(\sigma )= - ({\frac {\partial }{\partial t}}\,
\mathrm{F\_107}(t, \,\nu )) - w\,\mathrm{F\_97}(\nu ), {\eta _{p}
}(\sigma )= - 2\,p\,\mathrm{F\_97}(\nu ) - \mathrm{F\_51}(y, \,
\nu , \,U, \,P) \\
\mbox{} - 2\,P\,\mathrm{F\_97}(\nu ) + ({\frac {\partial ^{2}}{
\partial t^{2}}}\,\mathrm{F\_107}(t, \,\nu ))\,z - x\,({\frac {
\partial ^{2}}{\partial t^{2}}}\,\mathrm{F\_108}(t, \,\nu )) +
{\displaystyle \frac {\mathrm{F\_114}(t, \,\nu )}{\nu }}  \\
\mbox{} + {\displaystyle \frac {K\,x\,\mathrm{F\_71}(\nu )}{\nu }
}  + 3\,K\,x\,\mathrm{F\_97}(\nu ), \,{\eta _{U}}(\sigma )= -
\mathrm{F\_60}(y, \,\nu ) - U\,\mathrm{F\_97}(\nu ),  \\
{\eta _{P}}(\sigma )=\mathrm{F\_51}(y, \,\nu , \,U, \,P)], [
\mathrm{F\_51}(y, \,\nu , \,U, \,P), \,\mathrm{F\_114}(t, \,\nu )
, \,\mathrm{F\_60}(y, \,\nu ), \,\mathrm{F\_108}(t, \,\nu ),  \\
\mathrm{F\_107}(t, \,\nu ), \,\mathrm{F\_97}(\nu ), \,\mathrm{
F\_71}(\nu ), \,\mathrm{F\_110}(\nu ), \,\mathrm{F\_109}(\nu )]]
 }
}
\end{maplelatex}

\end{maplegroup}
\begin{maplegroup}
\begin{flushleft}
{\large Redefinition of group functions as in (2.10) and (2.11):}
\end{flushleft}

\end{maplegroup}
\begin{maplegroup}
\begin{mapleinput}
\mapleinline{active}{1d}{F_71(nu):=(2*a1(nu)-a4(nu))*nu; F_97(nu):=-(a1(nu)-a4(nu));
F_108(t,nu):=f1(t,nu); F_110(nu):=a3(nu); F_107(t,nu):=-f2(t,nu);
F_109(nu):=nu*a5(nu); F_60(y,nu):=-F(y,nu); F_114(t,nu):=nu*f3(t,nu);
F_51(y,nu,U,P):=2*(a1(nu)-a4(nu))*P+g2(y,nu,U,P);}{%
}
\end{mapleinput}

\mapleresult
\begin{maplelatex}
\mapleinline{inert}{2d}{F_71(nu) := (2*a1(nu)-a4(nu))*nu;}{%
\[
\mathrm{F\_71}(\nu ) := (2\,\mathrm{a1}(\nu ) - \mathrm{a4}(\nu )
)\,\nu
\]
}
\end{maplelatex}

\begin{maplelatex}
\mapleinline{inert}{2d}{F_97(nu) := -a1(nu)+a4(nu);}{%
\[
\mathrm{F\_97}(\nu ) :=  - \mathrm{a1}(\nu ) + \mathrm{a4}(\nu )
\]
}
\end{maplelatex}

\begin{maplelatex}
\mapleinline{inert}{2d}{F_108(t,nu) := f1(t,nu);}{%
\[
\mathrm{F\_108}(t, \,\nu ) := \mathrm{f1}(t, \,\nu )
\]
}
\end{maplelatex}

\begin{maplelatex}
\mapleinline{inert}{2d}{F_110(nu) := a3(nu);}{%
\[
\mathrm{F\_110}(\nu ) := \mathrm{a3}(\nu )
\]
}
\end{maplelatex}

\begin{maplelatex}
\mapleinline{inert}{2d}{F_107(t,nu) := -f2(t,nu);}{%
\[
\mathrm{F\_107}(t, \,\nu ) :=  - \mathrm{f2}(t, \,\nu )
\]
}
\end{maplelatex}

\begin{maplelatex}
\mapleinline{inert}{2d}{F_109(nu) := nu*a5(nu);}{%
\[
\mathrm{F\_109}(\nu ) := \nu \,\mathrm{a5}(\nu )
\]
}
\end{maplelatex}

\begin{maplelatex}
\mapleinline{inert}{2d}{F_60(y,nu) := -F(y,nu);}{%
\[
\mathrm{F\_60}(y, \,\nu ) :=  - \mathrm{F}(y, \,\nu )
\]
}
\end{maplelatex}

\begin{maplelatex}
\mapleinline{inert}{2d}{F_114(t,nu) := nu*f3(t,nu);}{%
\[
\mathrm{F\_114}(t, \,\nu ) := \nu \,\mathrm{f3}(t, \,\nu )
\]
}
\end{maplelatex}

\begin{maplelatex}
\mapleinline{inert}{2d}{F_51(y,nu,U,P) := 2*(a1(nu)-a4(nu))*P+g2(y,nu,U,P);}{%
\[
\mathrm{F\_51}(y, \,\nu , \,U, \,P) :=   2\,(\mathrm{a1}(\nu )
 - \mathrm{a4}(\nu ))\,P + \mathrm{g2}(y, \,\nu , \,U, \,P)
\]
}
\end{maplelatex}

\end{maplegroup}
\begin{maplegroup}
\begin{flushleft}
{\large Final Result (still identical to (2.10) and (2.11)):}
\end{flushleft}

\end{maplegroup}
\begin{maplegroup}
\begin{mapleinput}
\mapleinline{active}{1d}{simplify(sym[3,2]); simplify(sym[3,3]); simplify(sym[3,4]);
simplify(sym[3,1]); simplify(sym[3,5]); simplify(sym[3,6]);
simplify(sym[3,7]); simplify(sym[3,8]); simplify(sym[3,9]);
simplify(sym[3,10]); simplify(sym[3,11]);}{%
}
\end{mapleinput}

\mapleresult
\begin{maplelatex}
\mapleinline{inert}{2d}{xi[x](sigma) = f1(t,nu)+x*a1(nu);}{%
\[
{\xi _{x}}(\sigma )=\mathrm{f1}(t, \,\nu ) + x\,\mathrm{a1}(\nu )
\]
}
\end{maplelatex}

\begin{maplelatex}
\mapleinline{inert}{2d}{xi[y](sigma) = a3(nu)+y*a1(nu);}{%
\[
{\xi _{y}}(\sigma )=\mathrm{a3}(\nu ) + y\,\mathrm{a1}(\nu )
\]
}
\end{maplelatex}

\begin{maplelatex}
\mapleinline{inert}{2d}{xi[z](sigma) = f2(t,nu)+z*a1(nu);}{%
\[
{\xi _{z}}(\sigma )=\mathrm{f2}(t, \,\nu ) + z\,\mathrm{a1}(\nu )
\]
}
\end{maplelatex}

\begin{maplelatex}
\mapleinline{inert}{2d}{xi[t](sigma) = a5(nu)+t*a4(nu);}{%
\[
{\xi _{t}}(\sigma )=\mathrm{a5}(\nu ) + t\,\mathrm{a4}(\nu )
\]
}
\end{maplelatex}

\begin{maplelatex}
\mapleinline{inert}{2d}{xi[nu](sigma) = (2*a1(nu)-a4(nu))*nu;}{%
\[
{\xi _{\nu }}(\sigma )=(2\,\mathrm{a1}(\nu ) - \mathrm{a4}(\nu ))
\,\nu
\]
}
\end{maplelatex}

\begin{maplelatex}
\mapleinline{inert}{2d}{eta[u](sigma) = -F(y,nu)+diff(f1(t,nu),t)+u*a1(nu)-u*a4(nu);}{%
\[
{\eta _{u}}(\sigma )= - \mathrm{F}(y, \,\nu ) + ({\frac {
\partial }{\partial t}}\,\mathrm{f1}(t, \,\nu )) + u\,\mathrm{a1}
(\nu ) - u\,\mathrm{a4}(\nu )
\]
}
\end{maplelatex}

\begin{maplelatex}
\mapleinline{inert}{2d}{eta[v](sigma) = v*(a1(nu)-a4(nu));}{%
\[
{\eta _{v}}(\sigma )=v\,(\mathrm{a1}(\nu ) - \mathrm{a4}(\nu ))
\]
}
\end{maplelatex}

\begin{maplelatex}
\mapleinline{inert}{2d}{eta[w](sigma) = diff(f2(t,nu),t)+w*a1(nu)-w*a4(nu);}{%
\[
{\eta _{w}}(\sigma )=({\frac {\partial }{\partial t}}\,\mathrm{f2
}(t, \,\nu )) + w\,\mathrm{a1}(\nu ) - w\,\mathrm{a4}(\nu )
\]
}
\end{maplelatex}

\begin{maplelatex}
\mapleinline{inert}{2d}{eta[p](sigma) =
2*p*a1(nu)-2*p*a4(nu)-g2(y,nu,U,P)-diff(f2(t,nu)
,`$`(t,2))*z-x*diff(f1(t,nu),`$`(t,2))+f3(t,nu)-K*x*a1(nu)+2*K*x*a4(nu
);}{%
\maplemultiline{ {\eta _{p}}(\sigma )=2\,p\,\mathrm{a1}(\nu ) -
2\,p\,\mathrm{a4}( \nu ) - \mathrm{g2}(y, \,\nu , \,U, \,P) -
({\frac {\partial ^{2}}{
\partial t^{2}}}\,\mathrm{f2}(t, \,\nu ))\,z \\
\mbox{} - x\,({\frac {\partial ^{2}}{\partial t^{2}}}\,\mathrm{f1
}(t, \,\nu )) + \mathrm{f3}(t, \,\nu ) - K\,x\,\mathrm{a1}(\nu )
 + 2\,K\,x\,\mathrm{a4}(\nu ) }
}
\end{maplelatex}

\begin{maplelatex}
\mapleinline{inert}{2d}{eta[U](sigma) = F(y,nu)+U*a1(nu)-U*a4(nu);}{%
\[
{\eta _{U}}(\sigma )=\mathrm{F}(y, \,\nu ) + U\,\mathrm{a1}(\nu )
 - U\,\mathrm{a4}(\nu )
\]
}
\end{maplelatex}

\begin{maplelatex}
\mapleinline{inert}{2d}{eta[P](sigma) = 2*P*a1(nu)-2*P*a4(nu)+g2(y,nu,U,P);}{%
\[
{\eta _{P}}(\sigma )=  2\,P\,\mathrm{a1}(\nu ) - 2\,P\,\mathrm{
a4}(\nu ) + \mathrm{g2}(y, \,\nu , \,U, \,P)
\]
}
\end{maplelatex}

\end{maplegroup}
\begin{maplegroup}
\begin{mapleinput}
\end{mapleinput}

\end{maplegroup}

\bibliographystyle{jfm}
\bibliography{BibDaten}
%
% Manually written citations in the text, which need to be included
% PLEASE DO NOT DELETE OR COMMENT OUT THE NEXT LINE --- Thank you!
\phantom{\cite{Kolmogorov41.1,Kolmogorov41.2,Kolmogorov41.3,Kolmogorov62}}
\end{document}